\documentclass[longauth]{aa} 
\usepackage{graphicx}
\usepackage{txfonts}
\usepackage[colorlinks=true,citecolor=blue]{hyperref}
\usepackage{multirow}
\usepackage{amsmath,amstext}
\usepackage[T1]{fontenc}
\usepackage{color}
\usepackage{comment}
\usepackage{caption}
\DeclareCaptionFormat{cont}{#1 (cont.)#2#3\par}
\usepackage{tabularx}
\usepackage{subcaption}
\usepackage{dblfloatfix}

\DeclareRobustCommand{\ion}[2]{%
\relax\ifmmode
\ifx\testbx\f@series
{\mathbf{#1\,\mathsc{#2}}}\else
{\mathrm{#1\,\mathsc{#2}}}\fi
\else\textup{#1\,{\mdseries\textsc{#2}}}%
\fi}

\newcommand{\eg}{e.g.,\ }

\newcommand{\OIII}{O~{\sc iii}}
\newcommand{\CII}{C~{\sc ii}}

\newcommand{\CaII}{Ca~{\sc ii}}

\newcommand{\FeIII}{Fe~{\sc iii}}

\newcommand{\Nifs}{$^{56}$Ni}

\def\OIII5007Hb{[{\ion{O}{iii}}] $\lambda5007$/H$\beta$}
\def\ratioR23{([\ion{O}{ii}] $\lambda$3727 +[\ion{O}{iii}] $\lambda\lambda$4959,5007)/H$\beta$}
\def\R23{${\rm R}_{23}$}
\def\dS23{${\rm S}_{23}$}

\def\ratioS23{([\ion{S}{2}] $\lambda \lambda$6717,31 +[\ion{S}{3}] $\lambda\lambda$9069,9532)/H$\beta$}

\def\OIIId{[{\ion{O}{iii}}] $\lambda$5007}

\def\OIII{[{\ion{O}{iii}}]}

\def\NIId{[{\ion{N}{ii}}] $\lambda$6584}

\newcommand{\Had}{\rm{H}$\alpha$}

\newcommand{\Hbd}{\rm{H}$\beta$}

\newcommand{\SCSN}{2003fg-like SNe}
\newcommand{\DmB}{$\Delta${\rm{m}$_{15}$($B$)}}
\newcommand{\sBV}{s$_{BV}$}

\newcommand{\orcid}[1]{\href{https://orcid.org/#1}{\includegraphics[width=10pt]{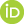}}}

\usepackage[normalem]{ulem}
\newcommand{\revdel}[1]{\textcolor{red}{}}
\newcommand{\revadd}[1]{#1}

\begin{document}

\title{The host galaxies of 2003fg-like type Ia supernovae} 
\titlerunning{2003fg-like SN Ia host galaxies}

\author{Llu\'is Galbany\inst{1,2}\fnmsep\thanks{\email{lgalbany@ice.csic.es}}\orcid{0000-0002-1296-6887},
Chris Ashall\inst{3}\orcid{0000-0002-5221-7557},
Jing Lu\inst{4},
Jason T. Hinkle\inst{5,6,7}\orcid{0000-0001-9668-2920},
Willem B. Hoogendam\inst{3}\orcid{0000-0003-3953-9532},\\
Benjamin J. Shappee\inst{3}\orcid{0000-0003-4631-1149},
Alaa Alburai\inst{1,2},
Ramon Sanfeliu\inst{1,2},
Joseph P. Anderson\inst{8}\orcid{0000-0003-0227-3451},
Eddie Baron\inst{9,10},\\
Chris Burns\inst{11}\orcid{0000-0003-4625-6629},
Georgios Dimitriadis\inst{12}\orcid{0000-0001-9494-179X},
Inma Dom\'inguez\inst{13}\thanks{$\dagger$ Deceased},
Peter Hoeflich\inst{14}\orcid{0000-0002-4338-6586},
Eric Y. Hsiao\inst{14}\orcid{0000-0003-1039-2928},\\
Hanindyo Kuncarayakti\inst{15,16},
Joseph D. Lyman\inst{17}\orcid{0000-0002-3464-0642},
Nidia Morrell\inst{18},
Mark M. Phillips\inst{18},\\
Sebasti\'an F. S\'anchez\inst{19},
Maximilian D. Stritzinger\inst{20}\orcid{0000-0002-5571-1833},
\mbox{Nicholas B. Suntzeff\inst{21}}
}
\authorrunning{Galbany et al.}

\institute{
Institute of Space Sciences (ICE, CSIC), Campus UAB, Carrer de Can Magrans, s/n, E-08193 Barcelona, Spain.
\and Institut d’Estudis Espacials de Catalunya (IEEC), E-08034 Barcelona, Spain.
\and Institute for Astronomy, University of Hawai`i at M\={a}noa, 2680 Woodlawn Dr., Honolulu, HI 96822, USA.
\and Department of Physics and Astronomy, Michigan State University, East Lansing, MI 48824, USA.
\and NHFP Einstein Fellow
\and Department of Astronomy, University of Illinois Urbana-Champaign, 1002 West Green Street, Urbana, IL 61801, USA 
\and NSF-Simons AI Institute for the Sky (SkAI), 172 E. Chestnut St., Chicago, IL 60611, USA
\and European Southern Observatory, Alonso de C\'ordova 3107, Vitacura, Casilla 19001, Santiago, Chile
\and Planetary Science Institute, 1700 East Fort Lowell Road, Suite 106, Tucson, AZ 85719-2395, USA
\and Hamburger Sternwarte, Gojenbergsweg 112, D-21029 Hamburg, Germany.
\and The Observatories of the Carnegie Institution for Science, 813 Santa Barbara Street, Pasadena, CA 91101, USA.
\and School of Physics and Astronomy, Lancaster University, Lancaster, LA1 4YB, UK
\and Departamento de F\'isica Te\'orica y del Cosmos, Universidad de Granada, E-18071 Granada, Spain.
\and Department of Physics, Florida State University, 77 Chieftan Way, Tallahassee, FL 32306, USA.
\and Tuorla Observatory, Department of Physics and Astronomy, FI-20014 University of Turku, Finland.
\and Finnish Centre for Astronomy with ESO (FINCA), FI-20014, University of Turku, Finland.
\and Department of Physics, University of Warwick, Coventry, CV4 7AL, UK.
\and Carnegie Observatories, Las Campanas Observatory, Casilla 601, La Serena, Chile.
\and Instituto de Astronom\'ia, Universidad Nacional Aut\'onoma de M\'exico, A.P. 70-264, 04510 México, D.F., Mexico.
\and Department of Physics and Astronomy, Aarhus University, Ny Munkegade 120, DK-8000 Aarhus C, Denmark.
\and George P. and Cynthia Woods Mitchell Institute for Fundamental Physics \& Astronomy, Texas A\&M University, Department of Physics and Astronomy, 4242 TAMU, College Station, TX 77843.
}

\date{Received \today; accepted XXX}

\abstract{
SN~2003fg-like events are a peculiar type Ia supernova (SN~Ia) subtype characterized by broader light curves, higher near-infrared luminosities, and stronger carbon absorptions at early times.
Here we present observations of the largest compilation of 2003fg-like SN Ia host galaxies to date, obtained with Integral Field Spectroscopy (IFS).
For 20 objects, we study both the global host-galaxy properties and, for the first time for a sizeable sample, the local environment at the SN position.
Globally, 2003fg-like SNe~Ia occur in galaxies with lower stellar mass, lower oxygen abundance, and marginally higher specific star-formation rate (sSFR) than those of normal SNe~Ia, although their hosts are not as extreme in such properties as superluminous SNe, nor representative of metal-poor dwarf-galaxy samples.
Locally, the SN positions show lower star-formation-rate, stellar-mass surface densities\revadd{, and lower sSFR,} than normal SN~Ia environments, consistent with a significant preference for the outskirts of their hosts, while their \revdel{sSFR and }stellar age indicators are typical. The most distinctive local property is metallicity, with 2003fg-like SNe~Ia occupying the most metal-poor environments among SNe~Ia. 
We also find a tentative positive correlation between the light-curve width and the oxygen abundance for 2003fg-like events.
Our results imply that 2003fg-like SNe~Ia arise from the merger of two white dwarfs (WDs) or the core-degenerate scenario, but disfavor the single, rapidly rotating super-M$_{ch}$ C-O WD progenitor, as this channel requires a young stellar population that we do not observe at the SN positions.
The preference of 2003fg-like SNe~Ia for low-metallicity environments suggests that they may have been more common in the early Universe. Since they are overluminous after light-curve standardization (negative Hubble residuals), they will inevitably enter distant SN~Ia samples used for cosmology, introducing potential systematic uncertainties in future dark energy experiments.
A better understanding of these peculiar SNe~Ia is therefore crucial, not only to decipher their progenitors and explosion mechanism, but also to reduce systematic errors in cosmological parameters.
}

\keywords{supernovae: general}

\maketitle

\section{Introduction} 
\label{sect:intro}

Type Ia supernovae (SNe~Ia) are luminous standardizable candles that can be used as cosmic rulers to map out the expansion history of the cosmos \citep{Riess98,Perlmutter99}. They are known to arise from the thermonuclear disruption of at least one carbon-oxygen (C-O) white dwarf (WD) in a binary system \citep{Whelan73,Iben84,Hoeflich:Khokhlov:96}. Yet the exact nature of their progenitor system(s) and explosion mechanism(s) is still debated \citep[see \eg][]{Maoz14,Blondin17,Hoeflich17,Livio18}. 

Observationally, SNe~Ia follow empirical relationships such as the luminosity-width relationship (LWR; \citealt{Phillips93,Hamuy96,Phillips99}), which is fundamental for their use as cosmological distance indicators. To date, several subtypes of thermonuclear SNe~Ia have been discovered that fall within different regions of the LWR (for a detailed review see \citealt{Taubenberger17}). Understanding the physics and explosions of outliers from the LWR offers a chance to comprehend the explosion mechanisms in these systems (e.g. \citealt{Hoeflich17,2026A&A...706A.252B}), establish how they link back to normal SNe~Ia, and determine if these subtypes provide a possible contamination for future high-redshift cosmological experiments, such as the Nancy Grace Roman Space Telescope \citep{2021arXiv211103081R}. 

One of the rarest subtypes of SNe~Ia is 2003fg-like events. Several of these objects have also been referred to as ``super-Chandrasekhar mass ($M_{ch}$)'' SNe~Ia due to their high luminosity, as suggested by high \Nifs\ and ejecta masses compared to typical SNe~Ia \citep{2006Natur.443..308H}. Their rate is about 30$\pm20$ yr$^{-1}$ Gpc$^{-3}$ h$^3_{70}$, corresponding to $\approx$0.2\% of the total population of SNe~Ia \citep{2024MNRAS.530.5016D,2025A&A...694A..10D, Desai26_Ia_rates}. 
However, while the ``super-Chandrasekhar'' label assumes these objects are over-luminous, it has since become apparent that not all of them are, as many factors may affect their luminosity \citep[see \eg][]{2021ApJ...920..107L, Ashall21}. Since the defining characteristics of the class are spectroscopic rather than a common luminosity (see below), the ``super-Chandrasekhar'' designation is potentially misleading. We therefore follow the convention of naming the subclass after the first discovered SN, that is 2003fg.
To date, a few tens of \SCSN\ have been reported in the literature. These include SN~2003fg (aka SNLS-03D3bb; \citealt{2006Natur.443..308H}), SN~2006gz \citep{2007ApJ...669L..17H,2009ApJ...690.1745M}, SN~2007if \citep{Scalzo10,Yuan10}, SN~2009dc \citep{Tanaka10,Yamanaka09,Silverman11,2011MNRAS.412.2735T}, SN~2012dn \citep{Chakradhari14,Brown14,Parrent16,Nagao18,2019MNRAS.488.5473T}, LSQ14fmg \citep{2020ApJ...900..140H}, ASASSN-15pz \citep{2019ApJ...880...35C}, ASASSN-15hy \citep{2021ApJ...920..107L}, SN~2020esm \citep{2022ApJ...927...78D}, SN~2021qvo \citep{Paniagua2026}, SN~2022pul \citep{2024ApJ...960...88S}, as well as a further five which were observed by the \textit{Carnegie Supernova Project} (CSP; \citealt{2006PASP..118....2H,2019PASP..131a4001P}) presented in \cite{Ashall21}. 

Until recently, identifying a 2003fg-like SN had been difficult. Using maximum light spectra alone, they can be confused with other subtypes such as \mbox{1991T-like} SNe~Ia. However, it is now apparent that there are a few ubiquitous properties of 2003fg-like SNe. These are: weak or no $i$-band secondary maximum \citep{Ashall21}, a high early-time ultraviolet (UV) flux, persistently blue UV colors, and non-monotonic rising light curve bumps \citep{Hoogendam2024}, weak early-time \CaII\ lines, and low ionization in the nebular phase \citep{2019MNRAS.488.5473T}. Other properties include: a broad optical light curve shape (\DmB$\lesssim 1.3$~mag)\footnote{\DmB\ is the difference in $B$-band magnitude between maximum light and +15~d \citep{Phillips93}.}, the $i$-band peaks after the phase of $B$-band maximum, a lack of strong \FeIII\ features in the early spectra, a peak $H$-band absolute magnitude brighter than $-$19~mag, carbon absorption at early times ($-$10~d from maximum light), and no clear $H$-band break at +10~d from maximum light \citep{Ashall21}. For the purpose of this work, we use the criteria defined in \citet{2020ApJ...895L...3A} to determine our sample, where \DmB$\lesssim$ 1.3 mag or \sBV$\gtrsim$0.8\footnote{\sBV\ is the difference in time between the $B$-band maximum and the maximum in the $B-V$ color light-curve, normalized to 30~days \citep{Burns14}.}, and the time of peak brightness in the $i$-band happens after that in the $B$-band ($t_i^{max}-t_B^{max}>0$).

The origin of 2003fg-like SNe~Ia is still debated, with several progenitor scenarios and explosion mechanisms having been proposed.
These include: the explosion of a C-O WD that exceeds the classical non-rotating $M_{ch}$ limit due to rapid rotation or high magnetic fields \citep{Yoon05, Das13}; the violent merger of two WDs whose total mass may exceed the $M_{ch}$ \citep{Scalzo10}; and the explosion of a C-O WD inside dense H/He-free circumstellar material (CSM; \citealt{Hachinger12, Noebauer16}). This latter model is also referred to as the envelope model \citep{Hoeflich:Khokhlov:96} and shows some of the most promising results \citep{2020ApJ...900..140H, 2021ApJ...920..107L}. \citet{Ashall21} analysed a large homogeneous sample of \SCSN\ and determined a number of unique correlations (e.g., the correlation between the pseudo-equivalent width of the \CII\ feature and light curve width), suggesting that the envelope model is a promising route to be explored. It has been suggested that such an envelope model may be consistent with the explosion of a degenerate core of an Asymptotic Giant Branch (AGB) star which has lost its outer H/He envelope in a core degenerate scenario \citep{Kashi11, 2020ApJ...900..140H, 2017hsn..book.1151H, 2021ApJ...920..107L}.
Further evidence for this interpretation comes from continuum-polarization measurements of two objects, which are consistent with nearly spherical explosions \citep{Tanaka10,Cikota19}. In contrast, broadband imaging polarimetry suggests that some events may be intrinsically aspherical. However, because these measurements integrate the polarization across an entire photometric band, it is difficult to determine whether the signal arises from global ejecta asymmetry, an asymmetric distribution of specific chemical species, circumstellar dust, or interstellar polarization \citep{2024A&A...687L..19N}.

\begin{table*}
\caption{General properties of the 20 2003fg-like SNe Ia in our sample. The last two columns give the projected SN–host separation and the host-normalized separation $d_{\rm DLR}$ (separation in units of the host's directional light radius toward the SN), derived from the {\sc HOSTPHOT} elliptical apertures.}
\label{table:prop}
\resizebox{\textwidth}{!}{%
\begin{tabular}{lcccccccccc}
\hline\hline
SN & z$_{\rm helio}$ & RA (2000) & DEC (2000) & IFS source & Date of Obs. & Exposure [s] & Airmass & Spatial Res. ["] & Sep. ["] & d$_{\rm DLR}$\\
\hline
2003fg               & 0.2440 & 14:16:18.78 &  +52:14:55.39 & PISCO   & 2023-04-15 & 3$\times$1000 & 1.3 & 2.70 & 3.51 & 0.83 \\
2006gz               & 0.0235 & 18:10:26.33 &  +30:59:44.41 & PISCO   & 2019-05-09 & 3$\times$900  & 1.1 & 2.70 & 31.07 & 1.02 \\
2007if               & 0.0742 & 01:10:51.37 &  +15:27:39.89 & AMUSING & 2020-10-27 & 4$\times$612  & 1.6 & 0.89 & 0.66 & 0.68 \\
2009dc               & 0.0214 & 15:51:12.12 &  +25:42:28.01 & PISCO   & 2018-02-14 & 3$\times$1200 & 1.4 & 2.70 & 24.94 & 2.12 \\
LSQ12gpw                & 0.0506 & 03:12:58.24 &$-$11:42:40.07 & AMUSING & 2018-11-19 & 4$\times$618  & 1.6 & 1.05 & 8.47 & 0.61 \\
2012dn               & 0.0102 & 20:23:36.26 &$-$28:16:43.39 & AMUSING & 2026-04-16 & 4$\times$619  & 1.4 & 1.66 & 36.32 & 0.81 \\
2013ao               & 0.0435 & 11:44:44.74 &$-$20:31:41.09 & AMUSING & 2015-05-25 & 4$\times$698  & 1.2 & 0.97 & 1.27 & 0.49 \\
LSQ14fmg                & 0.0649 & 22:16:46.10 &  +15:21:14.15 & AMUSING & 2017-08-04 & 4$\times$701  & 1.3 & 1.40 & 0.73 & 0.13 \\
CSS140126-120307-010132 & 0.0772 & 12:03:06.90 &$-$01:01:31.92 & AMUSING & 2020-12-27 & 4$\times$619  & 1.5 & 1.24 & 1.56 & 0.65 \\ 
CSS140501-170414+174839 & 0.0797 & 17:04:13.69 &  +17:48:39.40 & PISCO   & 2017-05-26 & 3$\times$1200 & 1.1 & 2.70 & 19.92 & 1.63 \\
2015M                & 0.0231 & 13:00:32.30 &  +27:58:41.09 & AMUSING & 2021-01-26 & 4$\times$614  & 1.8 & 1.02 & 7.30 & 1.26 \\ 
ASASSN-15hy             & 0.0185 & 20:10:02.35 &$-$00:44:21.20 & AMUSING & 2017-07-19 & 4$\times$701  & 1.2 & 1.66 & 13.90 & 0.98 \\
ASASSN-15pz             & 0.0148 & 03:08:48.48 &$-$35:13:51.24 & AMUSING & 2024-10-08 & 4$\times$791  & 1.6 & 1.47 & 20.72 & 0.65 \\ 
2016gxp              & 0.0177 & 00:14:34.58 &  +48:15:08.03 & PISCO   & 2023-12-15 & 3$\times$900  & 1.1 & 2.70 & 12.89 & 0.75 \\
2020esm              & 0.0362 & 14:07:18.26 &$-$05:07:37.67 & PISCO   & 2023-04-15 & 3$\times$1000 & 1.4 & 2.70 & 16.05 & 0.99 \\
2020hvf              & 0.0058 & 11:21:26.45 &  +03:00:52.85 & PISCO   & 2023-04-15 & 3$\times$1000 & 1.3 & 2.70 & 22.03 & 1.07 \\
2020krv              & 0.0519 & 21:56:39.94 &  +41:54:25.41 & PISCO   & 2024-12-23 & 3$\times$900  & 1.2 & 2.70 & 4.87 & 0.63 \\
2020sme              & 0.0451 & 02:44:20.82 &  +14:55:16.68 & PISCO   & 2024-12-25 & 3$\times$1000 & 1.1 & 2.70 & 8.61 & 1.10 \\
2021zny              & 0.0266 & 02:03:35.80 &  +15:44:33.36 & PISCO   & 2023-12-17 & 3$\times$900  & 1.2 & 2.70 & 23.25 & 0.80 \\
2022pul              & 0.0030 & 12:26:48.85 &  +08:26:55.32 & AMUSING & 2026-04-06 & 4$\times$612  & 1.9 & 0.99 & 131.75 & 1.65 \\
\hline
\end{tabular}
}
\end{table*}

Unlike for a majority of other SNe~Ia, many \SCSN\ have strong rising light curve ``bumps'' (different from ``excess'' flux emission; see discussion in \citealp{Hoogendam2024}). Recent work has identified strong bump features in SNe 2020hvf \citep{2021ApJ...923L...8J}, 2021qvo \citep{Paniagua2026}, 2021zny \citep{2023MNRAS.521.1162D}, and 2022ilv \citep{2023ApJ...943L..20S}, along with the potentially related 2002es-like SNe~Ia. 
The 2002es-like SNe~Ia are a distinct, sub-luminous class with cool photospheres and low ejecta velocities \citep{2012ApJ...751..142G}, but they share with 2003fg-like events the presence of early light-curve bumps and are speculated to arise from related progenitor channels, possibly a common core-degenerate or double-degenerate origin \citep{Hoogendam2024}.
These few \SCSN\ with well-sampled early light curves show a prevalence of such features that is higher than in the normal SN~Ia population, and such early bumps provide otherwise unavailable insight into the explosion environment and progenitor scenario. Current possible models include the interaction between the SN ejecta and the surrounding material near the progenitor, possibly via collision with a detached shell of circumstellar matter expelled in prior mass-loss episodes \citep[e.g.][]{1982ApJ...258..790C, 2011ApJ...729L...6C, 2023MNRAS.522.6035M}, the presence of a thick disk or torus of material resulting from a previous binary interaction \citep{2015MNRAS.447.2803L} or from material thrown off during the merging process \citep{Inoue2026}. Distinguishing between these scenarios requires densely sampled multi-band photometry beginning within hours of explosion, as the colour evolution and timescale of the bump carry information about the geometry and composition of the surrounding material. The relatively high frequency of early bumps in \SCSN, compared to the general SN~Ia population \citep{Hoogendam2024}, lends additional support to the envelope or core-degenerate model \citep{2020ApJ...900..140H, 2021ApJ...920..107L}, in which residual material near the exploding WD is a natural expectation.

The host galaxy and local properties of SNe may encode information about the age, metallicity, and nature of the progenitor systems (e.g. \citealt{2018ApJ...855..107G, 2018A&A...613A..35K}). In the case of 2003fg-like SNe~Ia, there have been only a few studies of their host galaxies. \citet{2011ApJ...733....3C} studied SN~2007if and found that the host was metal-poor compared to normal SNe~Ia. 
\citet{2011ApJ...737L..24K} analysed the hosts of four \SCSN\ and, assuming radial metallicity gradients, argued that the metallicity at the SN locations is even lower than the global values of the host galaxies.
Finally, \citet{2011MNRAS.412.2735T} suggested that all over-luminous SNe~Ia may come from low-mass host galaxies. The previous indications that 2003fg-like SNe explode in low-metallicity environments \citep{2020ApJ...900..140H, 2021ApJ...920..107L} may suggest that \SCSN\ would have been more common in the early Universe, because as one goes to higher redshift the average metallicity of the Universe is lower \cite{2019A&ARv..27....3M}. In addition, as 2003fg-like SNe~Ia are also overluminous once their brightness is standardized (negative Hubble residuals; \citealt{Ashall21}), they have the potential to cause a bias in future dark energy experiments. Over the past decade, the number of \SCSN\ has increased, and there has been no study of the host galaxies of a large sample of \SCSN\ as a group. It is therefore timely to analyse the global and local properties of their host galaxies in bulk.

Here we present the global and local properties of twenty host galaxies of \SCSN\ from Integral Field Spectroscopy (IFS). In Sect. \ref{sect:sample} we present the sample and observations, followed by their analysis in Sect. \ref{sect:Analysis}. In Sect. \ref{sect:Results}, we present the global properties of the hosts, followed by the local properties. Finally, the discussion and conclusions are presented in Sect. \ref{sect:diss} and \ref{sect:conc}, respectively. 


\section{Sample selection and observations} \label{sect:sample}

\subsection{Sample selection}

As noted in \cite{2020ApJ...900..140H}, the simplest way of distinguishing members of the 2003fg-like SN group from slow-declining normal SNe~Ia or 1991T-like objects is the timing of the $i$-band primary maximum occurring after that in the $B$ band, and the lack of\revadd{,} or a weak\revadd{,} secondary maximum in the $i$ band. This was demonstrated in the color stretch ($s_{BV}$) vs. time of maximum in the $i$-band with respect to the $B$-band ($t_i^{max}-t_B^{max}$) diagram presented in \cite{2020ApJ...895L...3A}, where all 2003fg-like SNe~Ia clustered at $s_{BV}>0.8$ and $t_i^{max}-t_B^{max}>0$. Not surprisingly, it was also consistent with the results from \cite{2014ApJ...795..142G} who found, by trying to photometrically identify sub-luminous SNe~Ia through independently fitting blue ($Bg$) and red ($ri$) filters, that 2003fg-like SNe~Ia clustered at a different location of the $\chi^2_{blue}$-$\chi^2_{red}$ diagram. Since 2003fg-like SNe~Ia do not show the secondary maximum similar to 91bg-like, they measured smaller $\chi^2_{red}$ when fit with a 91bg-like template. In addition, both subtypes show the luminosity peak in the redder bands later than in the blue bands \citep{2020ApJ...895L...3A}.

Following this simple classification, a few objects previously classified as 2003fg-like SNe~Ia in the literature have been excluded in this work. As an example, although SN~2009dr was initially classified as a possible 2003fg-like SN Ia by \cite{2009ATel.2037....1Q}, it was later reclassified as SN~Ic \citep{2020ApJ...892..153M,2016MNRAS.458.2973P}. This object, SN~2004gu, and SNF20080723-012 \citep{2012ApJ...757...12S} were also included in the seven 2003fg-like SNe~Ia sample of \cite{2011MNRAS.412.2735T}, based on several observational properties, including a relatively flat \ion{Si}{ii} velocity evolution and a sudden drop in velocity around day 20 past maximum. SN~2004gu was similar in peak brightness, decline rate, and spectral features to the 2003fg-like SN~2006gz, but the $iYJH$-band light curves of SN~2004gu all showed strong secondary maxima, and it had a negative $t_i^{max}$-$t_B^{max}$. SN~2004gu shares more similarities with the 1999aa-like SN~Ia subgroup \citep{2022ApJ...938...47P, 2024ApJS..273...16P}. In this work, we adopt the method of \cite{2020ApJ...895L...3A} to identify \SCSN, and therefore none of these three objects are treated as 2003fg-like SNe~Ia. We note that they were similarly excluded in the 2003fg-like SN sample of \cite{2011ApJ...737L..24K}.

\begin{figure*}[t]
\centering
\includegraphics[width=\textwidth]{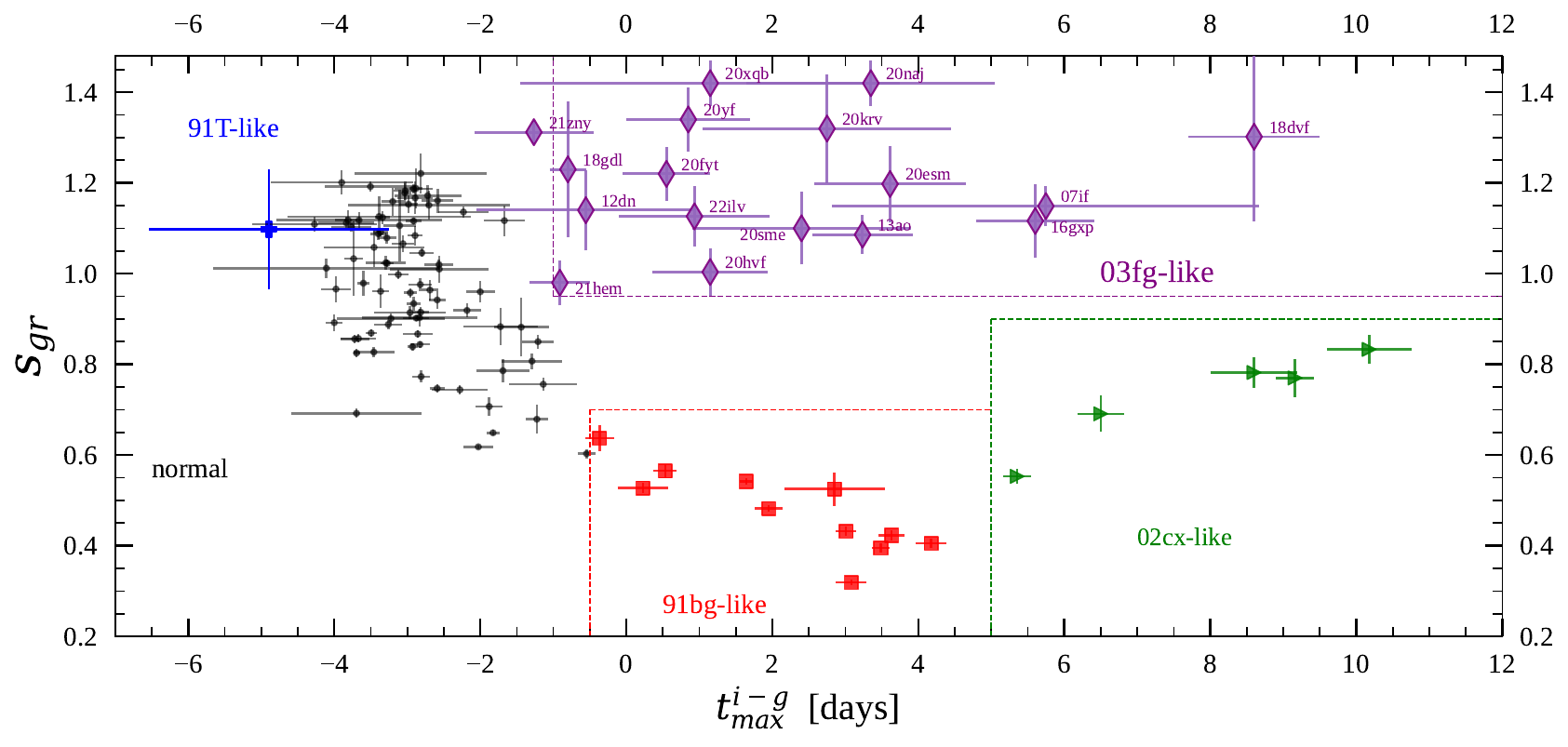}
\caption{$g$-band version of the time of the $i$-band maximum relative to the $g$-band maximum vs. $s_{gr}$ diagram presented in \cite{Ashall21} that defines regions where different SNe~Ia subtypes fall. All new candidates fall in the 2003fg-like SN (top-right) region except SN~2021zny, which is consistent with it within 1$\sigma$.}
\label{fig:03fg}
\end{figure*}

We obtained IFS of all 13 2003fg-like SN host galaxies presented in \cite{Ashall21}. Additionally, we determined whether some other objects reported through to the end of December 2022 could be included within the same definition of 2003fg-like SNe~Ia. We considered the 11 objects reported in the volume-limited ($z\lesssim 0.06$) sample in the Zwicky Transient Facility 2nd data release \citep{2025A&A...694A..10D} as resembling 2003fg-like SNe~Ia
(2018dvf, \citealt{2019ApJ...886..152Y}; 
2018gdl; 
2020fyt; 
2020hvf, \citealt{2021ApJ...923L...8J}; 
2020krv; 
2020naj; 
2020sme; 
2020xqb; 
2020yjf; 
2020abcu; 
2020esm, \citealt{2022ApJ...927...78D}), 
five more reported in TNS as SN~Ia-SC
(SN~2022ilv \citealt{2023ApJ...943L..20S};
SN~2021uhh; 
SN~2021zny, \citealt{2023MNRAS.521.1162D};
SN~2021aagz; 
SN~2022rge), 
SN~2022pul \citep{2024ApJ...960...88S},
the recently published SN~2021hem \citep{2026A&A...706A.252B},
and the unpublished SN~2016gxp (Stritzinger in prep.; priv. comm.).
Since most of these SNe were not observed in the $B$-band, we instead used the $s_{gr}$ color-stretch parameter and the difference between the $i$- and $g$-band times of maximum light ($t_{max}^{i-g}$).
For objects without available $i$-band observations, we used the time of maximum light in the ATLAS $o$-band\footnote{ATLAS light curves were obtained from the forced photometry service: \href{https://fallingstar-data.com/forcedphot/}{https://fallingstar-data.com/forcedphot/}.} and applied a conversion between $t^i_{max}$ and $t^o_{max}$. This relation was derived by fitting a sample of 100 SNe~Ia from ZTF DR2 \citep{2025A&A...694A...1R} with SNooPy\revadd{ \citep{2011AJ....141...19B}}, selecting those that had both $i$- and $o$-band data available in addition to $gr$. We identified a linear correlation in the $t_{max}^{i-g}$ vs $t_{max}^{o-g}$ plane and used it to estimate $t_{max}^{i-g}$ for our \SCSN\ without $i$-band. A systematic bias of $\sim$2.24 days was added to the estimated values, along with the measured uncertainty on $t_{max}^o$. As can be seen in Fig. \ref{fig:03fg}, we were able to include 14 of the 19 objects in the diagnostic, and all fall (within the errors) in the 2003fg-like SN region of the \cite{Ashall21} diagram, so we consider all of them as 2003fg-like SNe~Ia. 

Further IFS observations were acquired for six of these 14 objects: 
SN~2016gxp, 
SN~2020esm, 
SN~2020hvf, 
SN~2020krv,
SN~2020sme, and
SN~2021zny.
Additionally, we obtained observations at the location of SN~2022pul. Its light curve, reported by \cite{2024ApJ...960...88S}, does not cover epochs around maximum, making it difficult to measure the needed parameters reliably to add this object to Fig. \ref{fig:03fg}. 
However, we still consider SN~2022pul a 2003fg-like SN~Ia on the basis of its reported spectroscopic similarities to the subgroup \citep{2024ApJ...960...88S}. Its absence from Fig.~\ref{fig:03fg} reflects a lack of maximum-light photometry rather than a failure to meet the photometric criteria. Unlike SN~2004gu and SNF20080723-012, which are excluded because the measured criteria place them outside the 2003fg-like region, SN~2022pul simply cannot be tested with the available data.

\begin{figure*}[t]
\centering
\includegraphics[trim= 0cm 0cm 0cm 0cm,clip=True, width=0.97\textwidth]{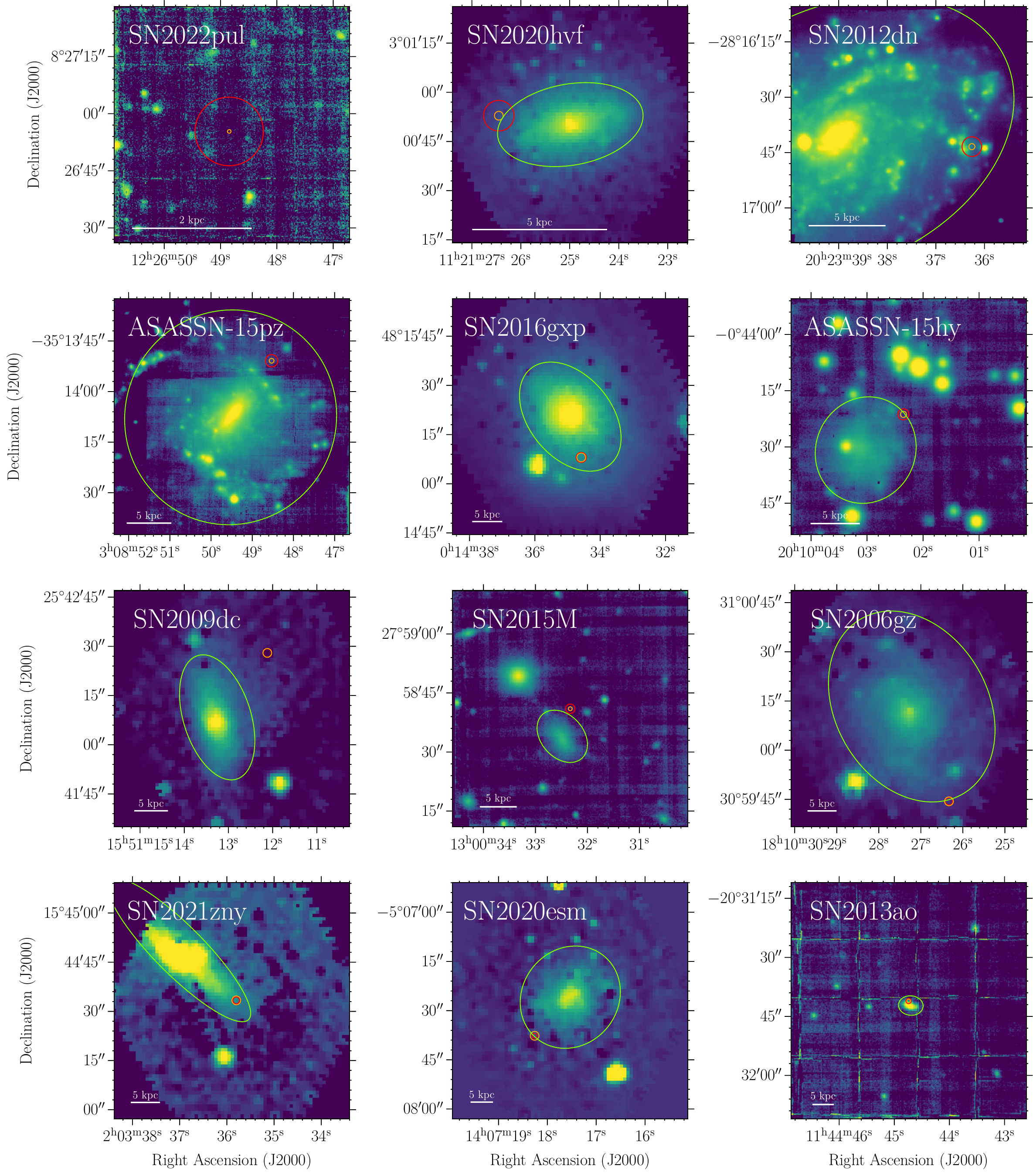}
\caption{Synthetic $r$-band images of a 1 arcmin$^2$ field-of-view of our 2003fg-like SN Ia host galaxies observed with IFS, sorted by redshift (lowest top left to highest bottom right). Green contours cover the full extent of the host galaxy, avoiding foreground stars when possible. Red and orange circles represent 1 kpc$^2$ apertures and an aperture of the spatial-resolution size, respectively, centered at the SN location.}
\label{fig:IFU}
\end{figure*}
\begin{figure*}\ContinuedFloat
\ContinuedFloat
\captionsetup{list=off,format=cont}
\centering
\includegraphics[trim= 0cm 0cm 0cm 0cm,clip=True, width=0.97\textwidth]{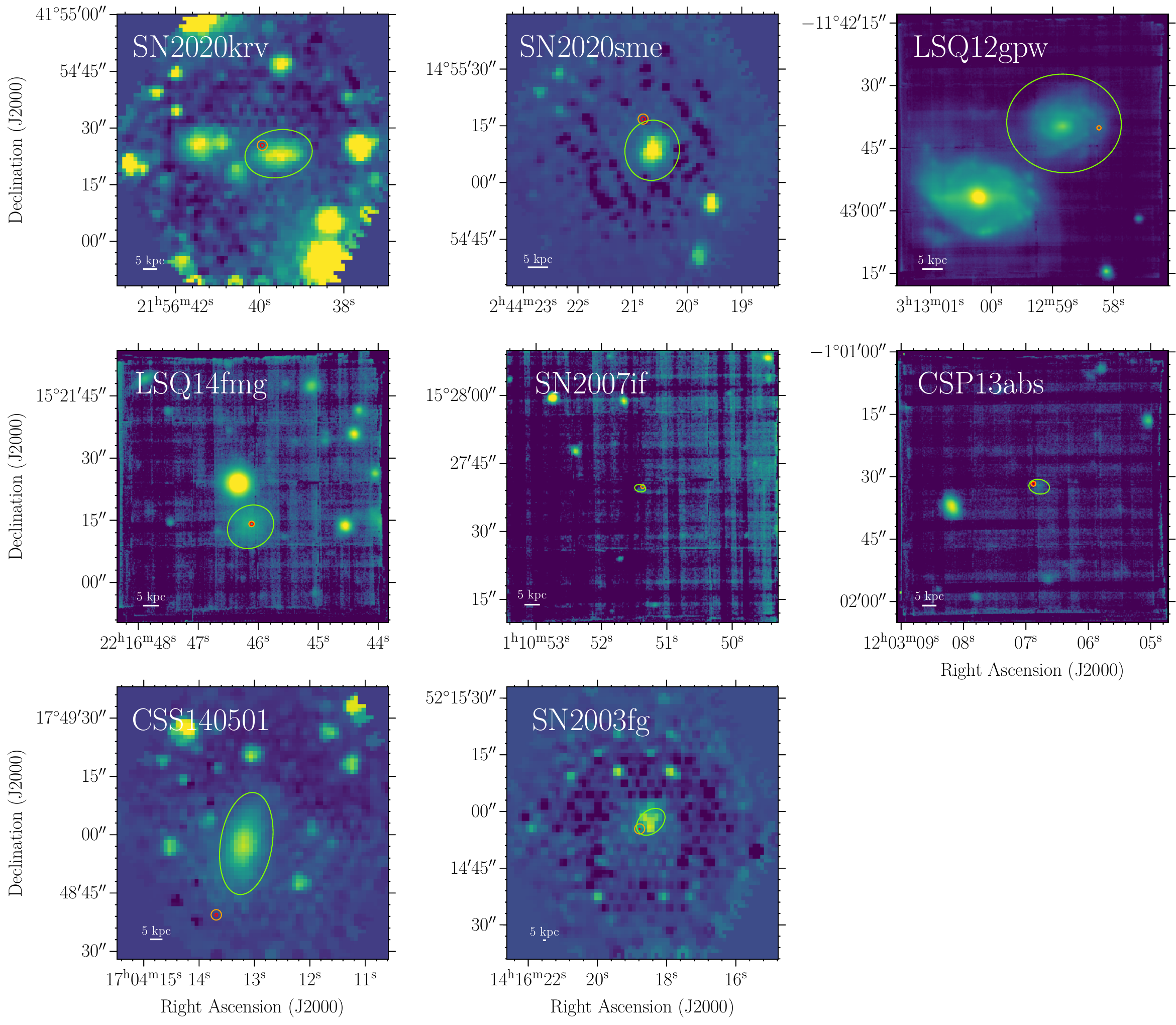}
\caption{}
\label{fig:IFU2}
\end{figure*}

A spectrum of the nearby candidate host galaxy of SN~2022ilv was obtained by \cite{2023ApJ...943L..20S}, revealing a redshift of 0.11, which is incompatible with the redshift of the SN. SN~2022ilv is located at 7.7 arcmin ($\sim$250 kpc) from the core of the luminous lenticular galaxy NGC 5872, suggesting it may have exploded either in the outskirts of the host or in a dwarf galaxy satellite. Consequently, this object is not included in our analysis; however, it is noteworthy that both scenarios would be indicative of low-metallicity environments. Also, due to the low surface brightness at the location of SN~2021hem ($>24$~mag arcsec$^{-2}$; \citealt{2026A&A...706A.252B}), it is deemed an unsuitable target for our study. In this work, we perform the analysis on 20 of the 32 2003fg-like SNe~Ia defined with this method. In Table \ref{table:prop} we summarize the main properties of all twenty objects, covering a redshift range of $0.003<z<0.244$.


\subsection{Observations}

Ten host galaxies were observed as part of the PMAS/PPak Integral-field Supernova hosts Compilation (PISCO; \citealt{2018ApJ...855..107G}). These data were obtained with the Potsdam Multi-Aperture Spectrophotometer (PMAS, \citealt{2005PASP..117..620R}) in the PPak fiber bundle mode \citep{2006PASP..118..129K} on the 3.5m telescope at the Calar Alto observatory, which provides 1$\times$1 arcsec$^2$ sampling within a hexagonal FoV of approximately 1 arcmin$^2$, spectral coverage from 3700 to 7300 \AA, and resolving power of R$\sim$500. Three exposures of either 900 or 1200~s, dithered by a few arcsec, were obtained for each galaxy. More information on the observing strategy and data reduction can be found in \cite{2016A&A...594A..36S}.

Another ten of the 2003fg-like SN Ia host galaxies were observed as part of the All-weather MUse Supernova Integral-field Nearby Galaxies (AMUSING; \citealt{2016MNRAS.455.4087G}). Data were obtained with the Multi Unit Spectroscopic Explorer (MUSE; \citealt{2010SPIE.7735E..08B}) mounted on the 8.2m Very Large Telescope at Cerro Paranal, which provides full spatial coverage within a field-of-view (FoV) of 1 arcmin$^2$, a spatial sampling of 0.2$\times$0.2 arcsec$^2$, wavelength coverage from 4700 to 9300 \AA, and an average resolving power of R$\sim$3000. Typically, four exposures of 600--700~s, rotated 90 degrees and with dithers of a few arcsec, were obtained for each galaxy. Details on the data reduction can be found in \cite{2016MNRAS.455.4087G} and \cite{2017A&A...602A..85K}.

Table \ref{table:prop} lists information on the observations for each host galaxy, such as the date of observation, exposure times, airmass during the observation, and final spatial resolution. Figure \ref{fig:IFU} shows $r$-band synthetic images of the 20 2003fg-like SN host galaxies, sorted from low to high redshift, produced by the convolution of the CSP $r$-band filter transmission curve with the spectra in the IFS datacubes. 


\section{Analysis}\label{sect:Analysis}

To study the host-galaxy global properties and the \revdel{most nearby}\revadd{immediate} environment of our sample of \SCSN, we conduct a similar analysis to that presented in \cite{2014A&A...572A..38G,2016A&A...591A..48G,2026arXiv260622173G} with a few modifications. First, we use {\sc HOSTPHOT} \citep{2022JOSS....7.4508M} to fit an elliptical aperture to the light profiles of the galaxies in combined $gri$ PanSTARRS or $grz$ Legacy Survey images (green contours in Fig. \ref{fig:IFU}). These ellipses are then projected into the IFS datacubes, and spectra within the ellipses are integrated to obtain a global spectrum for each galaxy. In addition, we also extract 1 kpc$^2$ spectra centered at the SN location to study the local environment (red circles in Fig. \ref{fig:IFU}). We adopted a \(1\,\mathrm{kpc}^2\) aperture as the baseline for the local analysis, while using the spatial resolution of each IFS cube as a lower limit on the aperture size. Thus, when the resolution element was larger than \(1\,\mathrm{kpc}^2\), we increased the aperture to match it (orange circles in Fig. \ref{fig:IFU}).

After correcting the obtained spectroscopy for Milky Way (MW) foreground reddening using the maps of \cite{2011ApJ...737..103S} and an $R_V$=3.1 \cite{1999PASP..111...63F} extinction law, we use STARLIGHT \citep{2005MNRAS.358..363C} to fit the stellar continuum in these spectra with a set of different simple stellar population (SSP) models. We choose a selection of 66 components (see \citealt{S12}) with 17 different ages (from 1 Myr to 18 Gyr) and four metallicities (0.2, 0.4, 1.0 and 2.5 Z$_\odot$, where Z$_\odot$=0.02; \citealt{2009ARA&A..47..481A}) coming from a slightly modified version of the models of \cite{2003MNRAS.344.1000B}\footnote{See \cite{2007ASPC..374..303B} for more information.}, replacing STELIB by the MILES spectral library \citep{2006MNRAS.371..703S}, Padova 1994 evolutionary tracks, \cite{2003PASP..115..763C} initial mass function (IMF) truncated at 0.1 and 100 M$_\odot$, with calculations of the TP-AGB evolutionary phase for stars of different mass and metallicity by \cite{2007A&A...469..239M} and \cite{2008A&A...482..883M}.

Although the MUSE wavelength coverage does not include data blueward of $\sim$4700\AA, we keep the stellar mass estimate from the best STARLIGHT fit, as the other stellar parameters may be less reliably constrained. On the other hand, gas-phase parameters are not strongly affected by the choice of stellar bases and the range of fitted stellar continuum \citep{2014A&A...572A..38G}. Statistical uncertainties on the stellar mass were measured by performing a Monte Carlo simulation, generating a sample of 1000 spectra by varying the observed flux within the errors, repeating the SSP fit, and calculating the 1$\sigma$ bounds on the distribution of the 1000 recovered masses.

The best STARLIGHT SSP fit is then subtracted from the observed spectra to obtain the pure emission gas spectra and to measure the flux of the most prominent emission lines (H$\beta$, \mbox{[\ion{O}{iii}] $\lambda$5007}, H$\alpha$, [\ion{N}{ii}] $\lambda$6584, and [\ion{S}{ii}] $\lambda\lambda$6716,31) by fitting Gaussian profiles. After correcting the measured fluxes for dust reddening from the host galaxy using the Balmer decrement, we estimate the star-formation rate (SFR) from the H$\alpha$ luminosity following \citet{1998ARA&A..36..189K}, converted to a \citet{2003PASP..115..763C} IMF, and the oxygen abundance through the \cite{2004MNRAS.348L..59P} O3N2 and the \cite{2016Ap&SS.361...61D} D16 calibrations. The specific star-formation rate (sSFR) is also estimated from the stellar mass and the H$\alpha$-based SFR. We repeat the same procedure on the local spectra, with the difference that the SFR and the stellar mass are reported as surface densities, $\Sigma_{\rm SFR}$ (in $M_\odot\,{\rm yr}^{-1}\,{\rm kpc}^{-2}$) and $\Sigma_{M_\star}$ (in $M_\odot\,{\rm kpc}^{-2}$), as they are measured within a fixed physical area centred on the SN location. For the larger, seeing-limited apertures, the measurements are renormalized to $1~{\rm kpc}^2$ to allow a fair comparison between objects.

In the global spectra, we have been able to measure gas-phase emission lines in all but the following cases:
(i) the global aperture of SN~2022pul, SN~2012dn, and SN~2021zny extends beyond the IFS field-of-view, so the datacubes do not cover the full extent of the host galaxy. For SN~2022pul, the covered region is not representative, and no reliable measurement can be obtained, so it is excluded from the global analysis. For SN~2012dn and SN~2021zny the covered emission is more than $50\%$ of the total, so a non-negligible fraction of the flux still falls outside the field of view. We therefore report the extensive quantities that scale with the missing flux (emission-line fluxes, SFR, and stellar mass) as lower limits, whereas the intensive quantities (sSFR, H$\alpha$EW, and the O3N2 and D16 oxygen abundances), being ratios of co-spatial measurements, are less affected and remain useful. These two objects are listed in Tables~\ref{tab:global_fluxes} and~\ref{tab:global_properties} and shown for reference in the figures to place them in context, but they are not included in the sample averages;
(ii) SN~2003fg is at a higher redshift (z=0.244), and our observations do not cover the H$\alpha$ and [\ion{S}{ii}] lines, so this object is excluded from most of the remainder of the global analysis; and
(iii) in the host spectra of SN~2007if and SN~2015M, the [\ion{N}{ii}] and [\ion{S}{ii}] lines are very weak or undetected, so for both objects we report 3$\sigma$ upper limits on their fluxes and the corresponding limits on the line ratios and abundances in both calibrators.

Regarding the local environment spectra:
(i) the SN~2009dc spectrum shows no stellar continuum, so no STARLIGHT fit is possible, and H$\alpha$EW, stellar mass, and sSFR cannot be measured. Emission-line fluxes are instead measured directly on the observed spectrum. Only H$\alpha$ is detected, and 3$\sigma$ upper limits are reported for all other lines. Since the underlying stellar absorption is not corrected for, the Balmer-line fluxes (and hence the derived SFR) are lower limits to the true emission;
(ii) similarly, we find no continuum signal in the SN~2022pul environment; no lines are detected, and we report 3$\sigma$ upper limits measured on the observed spectrum;
(iii) for SN~2015M we detect H$\alpha$, [\ion{N}{ii}], and the [\ion{O}{iii}] $\lambda$5007 line, and report 3$\sigma$ upper limits for the rest; and
(iv) in the local spectrum of SN~2007if only the Balmer lines are detected, and 3$\sigma$ upper limits are reported for the rest.
All measured fluxes and upper limits are reported in Appendix \ref{sec:tables} in Tables \ref{tab:global_fluxes} and \ref{tab:local_fluxes} for the global and local aperture, respectively.

\begin{figure}
\centering
\includegraphics[width=0.95\columnwidth]{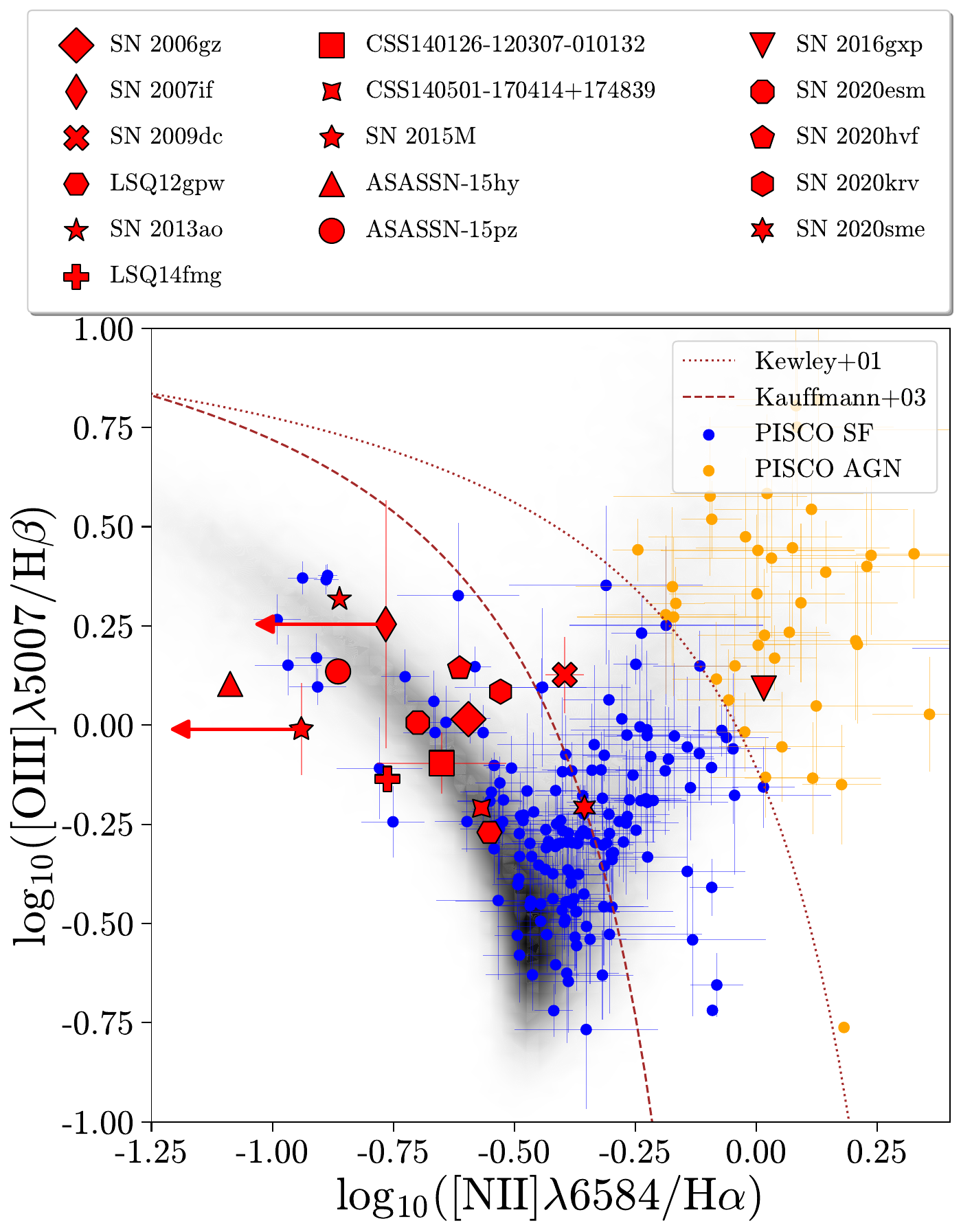}
\caption{[\ion{O}{iii}]/H$\beta$ vs [\ion{N}{ii}]/H$\alpha$ BPT diagram of the sixteen galaxies in our 2003fg-like SN sample (all but SN~2022pul, 2021zny, 2012dn, and 2003fg hosts), compared to the 222 SN Ia host galaxies observed with IFS from PISCO, and the background SDSS galaxies from \cite{2009ApJS..182..543A}. All but two 2003fg-like SN hosts are in the region below the \cite{2003MNRAS.346.1055K} dashed line, indicating their ionizing source is star formation. In the following, we excluded the SN~2016gxp host galaxy and 49 galaxies of the PISCO sample (in orange) that fall in the region above the dotted line by \cite{2001ApJ...556..121K}, which indicates that the ionization source is not underlying star formation but an active galactic nucleus (AGN).}
\label{fig:bpt}
\end{figure}

\section{Results}\label{sect:Results}

\subsection{Global properties of 2003fg-like SNe~Ia host galaxies}\label{sect:global}

A summary of the host-galaxy parameters measured in the global spectra is listed in Table \ref{tab:global_properties}. To put them in context with other sets of SN host galaxies, we will use samples of the following sources:
(i) 229 SN Ia star-forming host galaxies at $z<0.2$ of the total 589 from the SDSS-II/SNe survey included in \cite{2022A&A...659A..89G};
(ii) 222 SN Ia host galaxies observed with IFS in the PISCO sample \citep{2018ApJ...855..107G};
(iii) a compilation of superluminous SN (SLSNe) host galaxies restricted to redshifts $z<0.2$ from \cite{2015MNRAS.449..917L} and \cite{2017MNRAS.470.3566C}, since they have been found to explode in extreme metal-poor galaxies with high sSFR, and may be a good reference to see up to which point 2003fg-like SN hosts are metal-poor and efficient at creating stars;
(iv) the sample of low-mass metal-poor galaxies also known as {\it Blueberries} from \cite{2017ApJ...847...38Y};
(v) the sample of {\it Green Peas}, emission line galaxies, with a peculiar bright green colour and small size, from \cite{2009MNRAS.399.1191C};
(vi) a low-luminosity sample of 42 field galaxies from the Spitzer Local Volume Legacy survey by \cite{2012ApJ...754...98B} with global H$\alpha$ luminosities reported in \cite{2008ApJS..178..247K}; and
(vii) a sample of 45 low-metallicity blue compact dwarf galaxies from \cite{2018ApJ...863..134H}.
All masses and SFRs of these comparison samples have been corrected accordingly to match a \cite{2003PASP..115..763C} IMF, the one used in our analysis. Similarly, when possible\footnote{For \cite{2017ApJ...847...38Y}, \cite{2012ApJ...754...98B}, and \cite{2018ApJ...863..134H} samples the calibrators used were not included in the corrections provided by \cite{2008ApJ...681.1183K}.}, oxygen abundances have been rescaled to the O3N2 \cite{2004MNRAS.348L..59P} scale using transformations by \cite{2008ApJ...681.1183K}.

\begin{figure}[!t]
\centering
\includegraphics[width=0.92\columnwidth]{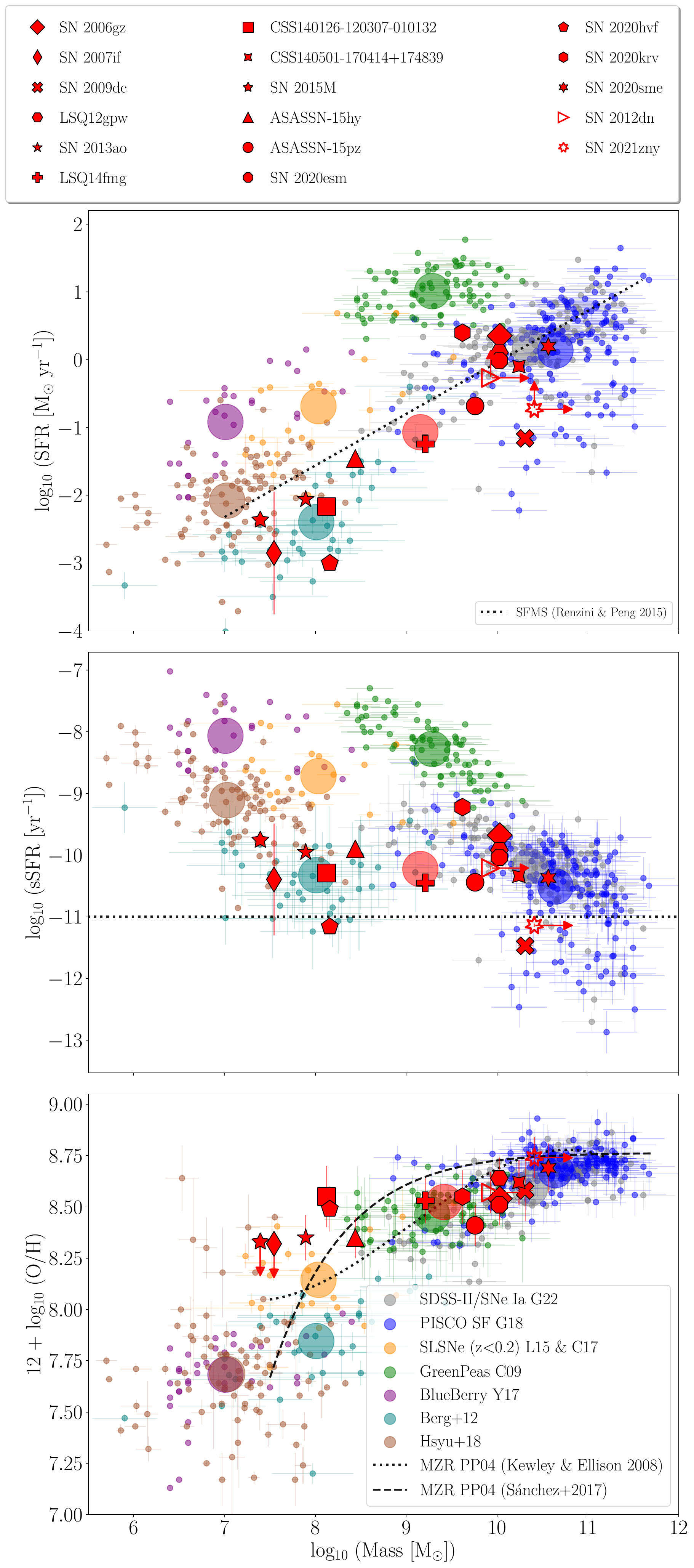}
\caption{SFR, sSFR, and oxygen abundance as a function of stellar mass for our sample of 2003fg-like SNe~Ia host galaxies (in red). We included the comparison samples of SNe~Ia, SLSNe, and low-luminosity hosts described in section \ref{sect:global}. Large dots correspond to the average values for each sample. The dotted lines correspond to a Main Sequence of Star-forming Galaxies (top panel; \citealt{2015ApJ...801L..29R}), a division between actively star-forming galaxies and passive galaxies at $\log_{10} ({\rm sSFR} ~yr^{-1}) < -11$ (as defined in \citealt{2019MNRAS.484.3785B}), and two mass-metallicity relations from \cite{2008ApJ...681.1183K} and \cite{2017MNRAS.469.2121S}. 
SN~2012dn and SN~2021zny are shown in open symbols as SFR and mass upper limits, for visual comparison. These two objects are not included in the sample mean (large red dot).
}
\label{fig:globsdss}
\end{figure}

To check whether the origin of the ionization source exclusively arises from the star formation, we used the BPT diagnostic diagram \citep{1981PASP...93....5B, 1987ApJS...63..295V}, a map of $O3\equiv\log_{10}\left(\frac{\textrm{\OIIId}}{\textrm{\Hbd}}\right)$ and $N2\equiv\log_{10}\left(\frac{\textrm{\NIId}}{\textrm{\Had}}\right)$ ratios, in which gas ionized by different sources occupies different areas. Two criteria commonly used to separate star-forming (SF) from AGN-dominated galaxies are the expressions in \cite{2001ApJ...556..121K} and \cite{2003MNRAS.346.1055K}, although we note that it has been demonstrated that {\it bona fide} {\sc \ion{H}{ii}} regions can be found as well in the composite area within the lines \citep{2014A&A...563A..49S}.

Figure \ref{fig:bpt} shows the BPT diagram populated by the full set of SDSS galaxies from \cite{2009ApJS..182..543A} in the background, all 222 star-forming PISCO SN Ia host galaxies, and our sixteen 2003fg-like SN hosts with global measurements (all but SN 2022pul, 2021zny, 2012dn, and 2003fg hosts). First, we find that most 2003fg-like SN hosts fall in the star-formation ionizing source region, dispersed from the bottom of the {\it V}-shape of all SDSS galaxies towards the left wing, i.e., towards the locus of lower-metallicity star-forming galaxies, as expected given the low gas-phase metallicities we measure for these hosts. The only exception is the host of SN~2016gxp, which falls above the \cite{2001ApJ...556..121K} line, meaning it is heavily affected by AGN emission and is therefore discarded from the following global analysis. PISCO hosts are spread over the diagram, and 49 galaxies have emission line ratios falling above the \cite{2001ApJ...556..121K} line, in the AGN-dominated region. These galaxies are excluded from the remainder of the global parameters comparison. 

Figure \ref{fig:globsdss} shows some typical SFR-mass, sSFR-mass, and mass-metallicity galaxy relations for the comparison samples and our sample of 2003fg-like SN Ia host galaxies. The large dots represent the average values measured across all objects in each sample. 
The three panels show that 2003fg-like SN hosts do not occupy the same region as the majority of SNe~Ia hosts. 2003fg-like SN hosts tend to fall in the less massive, less star-forming, and metal-poorer wing of the relations. In addition, those hosts seem to fall within the high-sSFR region among other SN Ia hosts.
When comparing 2003fg-like SN hosts with the hosts of SLSNe (in orange), the latter show average SFRs similar to those of 2003fg-like SN hosts, both at the low end of the normal SN~Ia host distribution. However, SLSN hosts are on average an order of magnitude less massive, which translates into significantly higher sSFR, and they are also clearly more metal-poor than 2003fg-like SN hosts.
The {\it Green Peas} (in green) are extreme star-forming galaxies, with the highest average SFR of all the samples considered here. Their stellar masses are, however, similar to those of 2003fg-like SN hosts, that is, at the low end of the SN~Ia host distribution, which again results in very high sSFR. Interestingly, their oxygen abundances are comparable to those of 2003fg-like SN hosts.
The low-luminosity field galaxies of \citet{2012ApJ...754...98B} (in teal) have low stellar masses, comparable to those of the SLSN hosts, but with lower SFRs, which results in sSFRs similar to those of 2003fg-like SN hosts. These galaxies have even lower metallicities than the SLSN hosts.
Finally, the {\it Blueberries} (in purple) and the blue compact dwarfs of \citet{2018ApJ...863..134H} (in brown) are the least massive and most metal-poor galaxies among the comparison samples. While the Blueberries have SFRs similar to those of 2003fg-like SN hosts, the \citet{2018ApJ...863..134H} galaxies show lower values. In both cases, their very low masses result in high sSFRs.

\begin{figure*}
\centering
\includegraphics[width=.99\textwidth]{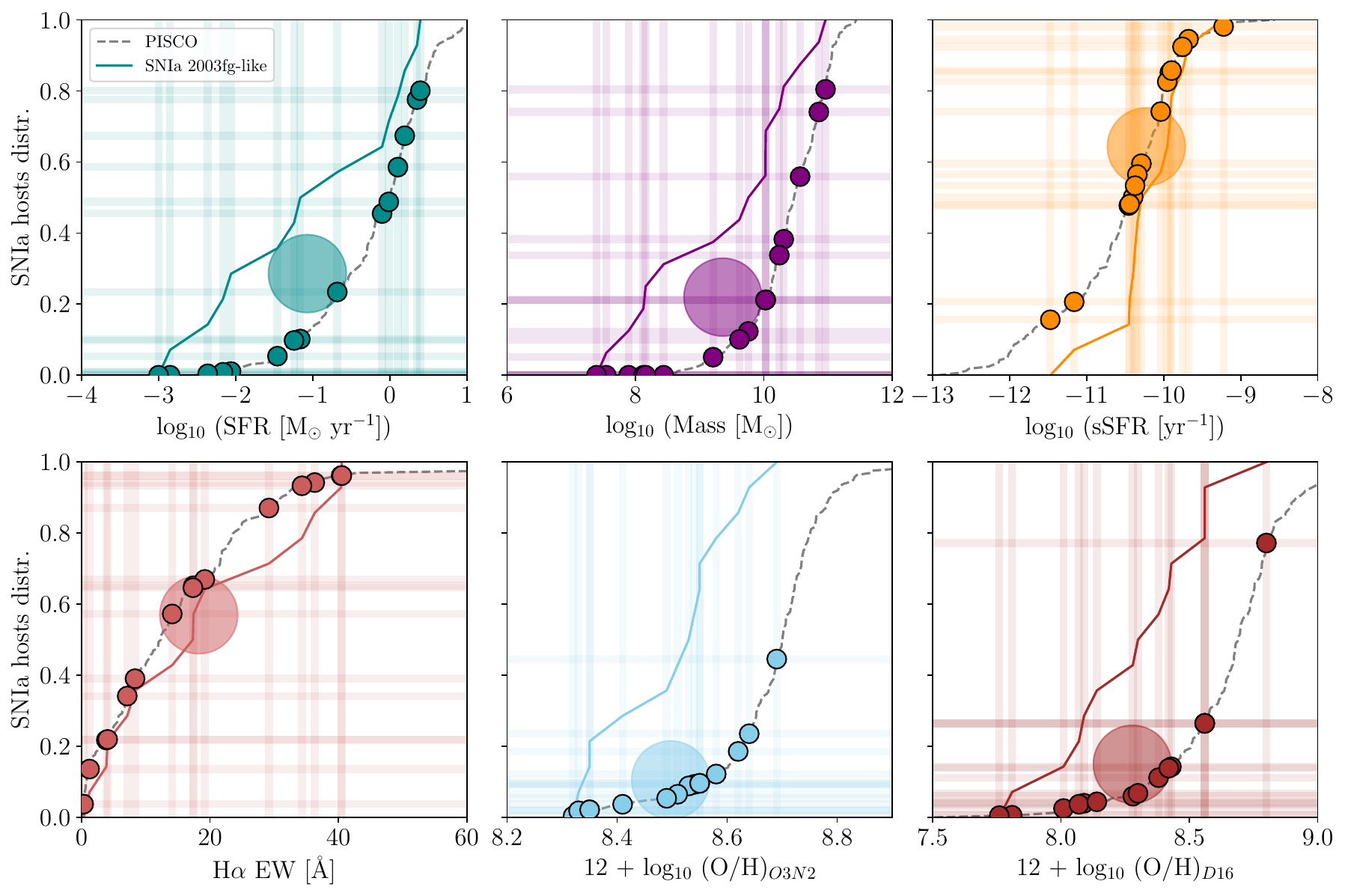}
\caption{Distributions of SFR, stellar mass, sSFR, H$\alpha$ equivalent width, and oxygen abundance in the O3N2 and D16 calibrators of all 173 SN Ia host galaxies in PISCO (in dashed grey) together with our 2003fg-like SN hosts presented in this work, which have been highlighted with colored dots on top of the PISCO distribution at the corresponding locations. Vertical and horizontal strips show the actual parameter value of each galaxy and the rank within PISCO. Coloured distributions correspond to the 2003fg-like SN hosts alone. The large colored dots represent the average value of all 2003fg-like SN hosts.}
\label{fig:globdist}
\end{figure*}

Figure \ref{fig:globdist} shows the normalized cumulative distribution of the six parameters described above and studied here, for our 2003fg-like SN hosts (colored) together with 173 star-forming SN Ia host galaxies in PISCO (dashed grey). Dots on top of the PISCO distributions, with their vertical and horizontal bars, indicate the value and percentile in which our \SCSN\ fall within the PISCO sample. The large dots represent the average value in the X and Y-axes of the 2003fg-like SN hosts. Compared to the PISCO galaxies, the 2003fg-like SN hosts are located at the lower end of the SFR, mass, and oxygen abundance distributions; however, they tend to have higher sSFR. On the other hand, their H$\alpha$EW is relatively well-distributed across the distribution. This is confirmed by comparing the PISCO and 2003fg-like SN distributions, in which all but the H$\alpha$EW and, marginally, the SFR and sSFR are clearly shifted. 
For each parameter we performed a two-sample Kolmogorov--Smirnov (K-S) test against the PISCO SN~Ia host sample, obtaining (K-S statistic $D$, $p$-value): 
$(0.43,\,0.009)_{\rm SFR}$, 
$(0.49,\,6\times10^{-4})_{\rm mass}$, 
$(0.35,\,0.060)_{\rm sSFR}$, 
$(0.20,\,0.59)_{\rm H\alpha EW}$, 
$(0.70,\,{<}5\times10^{-7})_{\rm O3N2}$, and 
$(0.67,\,{<}3\times10^{-6})_{\rm D16}$.
Given the number of K-S tests performed, a fixed threshold of $p<0.05$ would overstate the significance of individual comparisons. We therefore adopt a Bonferroni-corrected threshold of $\alpha = 0.05/N \simeq 8.3\times10^{-3}$, where $N=6$ is the number of host-galaxy parameters tested in each of the global and local analyses, and consider a difference significant only when the $p$-value falls below this value. This correction is conservative, since the tests are not independent: the same SN samples are compared across host-galaxy parameters that are themselves correlated (\eg stellar mass, metallicity, and sSFR through the mass--metallicity and star-forming main-sequence relations). Under this criterion, the differences in stellar mass and in both oxygen-abundance calibrators are highly significant, while the SFR sits marginally above the threshold.
H$\alpha$EW is the least important parameter differentiating 2003fg-like SN host galaxies from the sample of normal SNe~Ia hosts. H$\alpha$EW indicates how strong the ionization from recent star formation (young populations) is compared to the older population continuum emission. This indicates that the age of the young stellar population is the least important parameter distinguishing \SCSN\ and normal SN hosts. Conversely, metallicity seems to be the most important parameter.

\begin{figure*}[t]
\centering
\includegraphics[width=.99\textwidth]{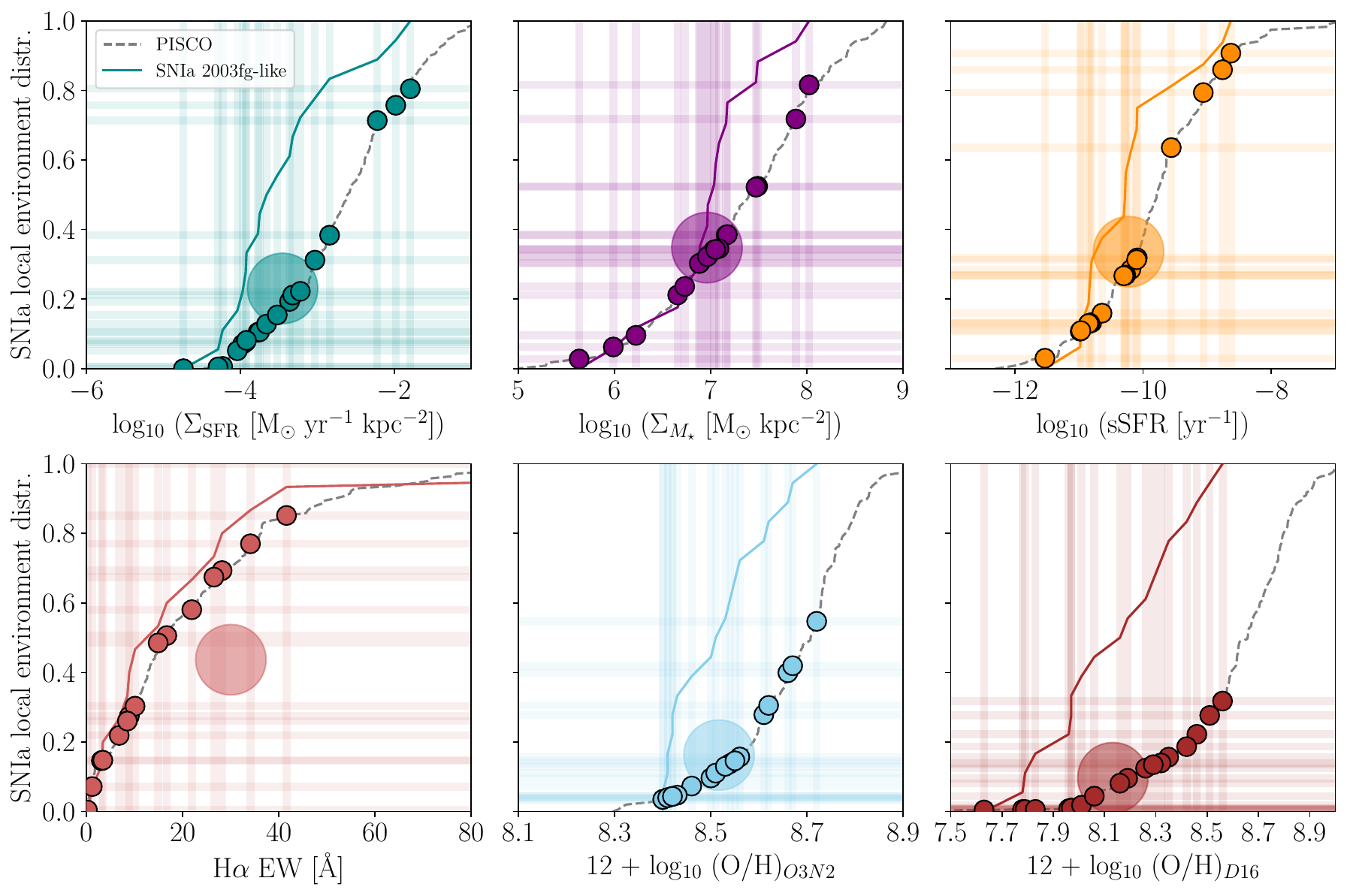}
\caption{Similarly to Fig. \ref{fig:globdist}, distributions of SFR and mass surface densities, sSFR, H$\alpha$ equivalent width, and oxygen abundance in the O3N2 and D16 calibrators measured at the local environment of all SN Ia included in PISCO (in grey), together with our 2003fg-like SNe~Ia presented in this work, which have been highlighted with colored dots on top of the distribution at the corresponding locations. Vertical and horizontal strips show the actual parameter value of the local environment and the rank within PISCO. Coloured distributions correspond to the 2003fg-like SN local environments alone. The large colored dots represent the average value of all 2003fg-like SN environments.}
\label{fig:IFU3}
\end{figure*}

\subsection{Local environments of 2003fg-like SNe~Ia}\label{sect:local}

A summary of the environmental parameters measured from the local spectra is listed in Table \ref{tab:local_properties}. Similar to what was done for the global parameters, we put the local environments of our 2003fg-like SNe~Ia in context using as a reference the 234 SN locations in the 222 PISCO SN~Ia host galaxies observed with IFS, taking for each parameter all locations with an available measurement. Figure \ref{fig:IFU3} shows the normalized cumulative distributions of the six parameters studied here, for all 2003fg-like SN local environments together with the SN Ia environments in PISCO. Two-sample K-S tests give
(K-S statistic $D$, $p$-value): 
$(0.52,\,1.2\times10^{-4})_{\Sigma_{SFR}}$, 
$(0.39,\,0.011)_{\Sigma_{\rm M_*}}$, 
$(0.45,\,0.004)_{\rm sSFR}$, 
$(0.20,\,0.56)_{\rm H\alpha EW}$, 
$(0.58,\,8.6\times10^{-6})_{\rm O3N2}$, and 
$(0.68,\,4.5\times10^{-8})_{\rm D16}$.
The $\Sigma_{\rm SFR}$ \revdel{and $\Sigma_{\rm M_*}$ }measured within the local aperture \revdel{are both }\revadd{is }significantly lower than at normal SN~Ia locations, consistent with the differences found for the global values. \revadd{$\Sigma_{\rm M_*}$ is also lower, although only significant at the uncorrected $0.05$ level}. \revdel{In contrast, the sSFR of 2003fg-like SN environments is unremarkable among SN~Ia locations. Although 2003fg-like SNe~Ia explode in regions of low stellar mass and low star formation, the amount of star formation per unit mass is typical of galaxies in PISCO.}\revadd{The local sSFR is likewise significantly lower than at normal SN~Ia locations, indicating that these regions form fewer stars per unit mass, consistent with the preference of 2003fg-like SNe~Ia for the older, quiescent outskirts of their hosts.} The H$\alpha$EW distribution is also compatible with that of the SNe~Ia, as found globally, indicating that the age of the stellar population at the SN position is not a distinctive property of 2003fg-like SNe~Ia. Finally, the most significant parameter differentiating the two samples is again the oxygen abundance, which is significantly lower for the 2003fg-like SN sample in both calibrators. 2003fg-like SNe~Ia do not only explode in more metal-poor galaxies than normal SNe~Ia but, within SN~Ia environments, they occupy the locations with the lowest metallicities.
\revdel{All these differences ($\Sigma_{\rm SFR}$, $\Sigma_{\rm M_*}$, and both abundance calibrators) remain significant under the Bonferroni-corrected threshold defined above, so our main result is unaffected by the stricter criterion.}\revadd{Under the Bonferroni-corrected threshold defined above, the differences in $\Sigma_{\rm SFR}$, sSFR, and both abundance calibrators remain significant, whereas $\Sigma_{\rm M_*}$ is significant only at the uncorrected $0.05$ level. Our main result, the systematically low local metallicity, is unaffected by the stricter criterion.}

Finally, we investigate the position of 2003fg-like SNe~Ia within their hosts through the host-normalized separation $d_{\rm DLR}$, the projected SN--host distance in units of the directional light radius of the galaxy in the direction of the SN, computed from the {\sc HOSTPHOT} elliptical apertures. The projected separations and $d_{\rm DLR}$ values of the twenty objects are listed in the last two columns of Table \ref{table:prop}. The median of our sample is $d_{\rm DLR}=0.82$, with 13 of the 20 objects (65\%) lying within one directional light radius. Compared to the 1086 normal SNe~Ia in the volume-limited ($z\leq0.06$) ZTF~DR2 sample \citep[median $d_{\rm DLR}=0.63$;][]{2025A&A...694A...1R}, our sample is shifted towards larger separations (K-S $D=0.39$, $p=0.004$). A simpler test is provided internally by ZTF~DR2, where the 25 SNe classified as 03fg-like \citep{2025A&A...694A..10D}, four of which (SNe~2020esm, 2020hvf, 2020krv, and 2020sme) are also part of our sample, differ from the normal SNe~Ia of the same catalogue with $D=0.32$, $p=0.010$ (median $d_{\rm DLR}=1.13$, with only 44\% within one directional light radius). We note that for the four objects in common, the ZTF~DR2 catalogue values are systematically larger than ours, reflecting the smaller light radii used in their host association, so our comparison against the ZTF normal-SN distribution is conservative. With respect to the other ZTF~DR2 subtypes, we find no significant difference with the 1991T-like ($D=0.27$, $p=0.10$) or 1991bg-like ($D=0.25$, $p=0.22$) samples, whereas the Iax sample differs significantly ($D=0.52$, $p=0.004$), showing the opposite tendency of being more centrally concentrated than 2003fg-like SNe~Ia, consistent with their preference for young, star-forming environments \citep{2018MNRAS.473.1359L}.

Taken together, 2003fg-like SNe~Ia show a significant tendency to explode in the outskirts of their hosts, which naturally explains the low $\Sigma_{\rm SFR}$ and $\Sigma_{M_*}$ surface densities measured at the SN positions, while sSFR and H$\alpha$EW remain typical, and are consistent with their low local metallicities given the negative abundance gradients of galaxies.

\subsection{Correlation between environmental and SN parameters}

\begin{figure}[!t]
\centering
\includegraphics[width=\columnwidth]{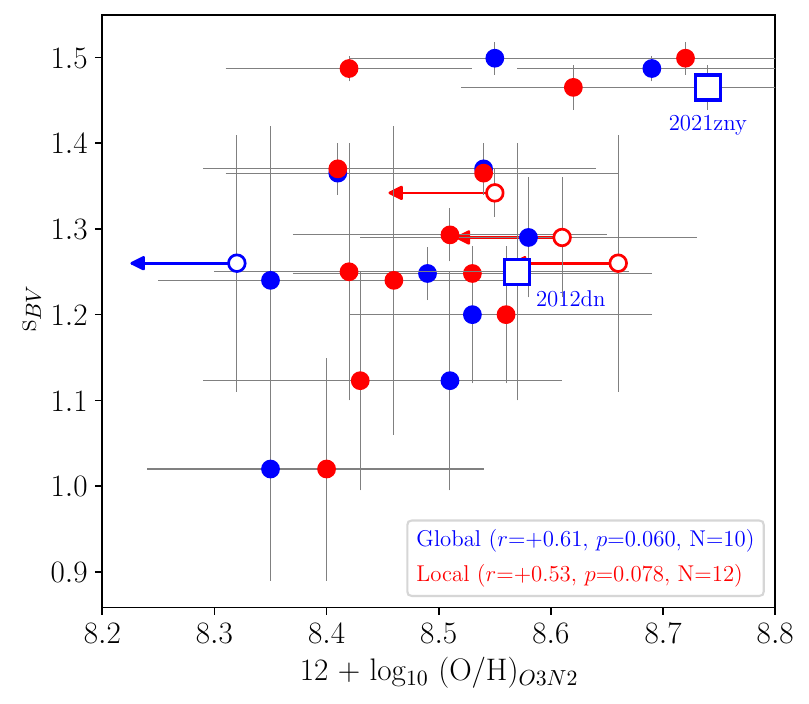}
\caption{Color stretch $s_{BV}$ as a function of the O3N2 gas-phase oxygen abundance measured in the global (blue) and local (red) host-galaxy spectra of the 2003fg-like SNe~Ia. Filled circles denote detections, open circles with arrows mark $3\sigma$ upper limits on the abundance, and open blue squares indicate the two objects (SN~2012dn and SN~2021zny) whose global aperture extends beyond the IFS field of view. 
The Pearson coefficients in the legend use detections only.}
\label{fig:localoh}
\end{figure}

We searched for correlations between the 2003fg-like SN light-curve parameters, color-stretch $s_{BV}$ and host reddening $E(B-V)_{\rm host}$, available for 15 and 19 of our objects, respectively, and all the global and local environmental parameters presented in this work. No significant correlation involving $E(B-V)_{\rm host}$ is found. For $s_{BV}$, the strongest trends are with the O3N2 oxygen abundance, both global (Pearson correlation coefficient $r=+0.61$, $p=0.060$, $N=10$) and local ($r=+0.53$, $p=0.078$, $N=12$; see Fig.~\ref{fig:localoh}), and with the global SFR ($r=+0.54$, $p=0.085$, $N=11$), in the sense that 2003fg-like SNe~Ia with broader light curves occur in more metal-rich and more star-forming galaxies. 
The global trend with O3N2 metallicity is robust to the addition of upper limits (adding the SN~2007if, $r=+0.56$, $p=0.072$, $N=11$) and the two incomplete hosts ($r=+0.65$, $p=0.022$, $N=12$). Including both gives $r=+0.61$, $p=0.026$, $N=13$.

\section{Discussion}\label{sect:diss}

\subsection{Comparison to previous global analyses}\label{sect:complit_global}

Four hosts in our sample have masses previously published in the literature, providing a useful consistency check of our measurements. \cite{2011MNRAS.412.2735T} concluded that 2003fg-like SNe~Ia show a tendency to explode in low-mass galaxies. However, we have convincingly argued that three of the seven objects in their sample are non-\SCSN\ \citep{2020ApJ...892..153M,2016MNRAS.458.2973P,2011ApJ...737L..24K,2020ApJ...895L...3A}, leaving four objects in common with our sample: SN~2006gz and SN~2009dc, whose host masses were measured in \cite{2011MNRAS.412.2735T}, and SN~2003fg and SN~2007if, for which they adopted measurements from \cite{2006Natur.443..308H} and \cite{2011ApJ...733....3C}, respectively.

The host of SN~2007if is the best-studied of the four and was shown to be a low-stellar-mass, metal-poor galaxy. \cite{2011ApJ...733....3C} presented deep Keck spectroscopic observations and reported a stellar mass of $\log_{10} (M_\ast/M_\odot) = 7.32 \pm 0.17$ dex, using the same IMF \citep{2003PASP..115..763C} as in this work but deriving the total mass from the $g-r$ color rather than from SSP synthesis\footnote{See Appendix~\ref{sec:07ifhighz} for an analysis of a high-$z$ background galaxy found very close to the SN~2007if host in the MUSE cube, already reported in \cite{2011ApJ...733....3C}.}. This value is in good agreement with the $7.30^{+0.31}_{-0.31}$ dex of \cite{2011MNRAS.412.2735T}, and both are consistent with our measurement of $7.54 \pm 0.09$ dex. \cite{2011ApJ...733....3C} also reported the nondetection of the two [\ion{N}{ii}]~$\lambda\lambda$6548,84 lines and noted that [\ion{O}{iii}]~$\lambda$5007 was relatively weak compared to H$\alpha$. As in their case, we adopt 3$\sigma$ upper limits for the two [\ion{N}{ii}] lines, which translate into an oxygen abundance of $12+\log_{10}(\mathrm{O/H}) \leq 8.30$ in the O3N2 scale. At face value, this seems at odds with the abundance reported by \cite{2011ApJ...733....3C}, 8.01~dex, one of the lowest values ever measured for a SN~Ia host. That value, however, was obtained with a different calibrator (KK04; \citealt{2004ApJ...617..240K}), and the KK04-to-O3N2 conversion of \cite{2008ApJ...681.1183K} is only valid down to $\mathrm{O/H}_{\mathrm{KK04}} = 8.2$~dex, below which the SN~2007if host falls. Indeed, recomputing the abundance directly from the fluxes reported in \cite{2011ApJ...733....3C}, assuming no reddening, yields 8.36~dex in the O3N2 scale, fully consistent with our measurement.

For the hosts of SN~2009dc and SN~2006gz measured by \cite{2011MNRAS.412.2735T} the agreement is reasonable. For the host of SN~2009dc we obtain $\log_{10} (M_\ast/M_\odot) = 10.31\pm0.09$, some 0.4~dex below their $10.68^{+0.07}_{-0.06}$, and for SN~2006gz host we find $10.03\pm0.09$ compared to their $10.28^{+0.01}_{-0.14}$. Offsets of this size are not unexpected between stellar masses derived with different methods: the \cite{2011MNRAS.412.2735T} values come from fitting galaxy photometry with ZPEG, whereas ours are based on STARLIGHT fits to the integrated IFS spectra, and differences in the assumed star-formation histories, dust treatment, and IMF can systematically shift $\log M_\ast$ by 0.2--0.3~dex. 
The disagreement is larger for SN~2003fg. \cite{2006Natur.443..308H} obtained $\log(M_\ast/M_\odot) = 8.93^{+0.81}_{-0.50}$ by fitting PEGASE.2 models to $ugriz$ photometry, about two orders of magnitude below our value of $10.96\pm0.09$. The origin of this difference is not methodological but a matter of host identification: their \textit{HST} imaging resolves the local environment of the SN, and their measurement refers to a small satellite galaxy of the larger system, whereas our ground-based IFS observations, which do not resolve the satellite, characterize the larger galaxy. Our near-solar oxygen abundance for this host in Sect.~\ref{sect:complit_local}, which falls on the mass--metallicity relation at $\log(M_\ast/M_\odot)\approx11$, confirms that our measurements consistently refer to the massive galaxy. If the true host is indeed the satellite, adopting the larger galaxy is a conservative choice that can only weaken our results: the satellite's much lower stellar mass would place the host in the low-mass region preferred by 2003fg-like SNe~Ia and, through the mass--metallicity relation, would imply a lower oxygen abundance, increasing the significance of our findings (although this object does not enter the sSFR and abundance distributions in any case, since H$\alpha$, [\ion{N}{ii}], and [\ion{S}{ii}] are outside our spectral coverage).
Finally, the global properties of the LSQ14fmg and ASASSN-15hy hosts were already presented, following the same methodology used here, in \cite{2020ApJ...900..140H} and \cite{2021ApJ...920..107L}, respectively, and we refer the reader to those works for a detailed discussion.

\subsection{Comparison to previous local estimates}\label{sect:complit_local}

Most of the previous works have focused on the global parameters of host galaxies. In a few cases, local parameters have been estimated indirectly. However, metallicity estimated through radial gradients is uncertain, especially at larger galactocentric distances, as it does not take into account the flattening occurring at distances of about 2 effective radii ($R_e$; see e.g., \citealt{2014A&A...563A..49S,2016A&A...587A..70S}). Similarly, the metallicities obtained from the galaxy core cannot always be considered as upper limits of SN local environments, because this approximation does not take into account the downturn in metallicity towards the center occurring in up to 30\% of all galaxies (see e.g. \citealt{2014A&A...563A..49S,2016A&A...591A..48G}).

For instance, the SN~2006gz local environment oxygen abundance was estimated by \cite{2011ApJ...737L..24K} to be 8.26 dex (in the N2 scale), fitting a linear metallicity gradient to four measurements obtained at four off-center \ion{H}{ii} regions present within the slit spectrum. The SN is at a projected distance of 14 kpc, which corresponds to $\sim$1.7 $R_e$ (from $R_e$ = 8.3 kpc; \citealt{2014A&A...563A..49S}), still within the linear gradient regime. We measure a local oxygen abundance of $8.41\pm0.12$ dex in the O3N2 scale and, independently, $8.41\pm0.17$ dex in the N2 scale of \cite{2004MNRAS.348L..59P}, consistent with their extrapolated value. For the same host, a metallicity of $8.35\pm0.03$~dex (in the O3N2 scale) was reported from an archival Subaru+FOCAS slit spectrum \citep{2009ApJ...690.1745M}, also in good agreement with our measurement at the SN location, while our integrated abundance is higher ($8.54$~dex). The difference between our global and local values follows the expected behaviour: the integrated spectrum is luminosity-weighted towards the inner, more metal-rich regions, and the $\sim$0.13~dex offset over the $\sim$1.7~$R_e$ separation of the SN corresponds to a typical disk abundance gradient \citep[e.g.][]{2014A&A...563A..49S}. The remaining small offset with respect to the slit measurement may be explained by the fact that we subtract the stellar continuum, accounting for the emission filling the underlying Balmer absorption. Without this correction, H$\beta$ is underestimated, and O3N2-based abundances are biased low. 

Similarly, \cite{2011ApJ...737L..24K} reported a central host-galaxy abundance of 8.57~dex for SN~2003fg, measured in the R$_{23}$ scale and converted to N2 using the \cite{2008ApJ...681.1183K} relations. Although H$\alpha$ and [\ion{N}{ii}] fall outside our spectral coverage at $z=0.244$, the [\ion{O}{ii}]~$\lambda$3727, H$\beta$, and [\ion{O}{iii}]~$\lambda\lambda$4959,5007 lines are all covered\footnote{The [\ion{O}{ii}] and [\ion{O}{iii}]~$\lambda$4959 fluxes used for the R$_{23}$ estimates are F([\ion{O}{ii}]~$\lambda3727) = 369.2\pm8.0$ and F([\ion{O}{iii}]~$\lambda4959) = 61.8\pm4.8$ from the global spectrum, and F([\ion{O}{ii}]~$\lambda3727) = 0.28\pm0.01$ and F([\ion{O}{iii}]~$\lambda4959) = 0.11\pm0.01$ at the SN location, all in units of $10^{-17}$ erg s$^{-1}$ cm$^{-2}$.}, allowing direct R$_{23}$ measurements: $\log R_{23} = 0.50\pm0.03$ from the global spectrum and $0.31\pm0.07$ at the SN position, corresponding to $12+\log{\rm (O/H)} = 9.01\pm0.02$ and $9.07\pm0.02$ in the KK04 upper-branch calibration (assuming negligible host reddening, which changes the results by $<0.03$~dex), or $\approx$8.6~dex in the N2 scale using the same conversion. Both measurements are in excellent agreement with their respective values and show no decline in abundance at the SN position.

The caveats on gradient extrapolations remain, however, relevant for the more remote locations in our sample. This is well illustrated by SN~2009dc, the object with the largest host-normalized separation in our sample ($d_{\rm DLR}=2.12$; Table~\ref{table:prop}). \cite{2011MNRAS.412.2735T} adopted for its host (UGC 10064) the metallicity reported by \cite{2008AJ....136....1W}, about three times solar (Z/Z$_\odot$=2.81), derived from stellar absorption features in a slit spectrum taken with the MODSPEC spectrograph at the 2.4~m Hiltner telescope. In contrast, from our integrated spectrum we measure a slightly sub-solar gas-phase abundance, $8.58\pm0.05$~dex in the O3N2 scale (i.e. Z/Z$_\odot\approx0.8$, adopting $12+\log{\rm (O/H)}_\odot=8.69$; \citealt{2009ARA&A..47..481A}), and at the SN location, where the low S/N of the spectrum only allows an upper limit, $<8.61$~dex. Part of the discrepancy likely reflects differences between tracers (stellar metallicities from absorption features probe the luminosity-weighted, older population on a different scale than gas-phase abundances) and aperture effects, since the slit samples the bright central regions far from the SN position. In any case, we find no indication of a gas-phase super-solar metallicity in this host, either globally or at the SN position.

Among the objects excluded from our global analysis, \cite{2019MNRAS.488.5473T} reported a local abundance of $8.57\pm0.05$~dex for SN~2012dn, averaging several calibrators. Our direct measurement at the SN position, available even though the datacube does not cover the full galaxy, gives $8.42\pm0.12$~dex in the O3N2 scale; the $\sim$0.15~dex offset is consistent with the systematic differences between abundance calibrators \citep{2008ApJ...681.1183K}. Either value lies within the lowest $\sim$20\% of the PISCO SN~Ia locations, so the absence of this object from the distributions of Fig.~\ref{fig:IFU3} does not affect our conclusions. For SN~2022pul, only flux limits can be derived at the SN position, and SN~2022ilv was not observed. Given their remote locations (2.2 arcmin, $\sim$9 kpc, $d_{\rm DLR}=1.65$ for SN~2022pul, and $\sim$250~kpc from the candidate host for SN~2022ilv), their environments are plausibly metal-poor, although, as discussed above, such gradient-based expectations should be treated with caution.

Finally, regarding the record-low metallicity claimed for the SN~2007if host by \cite{2011ApJ...733....3C} and discussed in Sect.~\ref{sect:complit_global}, our data provide only an uninformative limit at the SN position ($<8.66$~dex), so the comparison for that object remains a global one. The most metal-poor local environments in our sample are those of SN~2013ao and CSS140501-170414+174839 ($8.40$~dex, the former with the smaller uncertainty), closely followed by SN~2006gz and SN~2020sme ($8.41$--$8.42$~dex), all of which lie within the lowest $\sim$5\% of the PISCO SN~Ia locations (Fig.~\ref{fig:IFU3}). The high excitation of the SN~2013ao environment is directly visible in its spectrum, with [\ion{O}{iii}]~$\lambda$5007 brighter than H$\alpha$.


\subsection{Implications for progenitor scenarios}

Here, we place the environmental properties of 2003fg-like SNe in the context of the various explosion scenarios. In particular, we will discuss whether the environment drives the peculiarities of 2003fg-like SN explosions.

In Sect. 4, we confirmed previous results that, on average, \SCSN\ are found in galaxies that are less massive, less enriched in metals, and with a higher specific star formation rate compared to normal SNe~Ia. Despite the small sample of 2003fg-like objects, these differences appear to be statistically significant. Moreover, we find that 2003fg-like SNe~Ia preferentially explode in the outskirts of their hosts (Sect.~\ref{sect:local}): the sample has a median host-normalised separation $d_{\rm DLR}=0.82$ and is shifted towards larger separations than normal SNe~Ia (K-S $D=0.39$, $p=0.004$ against the volume-limited ZTF~DR2 sample). Only two objects lie close to their host centres (LSQ14fmg at $d_{\rm DLR}=0.13$, and SN~2013ao at $0.49$), while all remaining objects are found beyond $\sim$0.6 directional light radii. Given the negative metallicity gradients of galaxies, this remote-location preference is a natural counterpart to the low local oxygen abundances measured at the SN positions, and together they reinforce the picture of 2003fg-like SN progenitors born in metal-poor conditions.

For the first time, we are able to measure the direct environment rather than the average abundances in the host (see Sect. 4.2). The most important result of the local properties is that \SCSN\ come from lower-metallicity environments compared to normal SNe~Ia. Low-metallicity is the only parameter linking the objects. In fact, the local properties show a spread in the H$\alpha$EW distribution and sSFR, suggesting that \SCSN\ do not necessarily originate from young stellar systems and that the progenitors need not be short-lived. This result is at odds with what was previously suggested; however, those results came from an analysis of the whole host galaxy and not the local properties \citep{2006Natur.443..308H,2011ApJ...737L..24K,2011MNRAS.412.2735T,2011ApJ...733....3C}. 

The local low metallicity and remote location may suggest that the progenitors are members of an old stellar population, commonly associated with halo stars, rather than having a wide range of evolutionary times typically associated with normal SNe~Ia. But note that the metallicity of a SN reflects the metallicity in the host galaxy when the progenitor star was born and not the metallicity at the time of explosion. We find a tentative positive correlation between \sBV\ and local metallicity, such that SNe with lower \sBV\ values occur in more metal-poor environments. Since lower \sBV\ corresponds to a narrower light curve and a shorter diffusion timescale, this relation suggests that metallicity directly regulates the ejecta opacity. If the ejecta mass, density structure, ionization state, and distribution of radioactive material are otherwise similar across the sample, then metallicity becomes the sole driver of the observed variation in diffusion timescale: lower-metallicity ejecta have reduced line opacity, allowing photons to escape more rapidly and producing narrower light curves. Such changes in metallicity are known to produce differences in diffusion \citep[e.g.,][]{2014MNRAS.445.4427A,2018MNRAS.477..153A}.

One proposed progenitor scenario for 2003fg-like SNe~Ia is the explosion of a single, rapidly rotating C--O WD whose mass exceeds the classical Chandrasekhar limit \citep{Yoon05,2006Natur.443..308H}. Strong magnetic fields have also been proposed as a means of supporting a highly super-$M_{\rm Ch}$ WD \citep{Das13}. If the extreme WD mass is inherited from a relatively massive stellar progenitor and the explosion follows the accretion phase with only a short additional delay, this channel should preferentially be associated with young stellar populations \citep{2006Natur.443..308H}. However, the local \revdel{sSFR and }H$\alpha$ equivalent-width \revdel{distributions }\revadd{distribution }of 2003fg-like SNe \revdel{are }\revadd{is }consistent with \revdel{those }\revadd{that }of normal SNe~Ia\revadd{, and their local sSFR is, if anything, lower}, showing that a young stellar population is not a defining property of the subclass. This result disfavors the prompt explosion of a single, rapidly rotating super-$M_{\rm Ch}$ WD as the dominant origin of 2003fg-like SNe. \revadd{This conclusion is independently supported by the radiative-transfer parameter study of \citet{2023ApJ...953...13F}, who found that magnetized single-WD models fail to reproduce the observed properties of these events, resembling instead very bright normal SNe~Ia, whereas WD-merger models match both the width--luminosity relation and, critically, the low ejecta velocities. Their spectral modelling therefore provides a fully independent, non-environmental argument against the single rotating or magnetized super-$M_{\rm Ch}$ WD channel in favor of a merger origin.} A prolonged spin-down phase could weaken the connection between the formation of the massive WD and the age of the population observed at explosion, although this would require spin-down times spanning a range comparable to the delay times of normal SNe~Ia. Metallicity can affect stellar mass loss, WD structure, and the initial-to-final mass relation \citep{Dominguez01}, but there is currently no clear prediction that low metallicity preferentially allows a single WD to grow to the extreme masses required for 2003fg-like explosions. The combination of normal SN~Ia age indicators, systematically low local metallicities, and the preference for remote, old-population outskirts of the hosts is therefore not naturally explained by this scenario.

The merger of two C--O WDs provides a more direct route to a progenitor whose total mass exceeds $M_{\rm Ch}$ \citep{Iben84,Webbink84,Scalzo10}. The delay time in this scenario includes both the stellar-evolution time required to form the two WDs and the subsequent gravitational-wave inspiral, allowing double-degenerate systems to explode over a broad range of ages \citep{Maoz14}. The similarity between the local age indicators of 2003fg-like and normal SNe~Ia is therefore consistent with a merger origin and provides no evidence that these events require either an unusually prompt or unusually delayed population. Instead, the strong low-metallicity preference may indicate that metallicity influences which double-WD systems produce the 2003fg-like outcome. The close-binary fraction of solar-type stars increases strongly toward lower metallicity \citep{Moe2019}, potentially increasing the number of systems that undergo the binary interactions required to form close double WDs. Metallicity-dependent stellar evolution may also alter the component masses, WD structures, common-envelope evolution, and the amount of material lost or retained by the system \citep{Dominguez01}. These effects could preferentially produce mergers with a high total mass, a favorable mass ratio, or a substantial reservoir of C/O-rich merger debris. Such a configuration is consistent with the persistent carbon absorption, broad light curves, and early flux bumps observed in 2003fg-like SNe \citep{Ashall21,2022ApJ...927...78D,Hoogendam2024}. However, the merger scenario does not by itself guarantee the formation of the massive, relatively extended C/O-rich envelope inferred for several well-observed events, and the structure and geometry of the surrounding material depend sensitively on the merger configuration and the interval between merger and explosion. A double-degenerate origin therefore remains viable, but requires low metallicity to preferentially select the relatively rare subset of mergers capable of producing both the extreme ejecta mass and surrounding material characteristic of 2003fg-like SNe Ia.

The core-degenerate scenario is an alternative channel which can provide a natural explanation for the combined environmental and observational properties of the subclass. In this channel, a C--O WD merges with the hot degenerate C--O core of an AGB companion during or shortly after common-envelope evolution \citep{Kashi11}. The merger produces a massive, rapidly rotating remnant that may remain surrounded by a substantial C/O-rich envelope, a configuration capable of reproducing the broad light curves, low ejecta velocities, persistent carbon absorption, high gamma-ray trapping, and unusual near-infrared evolution of 2003fg-like SNe \citep{Hoeflich:Khokhlov:96,2020ApJ...900..140H,2021ApJ...920..107L,Ashall21}. Although the merger occurs during AGB evolution, rotational support can delay the thermonuclear explosion until after a spin-down phase \citep{2012MNRAS.419.1695I}. The normal local age indicators measured here, therefore, do not exclude this channel but imply that the explosion does not always occur immediately after common-envelope evolution. Instead, a distribution of post-merger delay times could allow core-degenerate systems to explode across stellar populations similar in age to those of normal SNe~Ia. The strong low-metallicity preference may then determine which core-degenerate systems develop the remnant and envelope structure required for a 2003fg-like explosion. Low metallicity can influence AGB mass loss, core growth, common-envelope evolution, and the amount of C/O-rich material retained around the remnant. In addition, a lower abundance of iron-group elements in the outer material reduces UV line blanketing and is consistent with the unusually high UV luminosities and persistently blue UV colors of the subclass \citep{2019MNRAS.488.5473T,2020ApJ...900..140H,2021ApJ...920..107L,Ashall21,Hoogendam2024}. In particular, \citet{Hoogendam2024} proposed that 2003fg-like SNe may arise from low-metallicity double-degenerate or core-degenerate systems enshrouded by carbon-rich material. The systematically low oxygen abundances measured directly at the SN locations provide independent environmental support for this interpretation. Although population-synthesis calculations predict that the total core-degenerate birthrate may increase with metallicity \citep{2017MNRAS.464.3965W}, 2003fg-like SNe may represent only a rare subset of this broader population for which low metallicity favors the required core mass, envelope mass, composition, or circumstellar structure. While the host environments alone cannot uniquely distinguish between a double-degenerate and core-degenerate origin, the ability of the core-degenerate scenario to naturally produce a massive remnant embedded within C/O-rich material makes it a plausible progenitor scenario when the environmental constraints are considered together with the defining photometric and spectroscopic properties of 2003fg-like SNe.

Overall, a more diverse set of explosion models is required to explore every realization, and early-time spectra to directly determine the metallicity in the outer envelope of \SCSN.  However, in these models a low-metallicity progenitor, relative to normal SNe~Ia, should be one of the main driving parameters.

\section{Conclusions}\label{sect:conc}

We have presented a systematic study of the largest sample of 2003fg-like SN~Ia host galaxies to date. IFS observations were obtained for twenty 2003fg-like SN hosts, from which we measured both global host properties and, for the first time for a sizeable sample, the properties of the local environment at the SN position, and compared them to normal SNe~Ia and other extreme galaxy samples.
Globally, 2003fg-like SNe~Ia explode in galaxies with lower stellar mass ($9.36\pm1.16$ dex), lower metallicity ($8.52\pm0.10$ dex, i.e. $\sim$0.68 Z$_\odot$), and marginally higher sSFR ($-10.23\pm0.54$ dex) than normal SNe~Ia, although not as extreme as the hosts of SLSNe or other metal-poor dwarf galaxies. This is consistent with previous analyses.
Locally, the SN positions show significantly lower SFR and stellar-mass surface densities than normal SN~Ia locations, consistent with their larger host-normalized separations (median $d_{\rm DLR}=0.82$ vs. 0.63 for the normal SNe~Ia in ZTF~DR2, K-S $p=0.004$; a preference for the outskirts shared by the independent ZTF~DR2 03fg-like sample), while their \revdel{sSFR and }H$\alpha$EW \revdel{are }\revadd{is }indistinguishable from \revdel{those }\revadd{that }of normal SNe~Ia\revadd{, and their local sSFR is, if anything, lower}. The most distinctive local property is again the low oxygen abundance ($8.51\pm0.10$~dex): within SN~Ia environments, 2003fg-like SNe~Ia occupy the most metal-poor locations. We also find a tentative positive correlation between the light-curve width $s_{BV}$ and the oxygen abundance, present in both the global and local measurements.

Many progenitor scenarios have been theorised for 2003fg-like SNe~Ia, such as the explosion of a rapidly rotating super-M$_{ch}$ WD, the merger of two WDs, and the core-degenerate scenario. Although our results do not allow us to distinguish between the latter two, they disfavour the rapidly rotating super-M$_{ch}$ WD channel: one of its predictions is that the main driving parameter is a young stellar age, whereas we find the age indicators at the SN positions to be typical of normal SNe~Ia. For the merger and core-degenerate scenarios to remain viable, more work is needed to determine the extent to which they require low-metallicity progenitors.

Previous work has shown that \SCSN\ appear to be overcorrected (negative Hubble residuals; brighter with respect to the Hubble flow) when their brightness is standardized using the Tripp relation \citep{1998A&A...331..815T}, and have the potential to cause biases in future dark energy experiments \citep{Ashall21}. Based on their preference for low-metallicity environments, we suggest that \SCSN\ may constitute a larger fraction of the SN~Ia population at $z\gtrsim1$, and particularly at $z\gtrsim2$, where galaxies have, on average, lower gas-phase metallicities than comparable galaxies in the local Universe. Future work should quantify how an evolving contribution from \SCSN\ to high-redshift SN~Ia samples could bias standardized distance measurements and the inferred dark-energy parameters. This will be particularly important for the Nancy Grace Roman Space Telescope, whose High Latitude Time Domain Survey is designed to use large samples of SNe~Ia extending to $z\gtrsim2$ to place precise constraints on the nature
of dark energy.

\begin{acknowledgements}
We dedicate this paper to the memory of our dear colleague I. Dom\'inguez. Her contributions to this work and to our collaboration were invaluable, but her impact reached far beyond her scientific achievements. She was a generous mentor who always found the time to guide, encourage, and inspire younger colleagues. Many of us owe an important part of our scientific development to her support, and her influence will remain with us long after this work.
L.G. acknowledges financial support from MCIN and AEI 10.13039/501100011033 under projects PID2023-151307NB-I00, CEX2020-001058-M, and by the MaX-CSIC Excellence Award MaX4-SOMMA-ICE.
C.A. and B.J.S. are supported by NASA grant 80NSSC19K1717 and NSF grants AST-1920392 and AST-1911074.
J.T.H. acknowledges support from NASA through the NASA Hubble Fellowship grant HST-HF2-51577.001-A, awarded by STScI. STScI is operated by the Association of Universities for Research in Astronomy, Incorporated, under NASA contract NAS5-26555.
W.B.H. acknowledges support from the National Science Foundation Graduate Research Fellowship Program under Grant Nos. 1842402 and 2236415. Any opinions, findings, conclusions, or recommendations expressed in this material are those of the author(s) and do not necessarily reflect the views of the National Science Foundation.
J.L. is supported by grant NSF-2206523.
P.H. is supported by grant NSF AST-230639.
H.K. was funded by the Academy of Finland projects 358691 and 358692.
JDL acknowledges support from a UK Research and Innovation Future Leaders Fellowship (grant references MR/T020784/1 and UKRI1062)
M.D.S. is funded by the Independent Research Fund Denmark (IRFD, grant number 10.46540/2032-00022B). 
Based on observations collected at the Centro Astron\'omico Hispano-Alem\'an (CAHA) at Calar Alto, operated jointly by Junta de Andaluc\'ia and Consejo Superior de Investigaciones Cient\'ificas (IAA-CSIC), under programmes: F17-3.5-001, F18-3.5-013, F19-3.5-001, 23A-3.5-003, 23B-3.5-004, 24B-3.5-001.
Based on observations collected at the European Southern Observatory under ESO programmes: 095.D-0091, 099.D-0022, 0102.D-0095, 106.2104, 114.26ZM, 116.28SK.

\end{acknowledgements}

\bibliographystyle{aa}
\bibliography{superch_host} 

@ARTICLE{2024ApJS..273...16P,
       author = {{Phillips}, M.~M. and {Ashall}, C. and {Brown}, Peter J. and {Galbany}, L. and {Tucker}, M.~A. and {Burns}, Christopher R. and {Contreras}, Carlos and {Hoeflich}, P. and {Hsiao}, E.~Y. and {Kumar}, S. and {Morrell}, Nidia and {Uddin}, Syed A. and {Baron}, E. and {Freedman}, Wendy L. and {Krisciunas}, Kevin and {Persson}, S.~E. and {Piro}, Anthony L. and {Shappee}, B.~J. and {Stritzinger}, Maximilian and {Suntzeff}, Nicholas B. and {Chakraborty}, Sudeshna and {Kirshner}, R.~P. and {Lu}, J. and {Marion}, G.~H. and {Polin}, Abigail and {Shahbandeh}, M.},
        title = "{1991T-like Supernovae}",
      journal = {\apjs},
         year = 2024,
        month = jul,
       volume = {273},
       number = {1},
          eid = {16},
        pages = {16},
          doi = {10.3847/1538-4365/ad4f7e},
archivePrefix = {arXiv},
       eprint = {2405.15027},
 primaryClass = {astro-ph.HE},
       adsurl = {https://ui.adsabs.harvard.edu/abs/2024ApJS..273...16P}
}

@ARTICLE{2024A&A...687L..19N,
       author = {{Nagao}, T. and {Maeda}, K. and {Mattila}, S. and {Kuncarayakti}, H. and {Guti{\'e}rrez}, C.~P. and {Cikota}, A.},
        title = "{The aspherical explosions of the 03fg-like Type Ia supernovae 2021zny and 2022ilv revealed by polarimetry}",
      journal = {\aap},
         year = 2024,
        month = jul,
       volume = {687},
          eid = {L19},
        pages = {L19},
          doi = {10.1051/0004-6361/202449999},
archivePrefix = {arXiv},
       eprint = {2406.18110},
 primaryClass = {astro-ph.HE},
       adsurl = {https://ui.adsabs.harvard.edu/abs/2024A&A...687L..19N}
}

@ARTICLE{Desai26_Ia_rates,
       author = {{Desai}, Dhvanil D. and {Shappee}, Benjamin J. and {Kochanek}, Christopher S. and {Stanek}, Krzysztof Z. and {Ashall}, Chris and {Beacom}, John F. and {Burns}, Christopher R. and {Do}, Aaron and {Dong}, Subo and {Hoogendam}, Willem B. and {Lu}, Jing and {Pessi}, Thallis and {Prieto}, Jose L. and {Thompson}, Todd A.},
        title = "{Supernova Rates and Luminosity Functions from ASAS-SN III: Over a Decade of Type Ia SNe and Their Subtypes}",
      journal = {arXiv e-prints},
         year = 2026,
        month = jan,
          eid = {arXiv:2602.00223},
        pages = {arXiv:2602.00223},
          doi = {10.48550/arXiv.2602.00223},
archivePrefix = {arXiv},
       eprint = {2602.00223},
 primaryClass = {astro-ph.HE},
       adsurl = {https://ui.adsabs.harvard.edu/abs/2026arXiv260200223D}
}

@ARTICLE{Inoue2026,
       author = {{Inoue}, Yusuke and {Maeda}, Keiichi and {Nagao}, Takashi and {Matsumoto}, Tatsuya},
        title = "{Formation of Circumstellar Material during Double-white-dwarf Mergers and the Early Excess Emissions in Type Ia Supernovae}",
      journal = {\apj},
         year = 2026,
        month = feb,
       volume = {997},
       number = {2},
          eid = {312},
        pages = {312},
          doi = {10.3847/1538-4357/ae2de7},
archivePrefix = {arXiv},
       eprint = {2512.10014},
 primaryClass = {astro-ph.HE},
       adsurl = {https://ui.adsabs.harvard.edu/abs/2026ApJ...997..312I}
}

@ARTICLE{Paniagua2026,
       author = {{Paniagua}, I.~A. Abreu and {Hoogendam}, W.~B. and {Jones}, D.~O. and {Dimitriadis}, G. and {Foley}, R.~J. and {Gall}, C. and {O'Brien}, J. and {Taggart}, K. and {Angus}, C.~R. and {Ashall}, C. and {Auchettl}, K. and {Coulter}, D.~A. and {Davis}, K.~W. and {de Boer}, T. and {Do}, A. and {Gao}, H. and {Izzo}, L. and {Lin}, C.-C. and {Lowe}, T.~B. and {Lai}, Z. and {Kaur}, R. and {Kong}, M.~Y. and {Rest}, A. and {Siebert}, M.~R. and {Yadavalli}, S.~K. and {Zenati}, Y. and {Wang}, Q.},
        title = "{The New Status Qvo? SN 2021qvo Is Another 2003fg-like Type Ia Supernova with a Rising Light-curve Bump}",
      journal = {\apj},
         year = 2026,
        month = feb,
       volume = {997},
       number = {2},
          eid = {261},
        pages = {261},
          doi = {10.3847/1538-4357/ae279b},
archivePrefix = {arXiv},
       eprint = {2508.13263},
 primaryClass = {astro-ph.HE},
       adsurl = {https://ui.adsabs.harvard.edu/abs/2026ApJ...997..261P}
}

@ARTICLE{2006PASP..118..129K,
       author = {{Kelz}, Andreas and {Verheijen}, Marc A.~W. and {Roth}, Martin M. and {Bauer}, Svend M. and {Becker}, Thomas and {Paschke}, Jens and {Popow}, Emil and {S{\'a}nchez}, Sebastian F. and {Laux}, Uwe},
        title = "{PMAS: The Potsdam Multi-Aperture Spectrophotometer. II. The Wide Integral Field Unit PPak}",
      journal = {\pasp},
         year = 2006,
        month = jan,
       volume = {118},
       number = {839},
        pages = {129-145},
          doi = {10.1086/497455},
archivePrefix = {arXiv},
       eprint = {astro-ph/0512557},
 primaryClass = {astro-ph},
       adsurl = {https://ui.adsabs.harvard.edu/abs/2006PASP..118..129K}
}

@ARTICLE{2019ApJ...886..152Y,
       author = {{Yao}, Yuhan and {Miller}, Adam A. and {Kulkarni}, S.~R. and {Bulla}, Mattia and {Masci}, Frank J. and {Goldstein}, Daniel A. and {Goobar}, Ariel and {Nugent}, Peter and {Dugas}, Alison and {Blagorodnova}, Nadia and {Neill}, James D. and {Rigault}, Mickael and {Sollerman}, Jesper and {Nordin}, J. and {Bellm}, Eric C. and {Cenko}, S. Bradley and {De}, Kishalay and {Dhawan}, Suhail and {Feindt}, Ulrich and {Fremling}, C. and {Gatkine}, Pradip and {Graham}, Matthew J. and {Graham}, Melissa L. and {Ho}, Anna Y.~Q. and {Hung}, T. and {Kasliwal}, Mansi M. and {Kupfer}, Thomas and {Laher}, Russ R. and {Perley}, Daniel A. and {Rusholme}, Ben and {Shupe}, David L. and {Soumagnac}, Maayane T. and {Taggart}, K. and {Walters}, Richard and {Yan}, Lin},
        title = "{ZTF Early Observations of Type Ia Supernovae. I. Properties of the 2018 Sample}",
      journal = {\apj},
         year = 2019,
        month = dec,
       volume = {886},
       number = {2},
          eid = {152},
        pages = {152},
          doi = {10.3847/1538-4357/ab4cf5},
archivePrefix = {arXiv},
       eprint = {1910.02967},
 primaryClass = {astro-ph.HE},
       adsurl = {https://ui.adsabs.harvard.edu/abs/2019ApJ...886..152Y}
}

@ARTICLE{2021ApJ...923L...8J,
       author = {{Jiang}, Ji-an and {Maeda}, Keiichi and {Kawabata}, Miho and {Doi}, Mamoru and {Shigeyama}, Toshikazu and {Tanaka}, Masaomi and {Tominaga}, Nozomu and {Nomoto}, Ken'ichi and {Niino}, Yuu and {Sako}, Shigeyuki and {Ohsawa}, Ryou and {Schramm}, Malte and {Yamanaka}, Masayuki and {Kobayashi}, Naoto and {Takahashi}, Hidenori and {Nakaoka}, Tatsuya and {Kawabata}, Koji S. and {Isogai}, Keisuke and {Aoki}, Tsutomu and {Kondo}, Sohei and {Mori}, Yuki and {Arimatsu}, Ko and {Kasuga}, Toshihiro and {Okumura}, Shin-ichiro and {Urakawa}, Seitaro and {Reichart}, Daniel E. and {Taguchi}, Kenta and {Arima}, Noriaki and {Beniyama}, Jin and {Uno}, Kohki and {Hamada}, Taisei},
        title = "{Discovery of the Fastest Early Optical Emission from Overluminous SN Ia 2020hvf: A Thermonuclear Explosion within a Dense Circumstellar Environment}",
      journal = {\apjl},
         year = 2021,
        month = dec,
       volume = {923},
       number = {1},
          eid = {L8},
        pages = {L8},
          doi = {10.3847/2041-8213/ac375f},
archivePrefix = {arXiv},
       eprint = {2111.09470},
 primaryClass = {astro-ph.HE},
       adsurl = {https://ui.adsabs.harvard.edu/abs/2021ApJ...923L...8J}
}

@ARTICLE{2023ApJ...943L..20S,
       author = {{Srivastav}, Shubham and {Smartt}, S.~J. and {Huber}, M.~E. and {Dimitriadis}, G. and {Chambers}, K.~C. and {Fulton}, Michael D. and {Moore}, Thomas and {Callan}, F.~P. and {Gillanders}, James H. and {Maguire}, K. and {Nicholl}, M. and {Shingles}, Luke J. and {Sim}, S.~A. and {Smith}, K.~W. and {Anderson}, J.~P. and {de Boer}, Thomas and {Chen}, Ting-Wan and {Gao}, Hua and {Young}, D.~R.},
        title = "{The Luminous Type Ia Supernova 2022ilv and Its Early Excess Emission}",
      journal = {\apjl},
         year = 2023,
        month = feb,
       volume = {943},
       number = {2},
          eid = {L20},
        pages = {L20},
          doi = {10.3847/2041-8213/acb2ce},
archivePrefix = {arXiv},
       eprint = {2211.10544},
 primaryClass = {astro-ph.HE},
       adsurl = {https://ui.adsabs.harvard.edu/abs/2023ApJ...943L..20S}
}

@ARTICLE{2022JOSS....7.4508M,
       author = {{M{\"u}ller-Bravo}, Tom{\'a}s and {Galbany}, Llu{\'\i}s},
        title = "{HostPhot: global and local photometry of galaxies hosting supernovae or other transients}",
      journal = {The Journal of Open Source Software},
         year = 2022,
        month = aug,
       volume = {7},
       number = {76},
          eid = {4508},
        pages = {4508},
          doi = {10.21105/joss.04508},
archivePrefix = {arXiv},
       eprint = {2208.08117},
 primaryClass = {astro-ph.CO},
       adsurl = {https://ui.adsabs.harvard.edu/abs/2022JOSS....7.4508M}
}

@ARTICLE{2022ApJ...938...47P,
       author = {{Phillips}, M.~M. and {Ashall}, C. and {Burns}, Christopher R. and {Contreras}, Carlos and {Galbany}, L. and {Hoeflich}, P. and {Hsiao}, E.~Y. and {Morrell}, Nidia and {Nugent}, Peter and {Uddin}, Syed A. and {Baron}, E. and {Freedman}, Wendy L. and {Harris}, Chelsea E. and {Krisciunas}, Kevin and {Kumar}, S. and {Lu}, J. and {Persson}, S.~E. and {Piro}, Anthony L. and {Polin}, Abigail and {Shahbandeh}, M. and {Stritzinger}, Maximilian and {Suntzeff}, Nicholas B.},
        title = "{The Absolute Magnitudes of 1991T-like Supernovae}",
      journal = {\apj},
         year = 2022,
        month = oct,
       volume = {938},
       number = {1},
          eid = {47},
        pages = {47},
          doi = {10.3847/1538-4357/ac9305},
archivePrefix = {arXiv},
       eprint = {2209.08031},
 primaryClass = {astro-ph.HE},
       adsurl = {https://ui.adsabs.harvard.edu/abs/2022ApJ...938...47P}
}

@ARTICLE{2023MNRAS.521.1162D,
       author = {{Dimitriadis}, Georgios and {Maguire}, Kate and {Karambelkar}, Viraj R. and {Lebron}, Ryan J. and {Liu}, Chang and {Kozyreva}, Alexandra and {Miller}, Adam A. and {Ridden-Harper}, Ryan and {Anderson}, Joseph P. and {Chen}, Ting-Wan and {Coughlin}, Michael and {Della Valle}, Massimo and {Drake}, Andrew and {Galbany}, Llu{\'\i}s and {Gromadzki}, Mariusz and {Groom}, Steven L. and {Guti{\'e}rrez}, Claudia P. and {Ihanec}, Nada and {Inserra}, Cosimo and {Johansson}, Joel and {M{\"u}ller-Bravo}, Tom{\'a}s E. and {Nicholl}, Matt and {Polin}, Abigail and {Rusholme}, Ben and {Schulze}, Steve and {Sollerman}, Jesper and {Srivastav}, Shubham and {Taggart}, Kirsty and {Wang}, Qinan and {Yang}, Yi and {Young}, David R.},
        title = "{SN 2021zny: an early flux excess combined with late-time oxygen emission suggests a double white dwarf merger event}",
      journal = {\mnras},
         year = 2023,
        month = may,
       volume = {521},
       number = {1},
        pages = {1162-1183},
          doi = {10.1093/mnras/stad536},
archivePrefix = {arXiv},
       eprint = {2302.08228},
 primaryClass = {astro-ph.HE},
       adsurl = {https://ui.adsabs.harvard.edu/abs/2023MNRAS.521.1162D}
}

@ARTICLE{2019ApJ...880...35C,
       author = {{Chen}, Ping and {Dong}, Subo and {Katz}, Boaz and {Kochanek}, C.~S. and {Kollmeier}, Juna A. and {Maguire}, K. and {Phillips}, M.~M. and {Prieto}, J.~L. and {Shappee}, B.~J. and {Stritzinger}, M.~D. and {Bose}, Subhash and {Brown}, Peter J. and {Holoien}, T.~W. -S. and {Galbany}, L. and {Milne}, Peter A. and {Morrell}, Nidia and {Piro}, Anthony L. and {Stanek}, K.~Z. and {Thompson}, Todd A. and {Young}, D.~R.},
        title = "{ASASSN-15pz: Revealing Significant Photometric Diversity among 2009dc-like, Peculiar SNe Ia}",
      journal = {\apj},
         year = 2019,
        month = jul,
       volume = {880},
       number = {1},
          eid = {35},
        pages = {35},
          doi = {10.3847/1538-4357/ab2630},
archivePrefix = {arXiv},
       eprint = {1904.03198},
 primaryClass = {astro-ph.HE},
       adsurl = {https://ui.adsabs.harvard.edu/abs/2019ApJ...880...35C}
}

@ARTICLE{2018ApJ...863..134H,
       author = {{Hsyu}, Tiffany and {Cooke}, Ryan J. and {Prochaska}, J. Xavier and {Bolte}, Michael},
        title = "{Searching for the Lowest-metallicity Galaxies in the Local Universe}",
      journal = {\apj},
         year = 2018,
        month = aug,
       volume = {863},
       number = {2},
          eid = {134},
        pages = {134},
          doi = {10.3847/1538-4357/aad18a},
       adsurl = {https://ui.adsabs.harvard.edu/abs/2018ApJ...863..134H}
}

@ARTICLE{2012ApJ...754...98B,
       author = {{Berg}, Danielle A. and {Skillman}, Evan D. and {Marble}, Andrew R. and {van Zee}, Liese and {Engelbracht}, Charles W. and {Lee}, Janice C. and {Kennicutt}, Robert C., Jr. and {Calzetti}, Daniela and {Dale}, Daniel A. and {Johnson}, Benjamin D.},
        title = "{Direct Oxygen Abundances for Low-luminosity LVL Galaxies}",
      journal = {\apj},
         year = 2012,
        month = aug,
       volume = {754},
       number = {2},
          eid = {98},
        pages = {98},
          doi = {10.1088/0004-637X/754/2/98},
archivePrefix = {arXiv},
       eprint = {1205.6782},
 primaryClass = {astro-ph.CO},
       adsurl = {https://ui.adsabs.harvard.edu/abs/2012ApJ...754...98B}
}

@ARTICLE{2008ApJS..178..247K,
       author = {{Kennicutt}, Robert C., Jr. and {Lee}, Janice C. and {Funes}, Jos{\'e} G. and {J.}, S. and {Sakai}, Shoko and {Akiyama}, Sanae},
        title = "{An H{\ensuremath{\alpha}} Imaging Survey of Galaxies in the Local 11 Mpc Volume}",
      journal = {\apjs},
         year = 2008,
        month = oct,
       volume = {178},
       number = {2},
        pages = {247-279},
          doi = {10.1086/590058},
archivePrefix = {arXiv},
       eprint = {0807.2035},
 primaryClass = {astro-ph},
       adsurl = {https://ui.adsabs.harvard.edu/abs/2008ApJS..178..247K}
}

@ARTICLE{2017ApJ...847...38Y,
       author = {{Yang}, Huan and {Malhotra}, Sangeeta and {Rhoads}, James E. and {Wang}, Junxian},
        title = "{Blueberry Galaxies: The Lowest Mass Young Starbursts}",
      journal = {\apj},
         year = 2017,
        month = sep,
       volume = {847},
       number = {1},
          eid = {38},
        pages = {38},
          doi = {10.3847/1538-4357/aa8809},
archivePrefix = {arXiv},
       eprint = {1706.02819},
 primaryClass = {astro-ph.GA},
       adsurl = {https://ui.adsabs.harvard.edu/abs/2017ApJ...847...38Y}
}

@INBOOK{Taubenberger17,
       author = {{Taubenberger}, Stefan},
        title = "{The Extremes of Thermonuclear Supernovae}",
    booktitle = {Handbook of Supernovae},
         year = 2017,
       editor = {{Alsabti}, Athem W. and {Murdin}, Paul},
        pages = {317},
          doi = {10.1007/978-3-319-21846-5\_37},
       adsurl = {https://ui.adsabs.harvard.edu/abs/2017hsn..book..317T}
}

@ARTICLE{Livio18,
       author = {{Livio}, Mario and {Mazzali}, Paolo},
        title = "{On the progenitors of Type Ia supernovae}",
      journal = {\physrep},
         year = 2018,
        month = mar,
       volume = {736},
        pages = {1-23},
          doi = {10.1016/j.physrep.2018.02.002},
archivePrefix = {arXiv},
       eprint = {1802.03125},
 primaryClass = {astro-ph.SR},
       adsurl = {https://ui.adsabs.harvard.edu/abs/2018PhR...736....1L}
}

@article{Moe2019,
  author  = {Moe, Maxwell and Kratter, Kaitlin M. and Badenes, Carles},
  title   = {The Close Binary Fraction of Solar-type Stars Is Strongly 
             Anticorrelated with Metallicity},
  journal = {The Astrophysical Journal},
  volume  = {875},
  number  = {1},
  pages   = {61},
  year    = {2019},
  doi     = {10.3847/1538-4357/ab0d88},
  eid     = {61},
  eprint  = {1808.02116},
  archivePrefix = {arXiv},
  primaryClass  = {astro-ph.SR}
}

@ARTICLE{Maoz14,
       author = {{Maoz}, Dan and {Mannucci}, Filippo and {Nelemans}, Gijs},
        title = "{Observational Clues to the Progenitors of Type Ia Supernovae}",
      journal = {\araa},
         year = 2014,
        month = aug,
       volume = {52},
        pages = {107-170},
          doi = {10.1146/annurev-astro-082812-141031},
archivePrefix = {arXiv},
       eprint = {1312.0628},
 primaryClass = {astro-ph.CO},
       adsurl = {https://ui.adsabs.harvard.edu/abs/2014ARA&A..52..107M}
}

@ARTICLE{Hoeflich17,
       author = {{Hoeflich}, P. and {Hsiao}, E.~Y. and {Ashall}, C. and {Burns}, C.~R. and {Diamond}, T.~R. and {Phillips}, M.~M. and {Sand}, D. and {Stritzinger}, M.~D. and {Suntzeff}, N. and {Contreras}, C. and {Krisciunas}, K. and {Morrell}, N. and {Wang}, L.},
        title = "{Light and Color Curve Properties of Type Ia Supernovae: Theory Versus Observations}",
      journal = {\apj},
         year = 2017,
        month = sep,
       volume = {846},
       number = {1},
          eid = {58},
        pages = {58},
          doi = {10.3847/1538-4357/aa84b2},
archivePrefix = {arXiv},
       eprint = {1707.05350},
 primaryClass = {astro-ph.SR},
       adsurl = {https://ui.adsabs.harvard.edu/abs/2017ApJ...846...58H}
}

@ARTICLE{Blondin17,
       author = {{Blondin}, St{\'e}phane and {Dessart}, Luc and {Hillier}, D. John and {Khokhlov}, Alexei M.},
        title = "{Evidence for sub-Chandrasekhar-mass progenitors of Type Ia supernovae at the faint end of the width-luminosity relation}",
      journal = {\mnras},
         year = 2017,
        month = sep,
       volume = {470},
       number = {1},
        pages = {157-165},
          doi = {10.1093/mnras/stw2492},
archivePrefix = {arXiv},
       eprint = {1706.01901},
 primaryClass = {astro-ph.SR},
       adsurl = {https://ui.adsabs.harvard.edu/abs/2017MNRAS.470..157B}
}

@ARTICLE{Webbink84,
       author = {{Webbink}, R.~F.},
        title = "{Double white dwarfs as progenitors of R Coronae Borealis stars and type I supernovae.}",
      journal = {\apj},
         year = 1984,
        month = feb,
       volume = {277},
        pages = {355-360},
          doi = {10.1086/161701},
       adsurl = {https://ui.adsabs.harvard.edu/abs/1984ApJ...277..355W}
}

@ARTICLE{Iben84,
       author = {{Iben}, I., Jr. and {Tutukov}, A.~V.},
        title = "{Supernovae of type I as end products of the evolution of binaries with components of moderate initial mass.}",
      journal = {\apjs},
         year = 1984,
        month = feb,
       volume = {54},
        pages = {335-372},
          doi = {10.1086/190932},
       adsurl = {https://ui.adsabs.harvard.edu/abs/1984ApJS...54..335I}
}

@ARTICLE{Perlmutter99,
       author = {{Perlmutter}, S. and {Aldering}, G. and {Goldhaber}, G. and {Knop}, R.~A. and {Nugent}, P. and {Castro}, P.~G. and {Deustua}, S. and {Fabbro}, S. and {Goobar}, A. and {Groom}, D.~E. and {Hook}, I.~M. and {Kim}, A.~G. and {Kim}, M.~Y. and {Lee}, J.~C. and {Nunes}, N.~J. and {Pain}, R. and {Pennypacker}, C.~R. and {Quimby}, R. and {Lidman}, C. and {Ellis}, R.~S. and {Irwin}, M. and {McMahon}, R.~G. and {Ruiz-Lapuente}, P. and {Walton}, N. and {Schaefer}, B. and {Boyle}, B.~J. and {Filippenko}, A.~V. and {Matheson}, T. and {Fruchter}, A.~S. and {Panagia}, N. and {Newberg}, H.~J.~M. and {Couch}, W.~J. and {Project}, The Supernova Cosmology},
        title = "{Measurements of {\ensuremath{\Omega}} and {\ensuremath{\Lambda}} from 42 High-Redshift Supernovae}",
      journal = {\apj},
         year = 1999,
        month = jun,
       volume = {517},
       number = {2},
        pages = {565-586},
          doi = {10.1086/307221},
archivePrefix = {arXiv},
       eprint = {astro-ph/9812133},
 primaryClass = {astro-ph},
       adsurl = {https://ui.adsabs.harvard.edu/abs/1999ApJ...517..565P}
}

@ARTICLE{Riess98,
       author = {{Riess}, Adam G. and {Filippenko}, Alexei V. and {Challis}, Peter and {Clocchiatti}, Alejandro and {Diercks}, Alan and {Garnavich}, Peter M. and {Gilliland}, Ron L. and {Hogan}, Craig J. and {Jha}, Saurabh and {Kirshner}, Robert P. and {Leibundgut}, B. and {Phillips}, M.~M. and {Reiss}, David and {Schmidt}, Brian P. and {Schommer}, Robert A. and {Smith}, R. Chris and {Spyromilio}, J. and {Stubbs}, Christopher and {Suntzeff}, Nicholas B. and {Tonry}, John},
        title = "{Observational Evidence from Supernovae for an Accelerating Universe and a Cosmological Constant}",
      journal = {\aj},
         year = 1998,
        month = sep,
       volume = {116},
       number = {3},
        pages = {1009-1038},
          doi = {10.1086/300499},
archivePrefix = {arXiv},
       eprint = {astro-ph/9805201},
 primaryClass = {astro-ph},
       adsurl = {https://ui.adsabs.harvard.edu/abs/1998AJ....116.1009R}
}

@ARTICLE{2011ApJ...737..103S,
       author = {{Schlafly}, Edward F. and {Finkbeiner}, Douglas P.},
        title = "{Measuring Reddening with Sloan Digital Sky Survey Stellar Spectra and Recalibrating SFD}",
      journal = {\apj},
         year = 2011,
        month = aug,
       volume = {737},
       number = {2},
          eid = {103},
        pages = {103},
          doi = {10.1088/0004-637X/737/2/103},
archivePrefix = {arXiv},
       eprint = {1012.4804},
 primaryClass = {astro-ph.GA},
       adsurl = {https://ui.adsabs.harvard.edu/abs/2011ApJ...737..103S}
}

@ARTICLE{2003MNRAS.344.1000B,
       author = {{Bruzual}, G. and {Charlot}, S.},
        title = "{Stellar population synthesis at the resolution of 2003}",
      journal = {\mnras},
         year = 2003,
        month = oct,
       volume = {344},
       number = {4},
        pages = {1000-1028},
          doi = {10.1046/j.1365-8711.2003.06897.x},
archivePrefix = {arXiv},
       eprint = {astro-ph/0309134},
 primaryClass = {astro-ph},
       adsurl = {https://ui.adsabs.harvard.edu/abs/2003MNRAS.344.1000B}
}

@ARTICLE{2009MNRAS.399.1191C,
       author = {{Cardamone}, Carolin and {Schawinski}, Kevin and {Sarzi}, Marc and {Bamford}, Steven P. and {Bennert}, Nicola and {Urry}, C.~M. and {Lintott}, Chris and {Keel}, William C. and {Parejko}, John and {Nichol}, Robert C. and {Thomas}, Daniel and {Andreescu}, Dan and {Murray}, Phil and {Raddick}, M. Jordan and {Slosar}, An{\v{z}}e and {Szalay}, Alex and {Vandenberg}, Jan},
        title = "{Galaxy Zoo Green Peas: discovery of a class of compact extremely star-forming galaxies}",
      journal = {\mnras},
         year = 2009,
        month = nov,
       volume = {399},
       number = {3},
        pages = {1191-1205},
          doi = {10.1111/j.1365-2966.2009.15383.x},
archivePrefix = {arXiv},
       eprint = {0907.4155},
 primaryClass = {astro-ph.CO},
       adsurl = {https://ui.adsabs.harvard.edu/abs/2009MNRAS.399.1191C}
}

@ARTICLE{2006MNRAS.371..703S,
       author = {{S{\'a}nchez-Bl{\'a}zquez}, P. and {Peletier}, R.~F. and {Jim{\'e}nez-Vicente}, J. and {Cardiel}, N. and {Cenarro}, A.~J. and {Falc{\'o}n-Barroso}, J. and {Gorgas}, J. and {Selam}, S. and {Vazdekis}, A.},
        title = "{Medium-resolution Isaac Newton Telescope library of empirical spectra}",
      journal = {\mnras},
         year = 2006,
        month = sep,
       volume = {371},
       number = {2},
        pages = {703-718},
          doi = {10.1111/j.1365-2966.2006.10699.x},
archivePrefix = {arXiv},
       eprint = {astro-ph/0607009},
 primaryClass = {astro-ph},
       adsurl = {https://ui.adsabs.harvard.edu/abs/2006MNRAS.371..703S}
}

@ARTICLE{2003PASP..115..763C,
       author = {{Chabrier}, Gilles},
        title = "{Galactic Stellar and Substellar Initial Mass Function}",
      journal = {\pasp},
         year = 2003,
        month = jul,
       volume = {115},
       number = {809},
        pages = {763-795},
          doi = {10.1086/376392},
archivePrefix = {arXiv},
       eprint = {astro-ph/0304382},
 primaryClass = {astro-ph},
       adsurl = {https://ui.adsabs.harvard.edu/abs/2003PASP..115..763C}
}

@ARTICLE{2007A&A...469..239M,
       author = {{Marigo}, P. and {Girardi}, L.},
        title = "{Evolution of asymptotic giant branch stars. I. Updated synthetic TP-AGB models and their basic calibration}",
      journal = {\aap},
         year = 2007,
        month = jul,
       volume = {469},
       number = {1},
        pages = {239-263},
          doi = {10.1051/0004-6361:20066772},
archivePrefix = {arXiv},
       eprint = {astro-ph/0703139},
 primaryClass = {astro-ph},
       adsurl = {https://ui.adsabs.harvard.edu/abs/2007A&A...469..239M}
}

@ARTICLE{2008A&A...482..883M,
       author = {{Marigo}, P. and {Girardi}, L. and {Bressan}, A. and {Groenewegen}, M.~A.~T. and {Silva}, L. and {Granato}, G.~L.},
        title = "{Evolution of asymptotic giant branch stars. II. Optical to far-infrared isochrones with improved TP-AGB models}",
      journal = {\aap},
         year = 2008,
        month = may,
       volume = {482},
       number = {3},
        pages = {883-905},
          doi = {10.1051/0004-6361:20078467},
archivePrefix = {arXiv},
       eprint = {0711.4922},
 primaryClass = {astro-ph},
       adsurl = {https://ui.adsabs.harvard.edu/abs/2008A&A...482..883M}
}

@ARTICLE{1999PASP..111...63F,
       author = {{Fitzpatrick}, Edward L.},
        title = "{Correcting for the Effects of Interstellar Extinction}",
      journal = {\pasp},
         year = 1999,
        month = jan,
       volume = {111},
       number = {755},
        pages = {63-75},
          doi = {10.1086/316293},
archivePrefix = {arXiv},
       eprint = {astro-ph/9809387},
 primaryClass = {astro-ph},
       adsurl = {https://ui.adsabs.harvard.edu/abs/1999PASP..111...63F}
}

@ARTICLE{2016A&A...594A..36S,
       author = {{S{\'a}nchez}, S.~F. and {Garc{\'\i}a-Benito}, R. and {Zibetti}, S. and {Walcher}, C.~J. and {Husemann}, B. and {Mendoza}, M.~A. and {Galbany}, L. and {Falc{\'o}n-Barroso}, J. and {Mast}, D. and {Aceituno}, J. and {Aguerri}, J.~A.~L. and {Alves}, J. and {Amorim}, A.~L. and {Ascasibar}, Y. and {Barrado-Navascues}, D. and {Barrera-Ballesteros}, J. and {Bekerait{\`e}}, S. and {Bland-Hawthorn}, J. and {Cano D{\'\i}az}, M. and {Cid Fernandes}, R. and {Cavichia}, O. and {Cortijo}, C. and {Dannerbauer}, H. and {Demleitner}, M. and {D{\'\i}az}, A. and {Dettmar}, R.~J. and {de Lorenzo-C{\'a}ceres}, A. and {del Olmo}, A. and {Galazzi}, A. and {Garc{\'\i}a-Lorenzo}, B. and {Gil de Paz}, A. and {Gonz{\'a}lez Delgado}, R. and {Holmes}, L. and {Igl{\'e}sias-P{\'a}ramo}, J. and {Kehrig}, C. and {Kelz}, A. and {Kennicutt}, R.~C. and {Kleemann}, B. and {Lacerda}, E.~A.~D. and {L{\'o}pez Fern{\'a}ndez}, R. and {L{\'o}pez S{\'a}nchez}, A.~R. and {Lyubenova}, M. and {Marino}, R. and {M{\'a}rquez}, I. and {Mendez-Abreu}, J. and {Moll{\'a}}, M. and {Monreal-Ibero}, A. and {Ortega Minakata}, R. and {Torres-Papaqui}, J.~P. and {P{\'e}rez}, E. and {Rosales-Ortega}, F.~F. and {Roth}, M.~M. and {S{\'a}nchez-Bl{\'a}zquez}, P. and {Schilling}, U. and {Spekkens}, K. and {Vale Asari}, N. and {van den Bosch}, R.~C.~E. and {van de Ven}, G. and {Vilchez}, J.~M. and {Wild}, V. and {Wisotzki}, L. and {Y{\i}ld{\i}r{\i}m}, A. and {Ziegler}, B.},
        title = "{CALIFA, the Calar Alto Legacy Integral Field Area survey. IV. Third public data release}",
      journal = {\aap},
         year = 2016,
        month = oct,
       volume = {594},
          eid = {A36},
        pages = {A36},
          doi = {10.1051/0004-6361/201628661},
archivePrefix = {arXiv},
       eprint = {1604.02289},
 primaryClass = {astro-ph.GA},
       adsurl = {https://ui.adsabs.harvard.edu/abs/2016A&A...594A..36S}
}

@ARTICLE{2017A&A...602A..85K,
       author = {{Kr{\"u}hler}, T. and {Kuncarayakti}, H. and {Schady}, P. and {Anderson}, J.~P. and {Galbany}, L. and {Gensior}, J.},
        title = "{Hot gas around SN 1998bw: Inferring the progenitor from its environment}",
      journal = {\aap},
         year = 2017,
        month = jun,
       volume = {602},
          eid = {A85},
        pages = {A85},
          doi = {10.1051/0004-6361/201630268},
archivePrefix = {arXiv},
       eprint = {1702.05430},
 primaryClass = {astro-ph.GA},
       adsurl = {https://ui.adsabs.harvard.edu/abs/2017A&A...602A..85K}
}

@ARTICLE{2016MNRAS.455.4087G,
       author = {{Galbany}, L. and {Anderson}, J.~P. and {Rosales-Ortega}, F.~F. and
         {Kuncarayakti}, H. and {Kr{\"u}hler}, T. and {S{\'a}nchez}, S.~F. and
         {Falc{\'o}n-Barroso}, J. and {P{\'e}rez}, E. and {Maureira}, J.~C. and
         {Hamuy}, M. and {Gonz{\'a}lez-Gait{\'a}n}, S. and {F{\"o}rster}, F. and
         {Moral}, V.},
        title = "{Characterizing the environments of supernovae with MUSE}",
      journal = {\mnras},
         year = 2016,
        month = feb,
       volume = {455},
       number = {4},
        pages = {4087-4099},
          doi = {10.1093/mnras/stv2620},
archivePrefix = {arXiv},
       eprint = {1511.01495},
 primaryClass = {astro-ph.GA},
       adsurl = {https://ui.adsabs.harvard.edu/abs/2016MNRAS.455.4087G}
}

@ARTICLE{2009ApJS..182..543A,
       author = {{Abazajian}, Kevork N. and {Adelman-McCarthy}, Jennifer K. and {Ag{\"u}eros}, Marcel A. and {Allam}, Sahar S. and {Allende Prieto}, Carlos and {An}, Deokkeun and {Anderson}, Kurt S.~J. and {Anderson}, Scott F. and {Annis}, James and {Bahcall}, Neta A. and {Bailer-Jones}, C.~A.~L. and {Barentine}, J.~C. and {Bassett}, Bruce A. and {Becker}, Andrew C. and {Beers}, Timothy C. and {Bell}, Eric F. and {Belokurov}, Vasily and {Berlind}, Andreas A. and {Berman}, Eileen F. and {Bernardi}, Mariangela and {Bickerton}, Steven J. and {Bizyaev}, Dmitry and {Blakeslee}, John P. and {Blanton}, Michael R. and {Bochanski}, John J. and {Boroski}, William N. and {Brewington}, Howard J. and {Brinchmann}, Jarle and {Brinkmann}, J. and {Brunner}, Robert J. and {Budav{\'a}ri}, Tam{\'a}s and {Carey}, Larry N. and {Carliles}, Samuel and {Carr}, Michael A. and {Castander}, Francisco J. and {Cinabro}, David and {Connolly}, A.~J. and {Csabai}, Istv{\'a}n and {Cunha}, Carlos E. and {Czarapata}, Paul C. and {Davenport}, James R.~A. and {de Haas}, Ernst and {Dilday}, Ben and {Doi}, Mamoru and {Eisenstein}, Daniel J. and {Evans}, Michael L. and {Evans}, N.~W. and {Fan}, Xiaohui and {Friedman}, Scott D. and {Frieman}, Joshua A. and {Fukugita}, Masataka and {G{\"a}nsicke}, Boris T. and {Gates}, Evalyn and {Gillespie}, Bruce and {Gilmore}, G. and {Gonzalez}, Belinda and {Gonzalez}, Carlos F. and {Grebel}, Eva K. and {Gunn}, James E. and {Gy{\"o}ry}, Zsuzsanna and {Hall}, Patrick B. and {Harding}, Paul and {Harris}, Frederick H. and {Harvanek}, Michael and {Hawley}, Suzanne L. and {Hayes}, Jeffrey J.~E. and {Heckman}, Timothy M. and {Hendry}, John S. and {Hennessy}, Gregory S. and {Hindsley}, Robert B. and {Hoblitt}, J. and {Hogan}, Craig J. and {Hogg}, David W. and {Holtzman}, Jon A. and {Hyde}, Joseph B. and {Ichikawa}, Shin-ichi and {Ichikawa}, Takashi and {Im}, Myungshin and {Ivezi{\'c}}, {\v{Z}}eljko and {Jester}, Sebastian and {Jiang}, Linhua and {Johnson}, Jennifer A. and {Jorgensen}, Anders M. and {Juri{\'c}}, Mario and {Kent}, Stephen M. and {Kessler}, R. and {Kleinman}, S.~J. and {Knapp}, G.~R. and {Konishi}, Kohki and {Kron}, Richard G. and {Krzesinski}, Jurek and {Kuropatkin}, Nikolay and {Lampeitl}, Hubert and {Lebedeva}, Svetlana and {Lee}, Myung Gyoon and {Lee}, Young Sun and {French Leger}, R. and {L{\'e}pine}, S{\'e}bastien and {Li}, Nolan and {Lima}, Marcos and {Lin}, Huan and {Long}, Daniel C. and {Loomis}, Craig P. and {Loveday}, Jon and {Lupton}, Robert H. and {Magnier}, Eugene and {Malanushenko}, Olena and {Malanushenko}, Viktor and {Mandelbaum}, Rachel and {Margon}, Bruce and {Marriner}, John P. and {Mart{\'\i}nez-Delgado}, David and {Matsubara}, Takahiko and {McGehee}, Peregrine M. and {McKay}, Timothy A. and {Meiksin}, Avery and {Morrison}, Heather L. and {Mullally}, Fergal and {Munn}, Jeffrey A. and {Murphy}, Tara and {Nash}, Thomas and {Nebot}, Ada and {Neilsen}, Eric H., Jr. and {Newberg}, Heidi Jo and {Newman}, Peter R. and {Nichol}, Robert C. and {Nicinski}, Tom and {Nieto-Santisteban}, Maria and {Nitta}, Atsuko and {Okamura}, Sadanori and {Oravetz}, Daniel J. and {Ostriker}, Jeremiah P. and {Owen}, Russell and {Padmanabhan}, Nikhil and {Pan}, Kaike and {Park}, Changbom and {Pauls}, George and {Peoples}, John, Jr. and {Percival}, Will J. and {Pier}, Jeffrey R. and {Pope}, Adrian C. and {Pourbaix}, Dimitri and {Price}, Paul A. and {Purger}, Norbert and {Quinn}, Thomas and {Raddick}, M. Jordan and {Re Fiorentin}, Paola and {Richards}, Gordon T. and {Richmond}, Michael W. and {Riess}, Adam G. and {Rix}, Hans-Walter and {Rockosi}, Constance M. and {Sako}, Masao and {Schlegel}, David J. and {Schneider}, Donald P. and {Scholz}, Ralf-Dieter and {Schreiber}, Matthias R. and {Schwope}, Axel D. and {Seljak}, Uro{\v{s}} and {Sesar}, Branimir and {Sheldon}, Erin and {Shimasaku}, Kazu and {Sibley}, Valena C. and {Simmons}, A.~E. and {Sivarani}, Thirupathi and {Allyn Smith}, J. and {Smith}, Martin C. and {Smol{\v{c}}i{\'c}}, Vernesa and {Snedden}, Stephanie A. and {Stebbins}, Albert and {Steinmetz}, Matthias and {Stoughton}, Chris and {Strauss}, Michael A. and {SubbaRao}, Mark and {Suto}, Yasushi and {Szalay}, Alexander S. and {Szapudi}, Istv{\'a}n and {Szkody}, Paula and {Tanaka}, Masayuki and {Tegmark}, Max and {Teodoro}, Luis F.~A. and {Thakar}, Aniruddha R. and {Tremonti}, Christy A. and {Tucker}, Douglas L. and {Uomoto}, Alan and {Vanden Berk}, Daniel E. and {Vandenberg}, Jan and {Vidrih}, S. and {Vogeley}, Michael S. and {Voges}, Wolfgang and {Vogt}, Nicole P. and {Wadadekar}, Yogesh and {Watters}, Shannon and {Weinberg}, David H. and {West}, Andrew A. and {White}, Simon D.~M. and {Wilhite}, Brian C. and {Wonders}, Alainna C. and {Yanny}, Brian and {Yocum}, D.~R. and {York}, Donald G. and {Zehavi}, Idit and {Zibetti}, Stefano and {Zucker}, Daniel B.},
        title = "{The Seventh Data Release of the Sloan Digital Sky Survey}",
      journal = {\apjs},
         year = 2009,
        month = jun,
       volume = {182},
       number = {2},
        pages = {543-558},
          doi = {10.1088/0067-0049/182/2/543},
archivePrefix = {arXiv},
       eprint = {0812.0649},
 primaryClass = {astro-ph},
       adsurl = {https://ui.adsabs.harvard.edu/abs/2009ApJS..182..543A}
}

@ARTICLE{2005MNRAS.358..363C,
   author = {{Cid Fernandes}, R. and {Mateus}, A. and {Sodr{\'e}}, L. and 
	{Stasi{\'n}ska}, G. and {Gomes}, J.~M.},
    title = "{Semi-empirical analysis of Sloan Digital Sky Survey galaxies - I. Spectral synthesis method}",
  journal = {\mnras},
   eprint = {astro-ph/0412481},
     year = 2005,
    month = apr,
   volume = 358,
    pages = {363-378},
      doi = {10.1111/j.1365-2966.2005.08752.x},
   adsurl = {http://adsabs.harvard.edu/abs/2005MNRAS.358..363C}
}

@ARTICLE{2004MNRAS.348L..59P,
   author = {{Pettini}, M. and {Pagel}, B.~E.~J.},
    title = "{[OIII]/[NII] as an abundance indicator at high redshift}",
  journal = {\mnras},
   eprint = {astro-ph/0401128},
     year = 2004,
    month = mar,
   volume = 348,
    pages = {L59-L63},
      doi = {10.1111/j.1365-2966.2004.07591.x},
   adsurl = {http://adsabs.harvard.edu/abs/2004MNRAS.348L..59P}
}

@ARTICLE{2016Ap&SS.361...61D,
   author = {{Dopita}, M.~A. and {Kewley}, L.~J. and {Sutherland}, R.~S. and 
	{Nicholls}, D.~C.},
    title = "{Chemical abundances in high-redshift galaxies: a powerful new emission line diagnostic}",
  journal = {\apss},
archivePrefix = "arXiv",
   eprint = {1601.01337},
     year = 2016,
    month = feb,
   volume = 361,
      eid = {61},
    pages = {61},
      doi = {10.1007/s10509-016-2657-8},
   adsurl = {http://adsabs.harvard.edu/abs/2016Ap%26SS.361...61D}
}

@ARTICLE{1998ARA&A..36..189K,
   author = {{Kennicutt}, Jr., R.~C.},
    title = "{Star Formation in Galaxies Along the Hubble Sequence}",
  journal = {\araa},
   eprint = {astro-ph/9807187},
     year = 1998,
   volume = 36,
    pages = {189-232},
      doi = {10.1146/annurev.astro.36.1.189},
   adsurl = {http://adsabs.harvard.edu/abs/1998ARA%26A..36..189K}
}

@ARTICLE{S12,
   author = {{Stanishev}, V. and {Rodrigues}, M. and {Mour{\~a}o}, A. and 
	{Flores}, H.},
    title = "{Type Ia supernova host galaxies as seen with IFU spectroscopy}",
  journal = {\aap},
archivePrefix = "arXiv",
   eprint = {1205.5183},
 primaryClass = "astro-ph.CO",
     year = 2012,
    month = sep,
   volume = 545,
      eid = {A58},
    pages = {A58},
      doi = {10.1051/0004-6361/201219188},
   adsurl = {http://adsabs.harvard.edu/abs/2012A%26A...545A..58S}
}

@INPROCEEDINGS{2007ASPC..374..303B,
   author = {{Bruzual}, G.},
    title = "{Stellar Populations: High Spectral Resolution Libraries. Improved TP-AGB Treatment}",
booktitle = {From Stars to Galaxies: Building the Pieces to Build Up the Universe},
     year = 2007,
   series = {Astronomical Society of the Pacific Conference Series},
   volume = 374,
   eprint = {arXiv:astro-ph/0702091},
   editor = {{Vallenari}, A. and {Tantalo}, R. and {Portinari}, L. and {Moretti}, A.
	},
    month = dec,
    pages = {303},
   adsurl = {http://adsabs.harvard.edu/abs/2007ASPC..374..303B}
}

@INPROCEEDINGS{2010SPIE.7735E..08B,
   author = {{Bacon}, R. and {Accardo}, M. and {Adjali}, L. and {Anwand}, H. and 
	{Bauer}, S. and {Biswas}, I. and {Blaizot}, J. and {Boudon}, D. and 
	{Brau-Nogue}, S. and {Brinchmann}, J. and {Caillier}, P. and 
	{Capoani}, L. and {Carollo}, C.~M. and {Contini}, T. and {Couderc}, P. and 
	{Daguis{\'e}}, E. and {Deiries}, S. and {Delabre}, B. and {Dreizler}, S. and 
	{Dubois}, J. and {Dupieux}, M. and {Dupuy}, C. and {Emsellem}, E. and 
	{Fechner}, T. and {Fleischmann}, A. and {Fran{\c c}ois}, M. and 
	{Gallou}, G. and {Gharsa}, T. and {Glindemann}, A. and {Gojak}, D. and 
	{Guiderdoni}, B. and {Hansali}, G. and {Hahn}, T. and {Jarno}, A. and 
	{Kelz}, A. and {Koehler}, C. and {Kosmalski}, J. and {Laurent}, F. and 
	{Le Floch}, M. and {Lilly}, S.~J. and {Lizon}, J.-L. and {Loupias}, M. and 
	{Manescau}, A. and {Monstein}, C. and {Nicklas}, H. and {Olaya}, J.-C. and 
	{Pares}, L. and {Pasquini}, L. and {P{\'e}contal-Rousset}, A. and 
	{Pell{\'o}}, R. and {Petit}, C. and {Popow}, E. and {Reiss}, R. and 
	{Remillieux}, A. and {Renault}, E. and {Roth}, M. and {Rupprecht}, G. and 
	{Serre}, D. and {Schaye}, J. and {Soucail}, G. and {Steinmetz}, M. and 
	{Streicher}, O. and {Stuik}, R. and {Valentin},, H. and {Vernet}, J. and 
	{Weilbacher}, P. and {Wisotzki}, L. and {Yerle}, N.},
    title = "{The MUSE second-generation VLT instrument}",
booktitle = {Ground-based and Airborne Instrumentation for Astronomy III},
     year = 2010,
   series = {\procspie},
   volume = 7735,
    month = jul,
      eid = {773508},
    pages = {773508},
      doi = {10.1117/12.856027},
   adsurl = {http://adsabs.harvard.edu/abs/2010SPIE.7735E..08B}
}

@ARTICLE{2005PASP..117..620R,
   author = {{Roth}, M.~M. and {Kelz}, A. and {Fechner}, T. and {Hahn}, T. and 
	{Bauer}, {S.-M.} and {Becker}, T. and {B{\"o}hm}, P. and {Christensen}, L. and 
	{Dionies}, F. and {Paschke}, J. and {Popow}, E. and {Wolter}, D. and 
	{Schmoll}, J. and {Laux}, U. and {Altmann}, W.},
    title = "{PMAS: The Potsdam Multi-Aperture Spectrophotometer. I. Design, Manufacture, and Performance}",
  journal = {\pasp},
   eprint = {arXiv:astro-ph/0502581},
     year = 2005,
    month = jun,
   volume = 117,
    pages = {620-642},
      doi = {10.1086/429877},
   adsurl = {http://adsabs.harvard.edu/abs/2005PASP..117..620R}
}

@ARTICLE{2024MNRAS.530.5016D,
       author = {{Desai}, D.~D. and {Kochanek}, C.~S. and {Shappee}, B.~J. and {Jayasinghe}, T. and {Stanek}, K.~Z. and {Holoien}, T.~W. -S. and {Thompson}, T.~A. and {Ashall}, C. and {Beacom}, J.~F. and {Do}, A. and {Dong}, Subo and {Prieto}, J.~L.},
        title = "{Supernova rates and luminosity functions from ASAS-SN I: 2014-2017 Type Ia SNe and their subtypes}",
      journal = {\mnras},
         year = 2024,
        month = jun,
       volume = {530},
       number = {4},
        pages = {5016-5029},
          doi = {10.1093/mnras/stae606},
archivePrefix = {arXiv},
       eprint = {2306.11100},
 primaryClass = {astro-ph.HE},
       adsurl = {https://ui.adsabs.harvard.edu/abs/2024MNRAS.530.5016D}
}

@ARTICLE{2022A&A...659A..89G,
       author = {{Galbany}, Llu{\'\i}s and {Smith}, Mat and {Duarte Puertas}, Salvador and {Gonz{\'a}lez-Gait{\'a}n}, Santiago and {Pessa}, Ismael and {Sako}, Masao and {Iglesias-P{\'a}ramo}, Jorge and {L{\'o}pez-S{\'a}nchez}, A.~R. and {Moll{\'a}}, Mercedes and {V{\'\i}lchez}, Jos{\'e} M.},
        title = "{Aperture-corrected spectroscopic type Ia supernova host galaxy properties}",
      journal = {\aap},
         year = 2022,
        month = mar,
       volume = {659},
          eid = {A89},
        pages = {A89},
          doi = {10.1051/0004-6361/202141568},
archivePrefix = {arXiv},
       eprint = {2112.02517},
 primaryClass = {astro-ph.GA},
       adsurl = {https://ui.adsabs.harvard.edu/abs/2022A&A...659A..89G}
}

@ARTICLE{2018A&A...613A..35K,
       author = {{Kuncarayakti}, H. and {Anderson}, J.~P. and {Galbany}, L. and {Maeda}, K. and {Hamuy}, M. and {Aldering}, G. and {Arimoto}, N. and {Doi}, M. and {Morokuma}, T. and {Usuda}, T.},
        title = "{Constraints on core-collapse supernova progenitors from explosion site integral field spectroscopy}",
      journal = {\aap},
         year = 2018,
        month = may,
       volume = {613},
          eid = {A35},
        pages = {A35},
          doi = {10.1051/0004-6361/201731923},
archivePrefix = {arXiv},
       eprint = {1711.05765},
 primaryClass = {astro-ph.SR},
       adsurl = {https://ui.adsabs.harvard.edu/abs/2018A&A...613A..35K}
}

@ARTICLE{2018ApJ...855..107G,
       author = {{Galbany}, L. and {Anderson}, J.~P. and {S{\'a}nchez}, S.~F. and
         {Kuncarayakti}, H. and {Pedraz}, S. and {Gonz{\'a}lez-Gait{\'a}n}, S. and
         {Stanishev}, V. and {Dom{\'\i}nguez}, I. and {Moreno-Raya}, M.~E. and
         {Wood-Vasey}, W.~M. and {Mour{\~a}o}, A.~M. and {Ponder}, K.~A. and
         {Badenes}, C. and {Moll{\'a}}, M. and {L{\'o}pez-S{\'a}nchez}, A.~R. and
         {Rosales-Ortega}, F.~F. and {V{\'\i}lchez}, J.~M. and
         {Garc{\'\i}a-Benito}, R. and {Marino}, R.~A.},
        title = "{PISCO: The PMAS/PPak Integral-field Supernova Hosts Compilation}",
      journal = {\apj},
         year = 2018,
        month = mar,
       volume = {855},
       number = {2},
          eid = {107},
        pages = {107},
          doi = {10.3847/1538-4357/aaaf20},
archivePrefix = {arXiv},
       eprint = {1802.01589},
 primaryClass = {astro-ph.GA},
       adsurl = {https://ui.adsabs.harvard.edu/abs/2018ApJ...855..107G}
}

@ARTICLE{1981PASP...93....5B,
   author = {{Baldwin}, J.~A. and {Phillips}, M.~M. and {Terlevich}, R.},
    title = "{Classification parameters for the emission-line spectra of extragalactic objects}",
  journal = {\pasp},
     year = 1981,
    month = feb,
   volume = 93,
    pages = {5-19},
      doi = {10.1086/130766},
   adsurl = {http://adsabs.harvard.edu/abs/1981PASP...93....5B}
}

@ARTICLE{2001ApJ...556..121K,
   author = {{Kewley}, L.~J. and {Dopita}, M.~A. and {Sutherland}, R.~S. and 
	{Heisler}, C.~A. and {Trevena}, J.},
    title = "{Theoretical Modeling of Starburst Galaxies}",
  journal = {\apj},
   eprint = {arXiv:astro-ph/0106324},
     year = 2001,
    month = jul,
   volume = 556,
    pages = {121-140},
      doi = {10.1086/321545},
   adsurl = {http://adsabs.harvard.edu/abs/2001ApJ...556..121K}
}

@ARTICLE{2014A&A...563A..49S,
   author = {{S{\'a}nchez}, S.~F. and {Rosales-Ortega}, F.~F. and {Iglesias-P{\'a}ramo}, J. and 
	{Moll{\'a}}, M. and {Barrera-Ballesteros}, J. and {Marino}, R.~A. and 
	{P{\'e}rez}, E. and {S{\'a}nchez-Blazquez}, P. and {Gonz{\'a}lez Delgado}, R. and 
	{Cid Fernandes}, R. and {de Lorenzo-C{\'a}ceres}, A. and {Mendez-Abreu}, J. and 
	{Galbany}, L. and {Falcon-Barroso}, J. and {Miralles-Caballero}, D. and 
	{Husemann}, B. and {Garc{\'{\i}}a-Benito}, R. and {Mast}, D. and 
	{Walcher}, C.~J. and {Gil de Paz}, A. and {Garc{\'{\i}}a-Lorenzo}, B. and 
	{Jungwiert}, B. and {V{\'{\i}}lchez}, J.~M. and {J{\'{\i}}lkov{\'a}}, L. and 
	{Lyubenova}, M. and {Cortijo-Ferrero}, C. and {D{\'{\i}}az}, A.~I. and 
	{Wisotzki}, L. and {M{\'a}rquez}, I. and {Bland-Hawthorn}, J. and 
	{Ellis}, S. and {van de Ven}, G. and {Jahnke}, K. and {Papaderos}, P. and 
	{Gomes}, J.~M. and {Mendoza}, M.~A. and {L{\'o}pez-S{\'a}nchez}, {\'A}.~R.
	},
    title = "{A characteristic oxygen abundance gradient in galaxy disks unveiled with CALIFA}",
  journal = {\aap},
archivePrefix = "arXiv",
   eprint = {1311.7052},
 primaryClass = "astro-ph.CO",
     year = 2014,
    month = mar,
   volume = 563,
      eid = {A49},
    pages = {A49},
      doi = {10.1051/0004-6361/201322343},
   adsurl = {http://adsabs.harvard.edu/abs/2014A%26A...563A..49S}
}

@ARTICLE{2022ApJ...927...78D,
       author = {{Dimitriadis}, Georgios and {Foley}, Ryan J. and {Arendse}, Nikki and {Coulter}, David A. and {Jacobson-Gal{\'a}n}, Wynn V. and {Siebert}, Matthew R. and {Izzo}, Luca and {Jones}, David O. and {Kilpatrick}, Charles D. and {Pan}, Yen-Chen and {Taggart}, Kirsty and {Auchettl}, Katie and {Gall}, Christa and {Hjorth}, Jens and {Kasen}, Daniel and {Piro}, Anthony L. and {Raimundo}, Sandra I. and {Ramirez-Ruiz}, Enrico and {Rest}, Armin and {Swift}, Jonathan J. and {Woosley}, Stan E.},
        title = "{A Carbon/Oxygen-dominated Atmosphere Days after Explosion for the ``Super-Chandrasekhar'' Type Ia SN 2020esm}",
      journal = {\apj},
         year = 2022,
        month = mar,
       volume = {927},
       number = {1},
          eid = {78},
        pages = {78},
          doi = {10.3847/1538-4357/ac4780},
archivePrefix = {arXiv},
       eprint = {2112.09930},
 primaryClass = {astro-ph.HE},
       adsurl = {https://ui.adsabs.harvard.edu/abs/2022ApJ...927...78D}
}

@ARTICLE{2003MNRAS.346.1055K,
   author = {{Kauffmann}, G. and {Heckman}, T.~M. and {Tremonti}, C. and 
	{Brinchmann}, J. and {Charlot}, S. and {White}, S.~D.~M. and 
	{Ridgway}, S.~E. and {Brinkmann}, J. and {Fukugita}, M. and 
	{Hall}, P.~B. and {Ivezi{\'c}}, {\v Z}. and {Richards}, G.~T. and 
	{Schneider}, D.~P.},
    title = "{The host galaxies of active galactic nuclei}",
  journal = {\mnras},
   eprint = {arXiv:astro-ph/0304239},
     year = 2003,
    month = dec,
   volume = 346,
    pages = {1055-1077},
      doi = {10.1111/j.1365-2966.2003.07154.x},
   adsurl = {http://adsabs.harvard.edu/abs/2003MNRAS.346.1055K}
}

@ARTICLE{1987ApJS...63..295V,
   author = {{Veilleux}, S. and {Osterbrock}, D.~E.},
    title = "{Spectral classification of emission-line galaxies}",
  journal = {\apjs},
     year = 1987,
    month = feb,
   volume = 63,
    pages = {295-310},
      doi = {10.1086/191166},
   adsurl = {http://adsabs.harvard.edu/abs/1987ApJS...63..295V}
}

@ARTICLE{2009ApJ...690.1745M,
       author = {{Maeda}, K. and {Kawabata}, K. and {Li}, W. and {Tanaka}, M. and {Mazzali}, P.~A. and {Hattori}, T. and {Nomoto}, K. and {Filippenko}, A.~V.},
        title = "{Subaru and Keck Observations of the Peculiar Type Ia Supernova 2006GZ at Late Phases}",
      journal = {\apj},
         year = 2009,
        month = jan,
       volume = {690},
       number = {2},
        pages = {1745-1752},
          doi = {10.1088/0004-637X/690/2/1745},
archivePrefix = {arXiv},
       eprint = {0808.0138},
 primaryClass = {astro-ph},
       adsurl = {https://ui.adsabs.harvard.edu/abs/2009ApJ...690.1745M}
}

@ARTICLE{2008AJ....136....1W,
       author = {{Wegner}, Gary and {Grogin}, Norman A.},
        title = "{Ages and Metallicities of Early-Type Void Galaxies from Line Strength Measurements}",
      journal = {\aj},
         year = 2008,
        month = jul,
       volume = {136},
       number = {1},
        pages = {1-17},
          doi = {10.1088/0004-6256/136/1/1},
archivePrefix = {arXiv},
       eprint = {0803.4519},
 primaryClass = {astro-ph},
       adsurl = {https://ui.adsabs.harvard.edu/abs/2008AJ....136....1W}
}

@ARTICLE{2016A&A...587A..70S,
       author = {{S{\'a}nchez-Menguiano}, L. and {S{\'a}nchez}, S.~F. and {P{\'e}rez}, I. and {Garc{\'\i}a-Benito}, R. and {Husemann}, B. and {Mast}, D. and {Mendoza}, A. and {Ruiz-Lara}, T. and {Ascasibar}, Y. and {Bland-Hawthorn}, J. and {Cavichia}, O. and {D{\'\i}az}, A.~I. and {Florido}, E. and {Galbany}, L. and {G{\'o}nzalez Delgado}, R.~M. and {Kehrig}, C. and {Marino}, R.~A. and {M{\'a}rquez}, I. and {Masegosa}, J. and {M{\'e}ndez-Abreu}, J. and {Moll{\'a}}, M. and {Del Olmo}, A. and {P{\'e}rez}, E. and {S{\'a}nchez-Bl{\'a}zquez}, P. and {Stanishev}, V. and {Walcher}, C.~J. and {L{\'o}pez-S{\'a}nchez}, {\'A}. R. and {Califa Collaboration}},
        title = "{Shape of the oxygen abundance profiles in CALIFA face-on spiral galaxies}",
      journal = {\aap},
         year = 2016,
        month = mar,
       volume = {587},
          eid = {A70},
        pages = {A70},
          doi = {10.1051/0004-6361/201527450},
archivePrefix = {arXiv},
       eprint = {1601.01542},
 primaryClass = {astro-ph.GA},
       adsurl = {https://ui.adsabs.harvard.edu/abs/2016A&A...587A..70S}
}

@ARTICLE{2016A&A...591A..48G,
       author = {{Galbany}, L. and {Stanishev}, V. and {Mour{\~a}o}, A.~M. and
         {Rodrigues}, M. and {Flores}, H. and {Walcher}, C.~J. and
         {S{\'a}nchez}, S.~F. and {Garc{\'\i}a-Benito}, R. and {Mast}, D. and
         {Badenes}, C. and {Gonz{\'a}lez Delgado}, R.~M. and {Kehrig}, C. and
         {Lyubenova}, M. and {Marino}, R.~A. and {Moll{\'a}}, M. and
         {Meidt}, S. and {P{\'e}rez}, E. and {van de Ven}, G. and
         {V{\'\i}lchez}, J.~M.},
        title = "{Nearby supernova host galaxies from the CALIFA survey. II. Supernova environmental metallicity}",
      journal = {\aap},
         year = 2016,
        month = jun,
       volume = {591},
          eid = {A48},
        pages = {A48},
          doi = {10.1051/0004-6361/201528045},
archivePrefix = {arXiv},
       eprint = {1603.07808},
 primaryClass = {astro-ph.GA},
       adsurl = {https://ui.adsabs.harvard.edu/abs/2016A&A...591A..48G}
}

@ARTICLE{2014A&A...572A..38G,
       author = {{Galbany}, L. and {Stanishev}, V. and {Mour{\~a}o}, A.~M. and
         {Rodrigues}, M. and {Flores}, H. and {Garc{\'\i}a-Benito}, R. and
         {Mast}, D. and {Mendoza}, M.~A. and {S{\'a}nchez}, S.~F. and
         {Badenes}, C. and {Barrera-Ballesteros}, J. and {Bland-Hawthorn}, J. and
         {Falc{\'o}n-Barroso}, J. and {Garc{\'\i}a-Lorenzo}, B. and
         {Gomes}, J.~M. and {Gonz{\'a}lez Delgado}, R.~M. and {Kehrig}, C. and
         {Lyubenova}, M. and {L{\'o}pez-S{\'a}nchez}, A.~R. and
         {de Lorenzo-C{\'a}ceres}, A. and {Marino}, R.~A. and {Meidt}, S. and
         {Moll{\'a}}, M. and {Papaderos}, P. and {P{\'e}rez-Torres}, M.~A. and
         {Rosales-Ortega}, F.~F. and {van de Ven}, G.},
        title = "{Nearby supernova host galaxies from the CALIFA Survey. I. Sample, data analysis, and correlation to star-forming regions}",
      journal = {\aap},
         year = 2014,
        month = dec,
       volume = {572},
          eid = {A38},
        pages = {A38},
          doi = {10.1051/0004-6361/201424717},
archivePrefix = {arXiv},
       eprint = {1409.1623},
 primaryClass = {astro-ph.GA},
       adsurl = {https://ui.adsabs.harvard.edu/abs/2014A&A...572A..38G}
}

@ARTICLE{2011ApJ...737L..24K,
       author = {{Khan}, Rubab and {Stanek}, K.~Z. and {Stoll}, R. and {Prieto}, J.~L.},
        title = "{Super-Chandrasekhar SNe Ia Strongly Prefer Metal-poor Environments}",
      journal = {\apjl},
         year = 2011,
        month = aug,
       volume = {737},
       number = {1},
          eid = {L24},
        pages = {L24},
          doi = {10.1088/2041-8205/737/1/L24},
archivePrefix = {arXiv},
       eprint = {1106.3071},
 primaryClass = {astro-ph.SR},
       adsurl = {https://ui.adsabs.harvard.edu/abs/2011ApJ...737L..24K}
}

@ARTICLE{2020ApJ...892..153M,
       author = {{Modjaz}, Maryam and {Bianco}, Federica B. and {Siwek}, Magdalena and {Huang}, Shan and {Perley}, Daniel A. and {Fierroz}, David and {Liu}, Yu-Qian and {Arcavi}, Iair and {Gal-Yam}, Avishay and {Filippenko}, Alexei V. and {Blagorodnova}, Nadia and {Cenko}, Bradley S. and {Kasliwal}, Mansi and {Kulkarni}, Shri and {Schulze}, Steve and {Taggart}, Kirsty and {Zheng}, Weikang},
        title = "{Host Galaxies of Type Ic and Broad-lined Type Ic Supernovae from the Palomar Transient Factory: Implications for Jet Production}",
      journal = {\apj},
         year = 2020,
        month = apr,
       volume = {892},
       number = {2},
          eid = {153},
        pages = {153},
          doi = {10.3847/1538-4357/ab4185},
archivePrefix = {arXiv},
       eprint = {1901.00872},
 primaryClass = {astro-ph.HE},
       adsurl = {https://ui.adsabs.harvard.edu/abs/2020ApJ...892..153M}
}

@ARTICLE{2009ATel.2037....1Q,
       author = {{Quimby}, R. and {Kasliwal}, M.~M. and {Nugent}, P. and {Howell}, D.~A. and {Rau}, A. and {Bhalerao}, V.},
        title = "{Palomar Transient Factory Discovers a Possible super-Chandrasekhar Type Ia Supernova}",
      journal = {The Astronomer's Telegram},
         year = 2009,
        month = apr,
       volume = {2037},
        pages = {1},
       adsurl = {https://ui.adsabs.harvard.edu/abs/2009ATel.2037....1Q}
}

@article{Kashi11,
 adsurl = {https://ui.adsabs.harvard.edu/abs/2011MNRAS.417.1466K},
 archiveprefix = {arXiv},
 author = {{Kashi}, Amit and {Soker}, Noam},
 doi = {10.1111/j.1365-2966.2011.19361.x},
 eprint = {1105.5698},
 journal = {\mnras},
 month = {October},
 number = {2},
 pages = {1466-1479},
 primaryclass = {astro-ph.SR},
 title = {{A circumbinary disc in the final stages of common envelope and the core-degenerate scenario for Type Ia supernovae}},
 volume = {417},
 year = {2011}
}

@article{Noebauer16,
 adsurl = {https://ui.adsabs.harvard.edu/abs/2016MNRAS.463.2972N},
 archiveprefix = {arXiv},
 author = {{Noebauer}, U.~M. and {Taubenberger}, S. and {Blinnikov}, S. and {Sorokina}, E. and {Hillebrandt}, W.},
 doi = {10.1093/mnras/stw2197},
 eprint = {1609.00241},
 journal = {\mnras},
 month = {December},
 number = {3},
 pages = {2972-2985},
 primaryclass = {astro-ph.HE},
 title = {{Type Ia supernovae within dense carbon- and oxygen-rich envelopes: a model for `Super-Chandrasekhar' explosions?}},
 volume = {463},
 year = {2016}
}

@article{Hachinger12,
 adsurl = {https://ui.adsabs.harvard.edu/abs/2012MNRAS.427.2057H},
 archiveprefix = {arXiv},
 author = {{Hachinger}, Stephan and {Mazzali}, Paolo A. and {Taubenberger}, Stefan and {Fink}, Michael and {Pakmor}, R{\"u}diger and {Hillebrandt}, Wolfgang and {Seitenzahl}, Ivo R.},
 doi = {10.1111/j.1365-2966.2012.22068.x},
 eprint = {1209.1339},
 journal = {\mnras},
 month = {December},
 number = {3},
 pages = {2057-2078},
 primaryclass = {astro-ph.SR},
 title = {{Spectral modelling of the 'super-Chandrasekhar' Type Ia SN 2009dc - testing a 2 M$_{{\ensuremath{\odot}}}$ white dwarf explosion model and alternatives}},
 volume = {427},
 year = {2012}
}

@article{Das13,
 adsurl = {https://ui.adsabs.harvard.edu/abs/2013PhRvL.110g1102D},
 archiveprefix = {arXiv},
 author = {{Das}, Upasana and {Mukhopadhyay}, Banibrata},
 doi = {10.1103/PhysRevLett.110.071102},
 eid = {071102},
 eprint = {1301.5965},
 journal = {\prl},
 month = {February},
 number = {7},
 pages = {071102},
 primaryclass = {astro-ph.SR},
 title = {{New Mass Limit for White Dwarfs: Super-Chandrasekhar Type Ia Supernova as a New Standard Candle}},
 volume = {110},
 year = {2013}
}

@ARTICLE{2023ApJ...953...13F,
       author = {{Fitz Axen}, Margot and {Nugent}, Peter},
        title = "{The Progenitors of Superluminous Type Ia Supernovae}",
      journal = {\apj},
         year = 2023,
        month = aug,
       volume = {953},
       number = {1},
          eid = {13},
        pages = {13},
          doi = {10.3847/1538-4357/acdd5d},
       eprint = {2306.07430},
 primaryclass = {astro-ph.HE},
       adsurl = {https://ui.adsabs.harvard.edu/abs/2023ApJ...953...13F}
}

@ARTICLE{2011AJ....141...19B,
       author = {{Burns}, Christopher R. and {Stritzinger}, Maximilian and {Phillips}, M.~M. and {Kattner}, ShiAnne and {Persson}, S.~E. and {Madore}, Barry F. and {Freedman}, Wendy L. and {Boldt}, Luis and {Campillay}, Abdo and {Contreras}, Carlos and {Folatelli}, Gaston and {Gonzalez}, Sergio and {Krzeminski}, Wojtek and {Morrell}, Nidia and {Salgado}, Francisco and {Suntzeff}, Nicholas B.},
        title = "{The Carnegie Supernova Project: Light-curve Fitting with SNooPy}",
      journal = {\aj},
         year = 2011,
        month = jan,
       volume = {141},
       number = {1},
          eid = {19},
        pages = {19},
          doi = {10.1088/0004-6256/141/1/19},
       eprint = {1010.4040},
 primaryclass = {astro-ph.CO},
       adsurl = {https://ui.adsabs.harvard.edu/abs/2011AJ....141...19B}
}

@ARTICLE{2017MNRAS.470.3566C,
       author = {{Chen}, Ting-Wan and {Smartt}, Stephen J. and {Yates}, Rob M. and {Nicholl}, Matt and {Kr{\"u}hler}, Thomas and {Schady}, Patricia and {Dennefeld}, Michel and {Inserra}, Cosimo},
        title = "{Superluminous supernova progenitors have a half-solar metallicity threshold}",
      journal = {\mnras},
         year = 2017,
        month = sep,
       volume = {470},
       number = {3},
        pages = {3566-3573},
          doi = {10.1093/mnras/stx1428},
archivePrefix = {arXiv},
       eprint = {1605.04925},
 primaryClass = {astro-ph.GA},
       adsurl = {https://ui.adsabs.harvard.edu/abs/2017MNRAS.470.3566C}
}

@article{Yoon05,
 adsurl = {https://ui.adsabs.harvard.edu/abs/2005A&A...435..967Y},
 archiveprefix = {arXiv},
 author = {{Yoon}, S. -C. and {Langer}, N.},
 doi = {10.1051/0004-6361:20042542},
 eprint = {astro-ph/0502133},
 journal = {\aap},
 month = {June},
 number = {3},
 pages = {967-985},
 primaryclass = {astro-ph},
 title = {{On the evolution of rapidly rotating massive white dwarfs towards supernovae or collapses}},
 volume = {435},
 year = {2005}
}

@ARTICLE{Hamuy96,
       author = {{Hamuy}, Mario and {Phillips}, M.~M. and {Suntzeff}, Nicholas B. and {Schommer}, Robert A. and {Maza}, Jose and {Aviles}, R.},
        title = "{The Absolute Luminosities of the Calan/Tololo Type IA Supernovae}",
      journal = {\aj},
         year = 1996,
        month = dec,
       volume = {112},
        pages = {2391},
          doi = {10.1086/118190},
archivePrefix = {arXiv},
       eprint = {astro-ph/9609059},
 primaryClass = {astro-ph},
       adsurl = {https://ui.adsabs.harvard.edu/abs/1996AJ....112.2391H}
}

@article{Phillips99,
 adsurl = {https://ui.adsabs.harvard.edu/abs/1999AJ....118.1766P},
 archiveprefix = {arXiv},
 author = {{Phillips}, M.~M. and {Lira}, Paulina and {Suntzeff}, Nicholas B. and {Schommer}, R.~A. and {Hamuy}, Mario and {Maza}, Jos{\'e}},
 doi = {10.1086/301032},
 eprint = {astro-ph/9907052},
 journal = {\aj},
 month = {October},
 number = {4},
 pages = {1766-1776},
 primaryclass = {astro-ph},
 title = {{The Reddening-Free Decline Rate Versus Luminosity Relationship for Type IA Supernovae}},
 volume = {118},
 year = {1999}
}

@article{Hoeflich:Khokhlov:96,
 adsurl = {https://ui.adsabs.harvard.edu/abs/1996ApJ...457..500H},
 archiveprefix = {arXiv},
 author = {{Hoeflich}, P. and {Khokhlov}, A.},
 doi = {10.1086/176748},
 eprint = {astro-ph/9602025},
 journal = {\apj},
 month = {February},
 pages = {500},
 primaryclass = {astro-ph},
 title = {{Explosion Models for Type IA Supernovae: A Comparison with Observed Light Curves, Distances, H 0, and Q 0}},
 volume = {457},
 year = {1996}
}

@article{Whelan73,
 adsurl = {https://ui.adsabs.harvard.edu/abs/1973ApJ...186.1007W},
 author = {{Whelan}, John and {Iben}, Icko, Jr.},
 doi = {10.1086/152565},
 journal = {\apj},
 month = {December},
 pages = {1007-1014},
 title = {{Binaries and Supernovae of Type I}},
 volume = {186},
 year = {1973}
}

@ARTICLE{Cikota19,
       author = {{Cikota}, Aleksandar and {Patat}, Ferdinando and {Wang}, Lifan and {Wheeler}, J. Craig and {Bulla}, Mattia and {Baade}, Dietrich and {H{\"o}flich}, Peter and {Cikota}, Stefan and {Clocchiatti}, Alejandro and {Maund}, Justyn R. and {Stevance}, Heloise F. and {Yang}, Yi},
        title = "{Linear spectropolarimetry of 35 Type Ia supernovae with VLT/FORS: an analysis of the Si II line polarization}",
      journal = {\mnras},
         year = 2019,
        month = nov,
       volume = {490},
       number = {1},
        pages = {578-599},
          doi = {10.1093/mnras/stz2322},
archivePrefix = {arXiv},
       eprint = {1908.07526},
 primaryClass = {astro-ph.HE},
       adsurl = {https://ui.adsabs.harvard.edu/abs/2019MNRAS.490..578C}
}

@ARTICLE{2019MNRAS.484.3785B,
       author = {{Brown}, J.~S. and {Stanek}, K.~Z. and {Holoien}, T.~W.-S. and {Kochanek}, C.~S. and {Shappee}, B.~J. and {Prieto}, J.~L. and {Dong}, S. and {Chen}, P. and {Thompson}, Todd A. and {Beacom}, J.~F. and {Stritzinger}, M.~D. and {Bersier}, D. and {Brimacombe}, J.},
        title = "{The relative specific Type Ia supernovae rate from three years of ASAS-SN}",
      journal = {\mnras},
         year = 2019,
        month = apr,
       volume = {484},
       number = {3},
        pages = {3785-3796},
          doi = {10.1093/mnras/stz258},
archivePrefix = {arXiv},
       eprint = {1810.00011},
 primaryClass = {astro-ph.GA},
       adsurl = {https://ui.adsabs.harvard.edu/abs/2019MNRAS.484.3785B}
}

@ARTICLE{Brown14,
       author = {{Brown}, Peter J. and {Kuin}, Paul and {Scalzo}, Richard and {Smitka}, Michael T. and {de Pasquale}, Massimiliano and {Holland}, Stephen and {Krisciunas}, Kevin and {Milne}, Peter and {Wang}, Lifan},
        title = "{Ultraviolet Observations of Super-Chandrasekhar Mass Type Ia Supernova Candidates with Swift UVOT}",
      journal = {\apj},
         year = 2014,
        month = may,
       volume = {787},
       number = {1},
          eid = {29},
        pages = {29},
          doi = {10.1088/0004-637X/787/1/29},
archivePrefix = {arXiv},
       eprint = {1404.0650},
 primaryClass = {astro-ph.HE},
       adsurl = {https://ui.adsabs.harvard.edu/abs/2014ApJ...787...29B}
}

@ARTICLE{Chakradhari14,
       author = {{Chakradhari}, N.~K. and {Sahu}, D.~K. and {Srivastav}, S. and {Anupama}, G.~C.},
        title = "{Supernova SN 2012dn: a spectroscopic clone of SN 2006gz}",
      journal = {\mnras},
         year = 2014,
        month = sep,
       volume = {443},
       number = {2},
        pages = {1663-1679},
          doi = {10.1093/mnras/stu1258},
archivePrefix = {arXiv},
       eprint = {1406.6139},
 primaryClass = {astro-ph.HE},
       adsurl = {https://ui.adsabs.harvard.edu/abs/2014MNRAS.443.1663C}
}

@ARTICLE{Parrent16,
       author = {{Parrent}, J.~T. and {Howell}, D.~A. and {Fesen}, R.~A. and {Parker}, S. and {Bianco}, F.~B. and {Dilday}, B. and {Sand}, D. and {Valenti}, S. and {Vink{\'o}}, J. and {Berlind}, P. and {Challis}, P. and {Milisavljevic}, D. and {Sanders}, N. and {Marion}, G.~H. and {Wheeler}, J.~C. and {Brown}, P. and {Calkins}, M.~L. and {Friesen}, B. and {Kirshner}, R. and {Pritchard}, T. and {Quimby}, R. and {Roming}, P.},
        title = "{Comparative analysis of SN 2012dn optical spectra: days -14 to +114}",
      journal = {\mnras},
         year = 2016,
        month = apr,
       volume = {457},
       number = {4},
        pages = {3702-3723},
          doi = {10.1093/mnras/stw239},
archivePrefix = {arXiv},
       eprint = {1603.03868},
 primaryClass = {astro-ph.HE},
       adsurl = {https://ui.adsabs.harvard.edu/abs/2016MNRAS.457.3702P}
}

@ARTICLE{Nagao18,
       author = {{Nagao}, Takashi and {Maeda}, Keiichi and {Yamanaka}, Masayuki},
        title = "{Polarization as a probe of dusty environments around Type Ia supernovae: radiative transfer models for SN 2012dn}",
      journal = {\mnras},
         year = 2018,
        month = jun,
       volume = {476},
       number = {4},
        pages = {4806-4813},
          doi = {10.1093/mnras/sty538},
archivePrefix = {arXiv},
       eprint = {1802.08954},
 primaryClass = {astro-ph.HE},
       adsurl = {https://ui.adsabs.harvard.edu/abs/2018MNRAS.476.4806N}
}

@ARTICLE{Silverman11,
       author = {{Silverman}, Jeffrey M. and {Ganeshalingam}, Mohan and {Li}, Weidong and {Filippenko}, Alexei V. and {Miller}, Adam A. and {Poznanski}, Dovi},
        title = "{Fourteen months of observations of the possible super-Chandrasekhar mass Type Ia Supernova 2009dc}",
      journal = {\mnras},
         year = 2011,
        month = jan,
       volume = {410},
       number = {1},
        pages = {585-611},
          doi = {10.1111/j.1365-2966.2010.17474.x},
archivePrefix = {arXiv},
       eprint = {1003.2417},
 primaryClass = {astro-ph.HE},
       adsurl = {https://ui.adsabs.harvard.edu/abs/2011MNRAS.410..585S}
}

@ARTICLE{Yamanaka09,
       author = {{Yamanaka}, M. and {Kawabata}, K.~S. and {Kinugasa}, K. and {Tanaka}, M. and {Imada}, A. and {Maeda}, K. and {Nomoto}, K. and {Arai}, A. and {Chiyonobu}, S. and {Fukazawa}, Y. and {Hashimoto}, O. and {Honda}, S. and {Ikejiri}, Y. and {Itoh}, R. and {Kamata}, Y. and {Kawai}, N. and {Komatsu}, T. and {Konishi}, K. and {Kuroda}, D. and {Miyamoto}, H. and {Miyazaki}, S. and {Nagae}, O. and {Nakaya}, H. and {Ohsugi}, T. and {Omodaka}, T. and {Sakai}, N. and {Sasada}, M. and {Suzuki}, M. and {Taguchi}, H. and {Takahashi}, H. and {Tanaka}, H. and {Uemura}, M. and {Yamashita}, T. and {Yanagisawa}, K. and {Yoshida}, M.},
        title = "{Early Phase Observations of Extremely Luminous Type Ia Supernova 2009dc}",
      journal = {\apjl},
         year = 2009,
        month = dec,
       volume = {707},
       number = {2},
        pages = {L118-L122},
          doi = {10.1088/0004-637X/707/2/L118},
archivePrefix = {arXiv},
       eprint = {0908.2059},
 primaryClass = {astro-ph.HE},
       adsurl = {https://ui.adsabs.harvard.edu/abs/2009ApJ...707L.118Y}
}

@ARTICLE{Tanaka10,
       author = {{Tanaka}, Masaomi and {Kawabata}, Koji S. and {Yamanaka}, Masayuki and {Maeda}, Keiichi and {Hattori}, Takashi and {Aoki}, Kentaro and {Nomoto}, Ken'ichi and {Iye}, Masanori and {Sasaki}, Toshiyuki and {Mazzali}, Paolo A. and {Pian}, Elena},
        title = "{Spectropolarimetry of Extremely Luminous Type Ia Supernova 2009dc: Nearly Spherical Explosion of Super-Chandrasekhar Mass White Dwarf}",
      journal = {\apj},
         year = 2010,
        month = may,
       volume = {714},
       number = {2},
        pages = {1209-1216},
          doi = {10.1088/0004-637X/714/2/1209},
archivePrefix = {arXiv},
       eprint = {0908.2057},
 primaryClass = {astro-ph.CO},
       adsurl = {https://ui.adsabs.harvard.edu/abs/2010ApJ...714.1209T}
}

@ARTICLE{Yuan10,
       author = {{Yuan}, F. and {Quimby}, R.~M. and {Wheeler}, J.~C. and {Vink{\'o}}, J. and {Chatzopoulos}, E. and {Akerlof}, C.~W. and {Kulkarni}, S. and {Miller}, J.~M. and {McKay}, T.~A. and {Aharonian}, F.},
        title = "{The Exceptionally Luminous Type Ia Supernova 2007if}",
      journal = {\apj},
         year = 2010,
        month = jun,
       volume = {715},
       number = {2},
        pages = {1338-1343},
          doi = {10.1088/0004-637X/715/2/1338},
archivePrefix = {arXiv},
       eprint = {1004.3329},
 primaryClass = {astro-ph.CO},
       adsurl = {https://ui.adsabs.harvard.edu/abs/2010ApJ...715.1338Y}
}

@ARTICLE{Scalzo10,
       author = {{Scalzo}, R.~A. and {Aldering}, G. and {Antilogus}, P. and {Aragon}, C. and {Bailey}, S. and {Baltay}, C. and {Bongard}, S. and {Buton}, C. and {Childress}, M. and {Chotard}, N. and {Copin}, Y. and {Fakhouri}, H.~K. and {Gal-Yam}, A. and {Gangler}, E. and {Hoyer}, S. and {Kasliwal}, M. and {Loken}, S. and {Nugent}, P. and {Pain}, R. and {P{\'e}contal}, E. and {Pereira}, R. and {Perlmutter}, S. and {Rabinowitz}, D. and {Rau}, A. and {Rigaudier}, G. and {Runge}, K. and {Smadja}, G. and {Tao}, C. and {Thomas}, R.~C. and {Weaver}, B. and {Wu}, C.},
        title = "{Nearby Supernova Factory Observations of SN 2007if: First Total Mass Measurement of a Super-Chandrasekhar-Mass Progenitor}",
      journal = {\apj},
         year = 2010,
        month = apr,
       volume = {713},
       number = {2},
        pages = {1073-1094},
          doi = {10.1088/0004-637X/713/2/1073},
archivePrefix = {arXiv},
       eprint = {1003.2217},
 primaryClass = {astro-ph.CO},
       adsurl = {https://ui.adsabs.harvard.edu/abs/2010ApJ...713.1073S}
}

@ARTICLE{2012ApJ...757...12S,
       author = {{Scalzo}, R. and {Aldering}, G. and {Antilogus}, P. and {Aragon}, C. and {Bailey}, S. and {Baltay}, C. and {Bongard}, S. and {Buton}, C. and {Canto}, A. and {Cellier-Holzem}, F. and {Childress}, M. and {Chotard}, N. and {Copin}, Y. and {Fakhouri}, H.~K. and {Gangler}, E. and {Guy}, J. and {Hsiao}, E.~Y. and {Kerschhaggl}, M. and {Kowalski}, M. and {Nugent}, P. and {Paech}, K. and {Pain}, R. and {Pecontal}, E. and {Pereira}, R. and {Perlmutter}, S. and {Rabinowitz}, D. and {Rigault}, M. and {Runge}, K. and {Smadja}, G. and {Tao}, C. and {Thomas}, R.~C. and {Weaver}, B.~A. and {Wu}, C. and {Nearby Supernova Factory}, The},
        title = "{A Search for New Candidate Super-Chandrasekhar-mass Type Ia Supernovae in the Nearby Supernova Factory Data Set}",
      journal = {\apj},
         year = 2012,
        month = sep,
       volume = {757},
       number = {1},
          eid = {12},
        pages = {12},
          doi = {10.1088/0004-637X/757/1/12},
archivePrefix = {arXiv},
       eprint = {1207.2695},
 primaryClass = {astro-ph.CO},
       adsurl = {https://ui.adsabs.harvard.edu/abs/2012ApJ...757...12S}
}

@ARTICLE{2019PASP..131a4001P,
       author = {{Phillips}, M.~M. and {Contreras}, Carlos and {Hsiao}, E.~Y. and {Morrell}, Nidia and {Burns}, Christopher R. and {Stritzinger}, Maximilian and {Ashall}, C. and {Freedman}, Wendy L. and {Hoeflich}, P. and {Persson}, S.~E. and {Piro}, Anthony L. and {Suntzeff}, Nicholas B. and {Uddin}, Syed A. and {Anais}, Jorge and {Baron}, E. and {Busta}, Luis and {Campillay}, Abdo and {Castell{\'o}n}, Sergio and {Corco}, Carlos and {Diamond}, T. and {Gall}, Christa and {Gonzalez}, Consuelo and {Holmbo}, Simon and {Krisciunas}, Kevin and {Roth}, Miguel and {Ser{\'o}n}, Jacqueline and {Taddia}, F. and {Torres}, Sim{\'o}n and {Anderson}, J.~P. and {Baltay}, C. and {Folatelli}, Gast{\'o}n and {Galbany}, L. and {Goobar}, A. and {Hadjiyska}, Ellie and {Hamuy}, Mario and {Kasliwal}, Mansi and {Lidman}, C. and {Nugent}, Peter E. and {Perlmutter}, S. and {Rabinowitz}, David and {Ryder}, Stuart D. and {Schmidt}, Brian P. and {Shappee}, B.~J. and {Walker}, Emma S.},
        title = "{Carnegie Supernova Project-II: Extending the Near-infrared Hubble Diagram for Type Ia Supernovae to z {\ensuremath{\sim}} 0.1}",
      journal = {\pasp},
         year = 2019,
        month = jan,
       volume = {131},
       number = {995},
        pages = {014001},
          doi = {10.1088/1538-3873/aae8bd},
archivePrefix = {arXiv},
       eprint = {1810.09252},
 primaryClass = {astro-ph.HE},
       adsurl = {https://ui.adsabs.harvard.edu/abs/2019PASP..131a4001P}
}

@ARTICLE{2020ApJ...895L...3A,
       author = {{Ashall}, C. and {Lu}, J. and {Burns}, C. and {Hsiao}, E.~Y. and {Stritzinger}, M. and {Suntzeff}, N.~B. and {Phillips}, M. and {Baron}, E. and {Contreras}, C. and {Davis}, S. and {Galbany}, L. and {Hoeflich}, P. and {Holmbo}, S. and {Morrell}, N. and {Karamehmetoglu}, E. and {Krisciunas}, K. and {Kumar}, S. and {Shahbandeh}, M. and {Uddin}, S.},
        title = "{Carnegie Supernova Project-II: A New Method to Photometrically Identify Sub-types of Extreme Type Ia Supernovae}",
      journal = {\apjl},
         year = 2020,
        month = may,
       volume = {895},
       number = {1},
          eid = {L3},
        pages = {L3},
          doi = {10.3847/2041-8213/ab8e37},
archivePrefix = {arXiv},
       eprint = {2003.11121},
 primaryClass = {astro-ph.HE},
       adsurl = {https://ui.adsabs.harvard.edu/abs/2020ApJ...895L...3A}
}

@ARTICLE{2014ApJ...795..142G,
       author = {{Gonz{\'a}lez-Gait{\'a}n}, S. and {Hsiao}, E.~Y. and {Pignata}, G. and {F{\"o}rster}, F. and {Guti{\'e}rrez}, C.~P. and {Bufano}, F. and {Galbany}, L. and {Folatelli}, G. and {Phillips}, M.~M. and {Hamuy}, M. and {Anderson}, J.~P. and {de Jaeger}, T.},
        title = "{Defining Photometric Peculiar Type Ia Supernovae}",
      journal = {\apj},
         year = 2014,
        month = nov,
       volume = {795},
       number = {2},
          eid = {142},
        pages = {142},
          doi = {10.1088/0004-637X/795/2/142},
archivePrefix = {arXiv},
       eprint = {1409.4811},
 primaryClass = {astro-ph.HE},
       adsurl = {https://ui.adsabs.harvard.edu/abs/2014ApJ...795..142G}
}

@ARTICLE{Phillips93,
       author = {{Phillips}, M.~M.},
        title = "{The Absolute Magnitudes of Type IA Supernovae}",
      journal = {\apjl},
         year = 1993,
        month = aug,
       volume = {413},
        pages = {L105},
          doi = {10.1086/186970},
       adsurl = {https://ui.adsabs.harvard.edu/abs/1993ApJ...413L.105P}
}

@ARTICLE{2021ApJ...920..107L,
       author = {{Lu}, J. and {Ashall}, C. and {Hsiao}, E.~Y. and {Hoeflich}, P. and {Galbany}, L. and {Baron}, E. and {Phillips}, M.~M. and {Contreras}, C. and {Burns}, C.~R. and {Suntzeff}, N.~B. and {Stritzinger}, M.~D. and {Anais}, J. and {Anderson}, J.~P. and {Brown}, P.~J. and {Busta}, L. and {Castell{\'o}n}, S. and {Davis}, S. and {Diamond}, T. and {Falco}, E. and {Gonzalez}, C. and {Hamuy}, M. and {Holmbo}, S. and {Holoien}, T.~W.-S. and {Krisciunas}, K. and {Kirshner}, R.~P. and {Kumar}, S. and {Kuncarayakti}, H. and {Marion}, G.~H. and {Morrell}, N. and {Persson}, S.~E. and {Piro}, A.~L. and {Prieto}, J.~L. and {Sand}, D.~J. and {Shahbandeh}, M. and {Shappee}, B.~J. and {Taddia}, F.},
        title = "{ASASSN-15hy: An Underluminous, Red 03fg-like Type Ia Supernova}",
      journal = {\apj},
         year = 2021,
        month = oct,
       volume = {920},
       number = {2},
          eid = {107},
        pages = {107},
          doi = {10.3847/1538-4357/ac1606},
archivePrefix = {arXiv},
       eprint = {2107.08150},
 primaryClass = {astro-ph.HE},
       adsurl = {https://ui.adsabs.harvard.edu/abs/2021ApJ...920..107L}
}

@ARTICLE{Ashall21,
       author = {{Ashall}, C. and {Lu}, J. and {Hsiao}, E.~Y. and {Hoeflich}, P. and {Phillips}, M.~M. and {Galbany}, L. and {Burns}, C.~R. and {Contreras}, C. and {Krisciunas}, K. and {Morrell}, N. and {Stritzinger}, M.~D. and {Suntzeff}, N.~B. and {Taddia}, F. and {Anais}, J. and {Baron}, E. and {Brown}, P.~J. and {Busta}, L. and {Campillay}, A. and {Castell{\'o}n}, S. and {Corco}, C. and {Davis}, S. and {Folatelli}, G. and {Forster}, F. and {Freedman}, W.~L. and {Gonzal{\'e}z}, C. and {Hamuy}, M. and {Holmbo}, S. and {Kirshner}, R.~P. and {Kumar}, S. and {Marion}, G.~H. and {Mazzali}, P. and {Morokuma}, T. and {Nugent}, P.~E. and {Persson}, S.~E. and {Piro}, A.~L. and {Roth}, M. and {Salgado}, F. and {Sand}, D.~J. and {Seron}, J. and {Shahbandeh}, M. and {Shappee}, B.~J.},
        title = "{Carnegie Supernova Project: The First Homogeneous Sample of ``Super-Chandrasekhar Mass''/2003fg-like Type Ia Supernova}",
      journal = {arXiv e-prints},
         year = 2021,
        month = jun,
          eid = {arXiv:2106.12140},
        pages = {arXiv:2106.12140},
archivePrefix = {arXiv},
       eprint = {2106.12140},
 primaryClass = {astro-ph.SR},
       adsurl = {https://ui.adsabs.harvard.edu/abs/2021arXiv210612140A}
}

@ARTICLE{1998A&A...331..815T,
       author = {{Tripp}, Robert},
        title = "{A two-parameter luminosity correction for Type IA supernovae}",
      journal = {\aap},
         year = 1998,
        month = mar,
       volume = {331},
        pages = {815-820},
       adsurl = {https://ui.adsabs.harvard.edu/abs/1998A&A...331..815T}
}

@ARTICLE{2009ARA&A..47..481A,
       author = {{Asplund}, Martin and {Grevesse}, Nicolas and {Sauval}, A. Jacques and {Scott}, Pat},
        title = "{The Chemical Composition of the Sun}",
      journal = {\araa},
         year = 2009,
        month = sep,
       volume = {47},
       number = {1},
        pages = {481-522},
          doi = {10.1146/annurev.astro.46.060407.145222},
archivePrefix = {arXiv},
       eprint = {0909.0948},
 primaryClass = {astro-ph.SR},
       adsurl = {https://ui.adsabs.harvard.edu/abs/2009ARA&A..47..481A}
}

@ARTICLE{2006PASP..118....2H,
       author = {{Hamuy}, Mario and {Folatelli}, Gast{\'o}n and {Morrell}, Nidia I. and {Phillips}, Mark M. and {Suntzeff}, Nicholas B. and {Persson}, S.~E. and {Roth}, Miguel and {Gonzalez}, Sergio and {Krzeminski}, Wojtek and {Contreras}, Carlos and {Freedman}, Wendy L. and {Murphy}, D.~C. and {Madore}, Barry F. and {Wyatt}, P. and {Maza}, Jos{\'e} and {Filippenko}, Alexei V. and {Li}, Weidong and {Pinto}, P.~A.},
        title = "{The Carnegie Supernova Project: The Low-Redshift Survey}",
      journal = {\pasp},
         year = 2006,
        month = jan,
       volume = {118},
       number = {839},
        pages = {2-20},
          doi = {10.1086/500228},
archivePrefix = {arXiv},
       eprint = {astro-ph/0512039},
 primaryClass = {astro-ph},
       adsurl = {https://ui.adsabs.harvard.edu/abs/2006PASP..118....2H}
}

@ARTICLE{2019MNRAS.488.5473T,
       author = {{Taubenberger}, S. and {Floers}, A. and {Vogl}, C. and {Kromer}, M. and {Spyromilio}, J. and {Aldering}, G. and {Antilogus}, P. and {Bailey}, S. and {Baltay}, C. and {Bongard}, S. and {Boone}, K. and {Buton}, C. and {Chotard}, N. and {Copin}, Y. and {Dixon}, S. and {Fouchez}, D. and {Fransson}, C. and {Gangler}, E. and {Gupta}, R.~R. and {Hachinger}, S. and {Hayden}, B. and {Hillebrandt}, W. and {Kim}, A.~G. and {Kowalski}, M. and {Leget}, P. -F. and {Leibundgut}, B. and {Mazzali}, P.~A. and {Noebauer}, U.~M. and {Nordin}, J. and {Pain}, R. and {Pakmor}, R. and {Pecontal}, E. and {Pereira}, R. and {Perlmutter}, S. and {Ponder}, K.~A. and {Rabinowitz}, D. and {Rigault}, M. and {Rubin}, D. and {Runge}, K. and {Saunders}, C. and {Smadja}, G. and {Tao}, C. and {Thomas}, R.~C.},
        title = "{SN 2012dn from early to late times: 09dc-like supernovae reassessed}",
      journal = {\mnras},
         year = 2019,
        month = oct,
       volume = {488},
       number = {4},
        pages = {5473-5488},
          doi = {10.1093/mnras/stz1977},
archivePrefix = {arXiv},
       eprint = {1907.06753},
 primaryClass = {astro-ph.HE},
       adsurl = {https://ui.adsabs.harvard.edu/abs/2019MNRAS.488.5473T}
}

@ARTICLE{Burns14,
       author = {{Burns}, Christopher R. and {Stritzinger}, Maximilian and {Phillips}, M.~M. and {Hsiao}, E.~Y. and {Contreras}, Carlos and {Persson}, S.~E. and {Folatelli}, Gaston and {Boldt}, Luis and {Campillay}, Abdo and {Castell{\'o}n}, Sergio and {Freedman}, Wendy L. and {Madore}, Barry F. and {Morrell}, Nidia and {Salgado}, Francisco and {Suntzeff}, Nicholas B.},
        title = "{The Carnegie Supernova Project: Intrinsic Colors of Type Ia Supernovae}",
      journal = {\apj},
         year = 2014,
        month = jul,
       volume = {789},
       number = {1},
          eid = {32},
        pages = {32},
          doi = {10.1088/0004-637X/789/1/32},
archivePrefix = {arXiv},
       eprint = {1405.3934},
 primaryClass = {astro-ph.CO},
       adsurl = {https://ui.adsabs.harvard.edu/abs/2014ApJ...789...32B}
}

@ARTICLE{2011MNRAS.412.2735T,
       author = {{Taubenberger}, S. and {Benetti}, S. and {Childress}, M. and {Pakmor}, R. and {Hachinger}, S. and {Mazzali}, P.~A. and {Stanishev}, V. and {Elias-Rosa}, N. and {Agnoletto}, I. and {Bufano}, F. and {Ergon}, M. and {Harutyunyan}, A. and {Inserra}, C. and {Kankare}, E. and {Kromer}, M. and {Navasardyan}, H. and {Nicolas}, J. and {Pastorello}, A. and {Prosperi}, E. and {Salgado}, F. and {Sollerman}, J. and {Stritzinger}, M. and {Turatto}, M. and {Valenti}, S. and {Hillebrandt}, W.},
        title = "{High luminosity, slow ejecta and persistent carbon lines: SN 2009dc challenges thermonuclear explosion scenarios}",
      journal = {\mnras},
         year = 2011,
        month = apr,
       volume = {412},
       number = {4},
        pages = {2735-2762},
          doi = {10.1111/j.1365-2966.2010.18107.x},
archivePrefix = {arXiv},
       eprint = {1011.5665},
 primaryClass = {astro-ph.SR},
       adsurl = {https://ui.adsabs.harvard.edu/abs/2011MNRAS.412.2735T}
}

@ARTICLE{2016MNRAS.458.2973P,
       author = {{Prentice}, S.~J. and {Mazzali}, P.~A. and {Pian}, E. and {Gal-Yam}, A. and {Kulkarni}, S.~R. and {Rubin}, A. and {Corsi}, A. and {Fremling}, C. and {Sollerman}, J. and {Yaron}, O. and {Arcavi}, I. and {Zheng}, W. and {Kasliwal}, M.~M. and {Filippenko}, A.~V. and {Cenko}, S.~B. and {Cao}, Y. and {Nugent}, P.~E.},
        title = "{The bolometric light curves and physical parameters of stripped-envelope supernovae}",
      journal = {\mnras},
         year = 2016,
        month = may,
       volume = {458},
       number = {3},
        pages = {2973-3002},
          doi = {10.1093/mnras/stw299},
archivePrefix = {arXiv},
       eprint = {1602.01736},
 primaryClass = {astro-ph.HE},
       adsurl = {https://ui.adsabs.harvard.edu/abs/2016MNRAS.458.2973P}
}

@ARTICLE{2008ApJ...681.1183K,
       author = {{Kewley}, Lisa J. and {Ellison}, Sara L.},
        title = "{Metallicity Calibrations and the Mass-Metallicity Relation for Star-forming Galaxies}",
      journal = {\apj},
         year = 2008,
        month = jul,
       volume = {681},
       number = {2},
        pages = {1183-1204},
          doi = {10.1086/587500},
archivePrefix = {arXiv},
       eprint = {0801.1849},
 primaryClass = {astro-ph},
       adsurl = {https://ui.adsabs.harvard.edu/abs/2008ApJ...681.1183K}
}

@ARTICLE{Dominguez01,
       author = {{Dom{\'\i}nguez}, Inma and {H{\"o}flich}, Peter and {Straniero}, Oscar},
        title = "{Constraints on the Progenitors of Type Ia Supernovae and Implications for the Cosmological Equation of State}",
      journal = {\apj},
         year = 2001,
        month = aug,
       volume = {557},
       number = {1},
        pages = {279-291},
          doi = {10.1086/321661},
archivePrefix = {arXiv},
       eprint = {astro-ph/0104257},
 primaryClass = {astro-ph},
       adsurl = {https://ui.adsabs.harvard.edu/abs/2001ApJ...557..279D}
}

@ARTICLE{2006Natur.443..308H,
       author = {{Howell}, D. Andrew and {Sullivan}, Mark and {Nugent}, Peter E. and {Ellis}, Richard S. and {Conley}, Alexander J. and {Le Borgne}, Damien and {Carlberg}, Raymond G. and {Guy}, Julien and {Balam}, David and {Basa}, Stephane and {Fouchez}, Dominique and {Hook}, Isobel M. and {Hsiao}, Eric Y. and {Neill}, James D. and {Pain}, Reynald and {Perrett}, Kathryn M. and {Pritchet}, Christopher J.},
        title = "{The type Ia supernova SNLS-03D3bb from a super-Chandrasekhar-mass white dwarf star}",
      journal = {\nat},
         year = 2006,
        month = sep,
       volume = {443},
       number = {7109},
        pages = {308-311},
          doi = {10.1038/nature05103},
archivePrefix = {arXiv},
       eprint = {astro-ph/0609616},
 primaryClass = {astro-ph},
       adsurl = {https://ui.adsabs.harvard.edu/abs/2006Natur.443..308H}
}

@ARTICLE{2011ApJ...733....3C,
       author = {{Childress}, M. and {Aldering}, G. and {Aragon}, C. and {Antilogus}, P. and {Bailey}, S. and {Baltay}, C. and {Bongard}, S. and {Buton}, C. and {Canto}, A. and {Chotard}, N. and {Copin}, Y. and {Fakhouri}, H.~K. and {Gangler}, E. and {Kerschhaggl}, M. and {Kowalski}, M. and {Hsiao}, E.~Y. and {Loken}, S. and {Nugent}, P. and {Paech}, K. and {Pain}, R. and {Pecontal}, E. and {Pereira}, R. and {Perlmutter}, S. and {Rabinowitz}, D. and {Runge}, K. and {Scalzo}, R. and {Thomas}, R.~C. and {Smadja}, G. and {Tao}, C. and {Weaver}, B.~A. and {Wu}, C.},
        title = "{Keck Observations of the Young Metal-poor Host Galaxy of the Super-Chandrasekhar-mass Type Ia Supernova SN 2007if}",
      journal = {\apj},
         year = 2011,
        month = may,
       volume = {733},
       number = {1},
          eid = {3},
        pages = {3},
          doi = {10.1088/0004-637X/733/1/3},
archivePrefix = {arXiv},
       eprint = {1103.2324},
 primaryClass = {astro-ph.CO},
       adsurl = {https://ui.adsabs.harvard.edu/abs/2011ApJ...733....3C}
}

@ARTICLE{2015MNRAS.449..917L,
       author = {{Leloudas}, G. and {Schulze}, S. and {Kr{\"u}hler}, T. and {Gorosabel}, J. and {Christensen}, L. and {Mehner}, A. and {de Ugarte Postigo}, A. and {Amor{\'\i}n}, R. and {Th{\"o}ne}, C.~C. and {Anderson}, J.~P. and {Bauer}, F.~E. and {Gallazzi}, A. and {He{\l}miniak}, K.~G. and {Hjorth}, J. and {Ibar}, E. and {Malesani}, D. and {Morell}, N. and {Vinko}, J. and {Wheeler}, J.~C.},
        title = "{Spectroscopy of superluminous supernova host galaxies. A preference of hydrogen-poor events for extreme emission line galaxies}",
      journal = {\mnras},
         year = 2015,
        month = may,
       volume = {449},
       number = {1},
        pages = {917-932},
          doi = {10.1093/mnras/stv320},
archivePrefix = {arXiv},
       eprint = {1409.8331},
 primaryClass = {astro-ph.GA},
       adsurl = {https://ui.adsabs.harvard.edu/abs/2015MNRAS.449..917L}
}

@ARTICLE{2007ApJ...669L..17H,
       author = {{Hicken}, M. and {Garnavich}, P.~M. and {Prieto}, J.~L. and {Blondin}, S. and {DePoy}, D.~L. and {Kirshner}, R.~P. and {Parrent}, J.},
        title = "{The Luminous and Carbon-rich Supernova 2006gz: A Double Degenerate Merger?}",
      journal = {\apjl},
         year = 2007,
        month = nov,
       volume = {669},
       number = {1},
        pages = {L17-L20},
          doi = {10.1086/523301},
archivePrefix = {arXiv},
       eprint = {0709.1501},
 primaryClass = {astro-ph},
       adsurl = {https://ui.adsabs.harvard.edu/abs/2007ApJ...669L..17H}
}

@ARTICLE{2020ApJ...900..140H,
       author = {{Hsiao}, E.~Y. and {Hoeflich}, P. and {Ashall}, C. and {Lu}, J. and {Contreras}, C. and {Burns}, C.~R. and {Phillips}, M.~M. and {Galbany}, L. and {Anderson}, J.~P. and {Baltay}, C. and {Baron}, E. and {Castell{\'o}n}, S. and {Davis}, S. and {Freedman}, Wendy L. and {Gall}, C. and {Gonzalez}, C. and {Graham}, M.~L. and {Hamuy}, M. and {Holoien}, T.~W. -S. and {Karamehmetoglu}, E. and {Krisciunas}, K. and {Kumar}, S. and {Kuncarayakti}, H. and {Morrell}, N. and {Moriya}, T.~J. and {Nugent}, P.~E. and {Perlmutter}, S. and {Persson}, S.~E. and {Piro}, A.~L. and {Rabinowitz}, D. and {Roth}, M. and {Shahbandeh}, M. and {Shappee}, B.~J. and {Stritzinger}, M.~D. and {Suntzeff}, N.~B. and {Taddia}, F. and {Uddin}, S.~A.},
        title = "{Carnegie Supernova Project II: The Slowest Rising Type Ia Supernova LSQ14fmg and Clues to the Origin of Super-Chandrasekhar/03fg-like Events}",
      journal = {\apj},
         year = 2020,
        month = sep,
       volume = {900},
       number = {2},
          eid = {140},
        pages = {140},
          doi = {10.3847/1538-4357/abaf4c},
archivePrefix = {arXiv},
       eprint = {2008.05614},
 primaryClass = {astro-ph.HE},
       adsurl = {https://ui.adsabs.harvard.edu/abs/2020ApJ...900..140H}
}

@INBOOK{2017hsn..book.1151H,
       author = {{Hoeflich}, Peter},
        title = "{Explosion Physics of Thermonuclear Supernovae and Their Signatures}",
    booktitle = {Handbook of Supernovae},
         year = 2017,
       editor = {{Alsabti}, Athem W. and {Murdin}, Paul},
        pages = {1151},
          doi = {10.1007/978-3-319-21846-5\_56},
       adsurl = {https://ui.adsabs.harvard.edu/abs/2017hsn..book.1151H}
}

@ARTICLE{2025A&A...694A...1R,
       author = {{Rigault}, M. and {Smith}, M. and {Goobar}, A. and {Maguire}, K. and {Dimitriadis}, G. and {Johansson}, J. and {Nordin}, J. and {Burgaz}, U. and {Dhawan}, S. and {Sollerman}, J. and {Regnault}, N. and {Kowalski}, M. and {Nugent}, P. and {Andreoni}, I. and {Amenouche}, M. and {Aubert}, M. and {Barjou-Delayre}, C. and {Bautista}, J. and {Bellm}, E. and {Betoule}, M. and {Bloom}, J.~S. and {Carreres}, B. and {Chen}, T.~X. and {Copin}, Y. and {Deckers}, M. and {de Jaeger}, T. and {Feinstein}, F. and {Fouchez}, D. and {Fremling}, C. and {Galbany}, L. and {Ginolin}, M. and {Graham}, M. and {Groom}, S.~L. and {Harvey}, L. and {Kasliwal}, M.~M. and {Kenworthy}, W.~D. and {Kim}, Y. -L. and {Kuhn}, D. and {Kulkarni}, S.~R. and {Lacroix}, L. and {Laher}, R.~R. and {Masci}, F.~J. and {M{\"u}ller-Bravo}, T.~E. and {Miller}, A. and {Osman}, M. and {Perley}, D. and {Popovic}, B. and {Purdum}, J. and {Qin}, Y. -J. and {Racine}, B. and {Reusch}, S. and {Riddle}, R. and {Rosnet}, P. and {Rosselli}, D. and {Ruppin}, F. and {Senzel}, R. and {Rusholme}, B. and {Schweyer}, T. and {Terwel}, J.~H. and {Townsend}, A. and {Tzanidakis}, A. and {Wold}, A. and {Yan}, L.},
        title = "{ZTF SN Ia DR2: Overview}",
      journal = {\aap},
         year = 2025,
        month = feb,
       volume = {694},
          eid = {A1},
        pages = {A1},
          doi = {10.1051/0004-6361/202450388},
archivePrefix = {arXiv},
       eprint = {2409.04346},
 primaryClass = {astro-ph.CO},
       adsurl = {https://ui.adsabs.harvard.edu/abs/2025A&A...694A...1R}
}

@ARTICLE{2026A&A...706A.252B,
       author = {{Bose}, S. and {Stritzinger}, M.~D. and {Malmgaard}, A. and {Miller}, C.~J. and {Elias-Rosa}, N. and {Fynbo}, J.~P.~U. and {Ashall}, C. and {Burns}, C.~R. and {DerKacy}, J.~M. and {Galbany}, L. and {Guti{\'e}rrez}, C.~P. and {Hoogendam}, W.~B. and {Hsiao}, E.~Y. and {Jensen}, E.~A.~M. and {Medler}, K. and {Alburai}, A. and {Anderson}, J. and {Baron}, E. and {Duarte}, J. and {Gromadzki}, M. and {Inserra}, C. and {Mazzali}, P.~A. and {M{\"u}ller-Bravo}, T.~E. and {Lundqvist}, P. and {Reguitti}, A. and {Salmaso}, I. and {Sand}, D.~J. and {Valerin}, G.},
        title = "{The Type Ia supernova 2021hem: A 2003fg-like event in an apparently hostless environment}",
      journal = {\aap},
         year = 2026,
        month = feb,
       volume = {706},
          eid = {A252},
        pages = {A252},
          doi = {10.1051/0004-6361/202558053},
archivePrefix = {arXiv},
       eprint = {2511.07529},
 primaryClass = {astro-ph.HE},
       adsurl = {https://ui.adsabs.harvard.edu/abs/2026A&A...706A.252B}
}

@ARTICLE{1982ApJ...258..790C,
       author = {{Chevalier}, R.~A.},
        title = "{Self-similar solutions for the interaction of stellar ejecta with an external medium.}",
      journal = {\apj},
         year = 1982,
        month = jul,
       volume = {258},
        pages = {790-797},
          doi = {10.1086/160126},
       adsurl = {https://ui.adsabs.harvard.edu/abs/1982ApJ...258..790C}
}

@ARTICLE{2023MNRAS.522.6035M,
       author = {{Moriya}, Takashi J. and {Mazzali}, Paolo A. and {Ashall}, Chris and {Pian}, Elena},
        title = "{Early excess emission in Type Ia supernovae from the interaction between supernova ejecta and their circumstellar wind}",
      journal = {\mnras},
         year = 2023,
        month = jul,
       volume = {522},
       number = {4},
        pages = {6035-6042},
          doi = {10.1093/mnras/stad1386},
archivePrefix = {arXiv},
       eprint = {2305.03363},
 primaryClass = {astro-ph.HE},
       adsurl = {https://ui.adsabs.harvard.edu/abs/2023MNRAS.522.6035M}
}

@ARTICLE{2011ApJ...729L...6C,
       author = {{Chevalier}, Roger A. and {Irwin}, Christopher M.},
        title = "{Shock Breakout in Dense Mass Loss: Luminous Supernovae}",
      journal = {\apjl},
         year = 2011,
        month = mar,
       volume = {729},
       number = {1},
          eid = {L6},
        pages = {L6},
          doi = {10.1088/2041-8205/729/1/L6},
archivePrefix = {arXiv},
       eprint = {1101.1111},
 primaryClass = {astro-ph.HE},
       adsurl = {https://ui.adsabs.harvard.edu/abs/2011ApJ...729L...6C}
}

@ARTICLE{2015MNRAS.447.2803L,
       author = {{Levanon}, Naveh and {Soker}, Noam and {Garc{\'\i}a-Berro}, Enrique},
        title = "{Constraining the double-degenerate scenario for Type Ia supernovae from merger ejected matter}",
      journal = {\mnras},
         year = 2015,
        month = mar,
       volume = {447},
       number = {3},
        pages = {2803-2809},
          doi = {10.1093/mnras/stu2580},
archivePrefix = {arXiv},
       eprint = {1408.1375},
 primaryClass = {astro-ph.SR},
       adsurl = {https://ui.adsabs.harvard.edu/abs/2015MNRAS.447.2803L}
}

@ARTICLE{Hoogendam2024,
       author = {{Hoogendam}, W.~B. and {Shappee}, B.~J. and {Brown}, P.~J. and {Tucker}, M.~A. and {Ashall}, C. and {Piro}, A.~L.},
        title = "{From out of the Blue: Swift Links 2002es-like, 2003fg-like, and Early Time Bump Type Ia Supernovae}",
      journal = {\apj},
         year = 2024,
        month = may,
       volume = {966},
       number = {1},
          eid = {139},
        pages = {139},
          doi = {10.3847/1538-4357/ad33ba},
archivePrefix = {arXiv},
       eprint = {2309.11563},
 primaryClass = {astro-ph.HE},
       adsurl = {https://ui.adsabs.harvard.edu/abs/2024ApJ...966..139H}
}

@ARTICLE{2026arXiv260622173G,
       author = {{Galbany}, Llu{\'\i}s and {Abelson}, Cullen and {Alburai}, Alaa and {Anderson}, Joseph P and {Ascasibar}, Yago and {Ashall}, Chris and {Badenes}, Carles and {Burns}, Chris and {Di{\'e}guez Gurn{\'e}s}, {\`E}lia and {Garc{\'\i}a Soto}, Albert and {Guti{\'e}rrez}, Claudia P. and {Hsiao}, Eric Y. and {Kuncarayakti}, Hanindyo and {Levan}, Andrew J. and {Lyman}, Jospeh and {Phillips}, Mark M. and {S{\'a}nchez}, Sebastian F. and {Sanfeliu}, Ramon and {Stritzinger}, Maximilian and {Steeghs}, Danny},
        title = "{The local ultraviolet signature of Type Ia supernova environments from HST and MUSE}",
      journal = {arXiv e-prints},
         year = 2026,
        month = jun,
          eid = {arXiv:2606.22173},
        pages = {arXiv:2606.22173},
          doi = {10.48550/arXiv.2606.22173},
archivePrefix = {arXiv},
       eprint = {2606.22173},
 primaryClass = {astro-ph.CO},
       adsurl = {https://ui.adsabs.harvard.edu/abs/2026arXiv260622173G}
}

@ARTICLE{2004ApJ...617..240K,
       author = {{Kobulnicky}, Henry A. and {Kewley}, Lisa J.},
        title = "{Metallicities of 0.3<z<1.0 Galaxies in the GOODS-North Field}",
      journal = {\apj},
         year = 2004,
        month = dec,
       volume = {617},
       number = {1},
        pages = {240-261},
          doi = {10.1086/425299},
archivePrefix = {arXiv},
       eprint = {astro-ph/0408128},
 primaryClass = {astro-ph},
       adsurl = {https://ui.adsabs.harvard.edu/abs/2004ApJ...617..240K}
}

@ARTICLE{2025A&A...694A..10D,
       author = {{Dimitriadis}, G. and {Burgaz}, U. and {Deckers}, M. and {Maguire}, K. and {Johansson}, J. and {Smith}, M. and {Rigault}, M. and {Frohmaier}, C. and {Sollerman}, J. and {Galbany}, L. and {Kim}, Y.-L. and {Liu}, C. and {Miller}, A.~A. and {Nugent}, P.~E. and {Alburai}, A. and {Chen}, P. and {Dhawan}, S. and {Ginolin}, M. and {Goobar}, A. and {Groom}, S.~L. and {Harvey}, L. and {Kenworthy}, W.~D. and {Kulkarni}, S.~R. and {Phan}, K. and {Popovic}, B. and {Riddle}, R.~L. and {Rusholme}, B. and {M{\"u}ller-Bravo}, T.~E. and {Nordin}, J. and {Terwel}, J.~H. and {Townsend}, A.},
        title = "{ZTF SN Ia DR2: The diversity and relative rates of the thermonuclear supernova population}",
      journal = {\aap},
         year = 2025,
        month = feb,
       volume = {694},
          eid = {A10},
        pages = {A10},
          doi = {10.1051/0004-6361/202451852},
archivePrefix = {arXiv},
       eprint = {2409.04200},
 primaryClass = {astro-ph.HE},
       adsurl = {https://ui.adsabs.harvard.edu/abs/2025A&A...694A..10D}
      
}

@ARTICLE{2014MNRAS.445.4427A,
       author = {{Ashall}, C. and {Mazzali}, P. and {Bersier}, D. and {Hachinger}, S. and {Phillips}, M. and {Percival}, S. and {James}, P. and {Maguire}, K.},
        title = "{Photometric and spectroscopic observations, and abundance tomography modelling of the Type Ia supernova SN 2014J located in M82}",
      journal = {\mnras},
         year = 2014,
        month = dec,
       volume = {445},
       number = {4},
        pages = {4427-4437},
          doi = {10.1093/mnras/stu1995},
archivePrefix = {arXiv},
       eprint = {1409.7066},
 primaryClass = {astro-ph.SR},
       adsurl = {https://ui.adsabs.harvard.edu/abs/2014MNRAS.445.4427A}
}

@ARTICLE{2018MNRAS.477..153A,
       author = {{Ashall}, C. and {Mazzali}, P.~A. and {Stritzinger}, M.~D. and {Hoeflich}, P. and {Burns}, C.~R. and {Gall}, C. and {Hsiao}, E.~Y. and {Phillips}, M.~M. and {Morrell}, N. and {Foley}, Ryan J.},
        title = "{On the type Ia supernovae 2007on and 2011iv: evidence for Chandrasekhar-mass explosions at the faint end of the luminosity-width relationship}",
      journal = {\mnras},
         year = 2018,
        month = jun,
       volume = {477},
       number = {1},
        pages = {153-174},
          doi = {10.1093/mnras/sty632},
archivePrefix = {arXiv},
       eprint = {1802.09460},
 primaryClass = {astro-ph.HE},
       adsurl = {https://ui.adsabs.harvard.edu/abs/2018MNRAS.477..153A}
}

@ARTICLE{2024ApJ...960...88S,
       author = {{Siebert}, Matthew R. and {Kwok}, Lindsey A. and {Johansson}, Joel and {Jha}, Saurabh W. and {Blondin}, St{\'e}phane and {Dessart}, Luc and {Foley}, Ryan J. and {Hillier}, D. John and {Larison}, Conor and {Pakmor}, R{\"u}diger and {Temim}, Tea and {Andrews}, Jennifer E. and {Auchettl}, Katie and {Badenes}, Carles and {Barna}, Barnabas and {Bostroem}, K. Azalee and {Brenner Newman}, Max J. and {Brink}, Thomas G. and {Bustamante-Rosell}, Mar{\'\i}a Jos{\'e} and {Camacho-Neves}, Yssavo and {Clocchiatti}, Alejandro and {Coulter}, David A. and {Davis}, Kyle W. and {Deckers}, Maxime and {Dimitriadis}, Georgios and {Dong}, Yize and {Farah}, Joseph and {Filippenko}, Alexei V. and {Fl{\"o}rs}, Andreas and {Fox}, Ori D. and {Garnavich}, Peter and {Padilla Gonzalez}, Estefania and {Graur}, Or and {Hambsch}, Franz-Josef and {Hosseinzadeh}, Griffin and {Howell}, D. Andrew and {Hughes}, John P. and {Kerzendorf}, Wolfgang E. and {Le Saux}, Xavier K. and {Maeda}, Keiichi and {Maguire}, Kate and {McCully}, Curtis and {Mihalenko}, Cassidy and {Newsome}, Megan and {O'Brien}, John T. and {Pearson}, Jeniveve and {Pellegrino}, Craig and {Pierel}, Justin D.~R. and {Polin}, Abigail and {Rest}, Armin and {Rojas-Bravo}, C{\'e}sar and {Sand}, David J. and {Schwab}, Michaela and {Shahbandeh}, Melissa and {Shrestha}, Manisha and {Smith}, Nathan and {Strolger}, Louis-Gregory and {Szalai}, Tam{\'a}s and {Taggart}, Kirsty and {Terreran}, Giacomo and {Terwel}, Jacco H. and {Tinyanont}, Samaporn and {Valenti}, Stefano and {Vink{\'o}}, J{\'o}zsef and {Wheeler}, J. Craig and {Yang}, Yi and {Zheng}, WeiKang and {Ashall}, Chris and {DerKacy}, James M. and {Galbany}, Llu{\'\i}s and {Hoeflich}, Peter and {Hsiao}, Eric and {de Jaeger}, Thomas and {Lu}, Jing and {Maund}, Justyn and {Medler}, Kyle and {Morrell}, Nidia and {Shappee}, Benjamin J. and {Stritzinger}, Maximilian and {Suntzeff}, Nicholas and {Tucker}, Michael and {Wang}, Lifan},
        title = "{Ground-based and JWST Observations of SN 2022pul. I. Unusual Signatures of Carbon, Oxygen, and Circumstellar Interaction in a Peculiar Type Ia Supernova}",
      journal = {\apj},
         year = 2024,
        month = jan,
       volume = {960},
       number = {1},
          eid = {88},
        pages = {88},
          doi = {10.3847/1538-4357/ad0975},
archivePrefix = {arXiv},
       eprint = {2308.12449},
 primaryClass = {astro-ph.HE},
       adsurl = {https://ui.adsabs.harvard.edu/abs/2024ApJ...960...88S}
}

@ARTICLE{2017MNRAS.464.3965W,
       author = {{Wang}, B. and {Zhou}, W.-H. and {Zuo}, Z.-Y. and {Li}, Y.-B. and {Luo}, X. and {Zhang}, J.-J. and {Liu}, D.-D. and {Wu}, C.-Y.},
        title = "{The core-degenerate scenario for the progenitors of Type Ia supernovae}",
      journal = {\mnras},
         year = 2017,
        month = feb,
       volume = {464},
       number = {4},
        pages = {3965-3971},
          doi = {10.1093/mnras/stw2646},
archivePrefix = {arXiv},
       eprint = {1610.03662},
 primaryClass = {astro-ph.SR},
       adsurl = {https://ui.adsabs.harvard.edu/abs/2017MNRAS.464.3965W}
}

\begin{appendix}

\section{A high-z galaxy next to the host galaxy of SN~2007if} \label{sec:07ifhighz}

\begin{figure}[!ht]
\centering
\includegraphics[width=0.9\columnwidth]{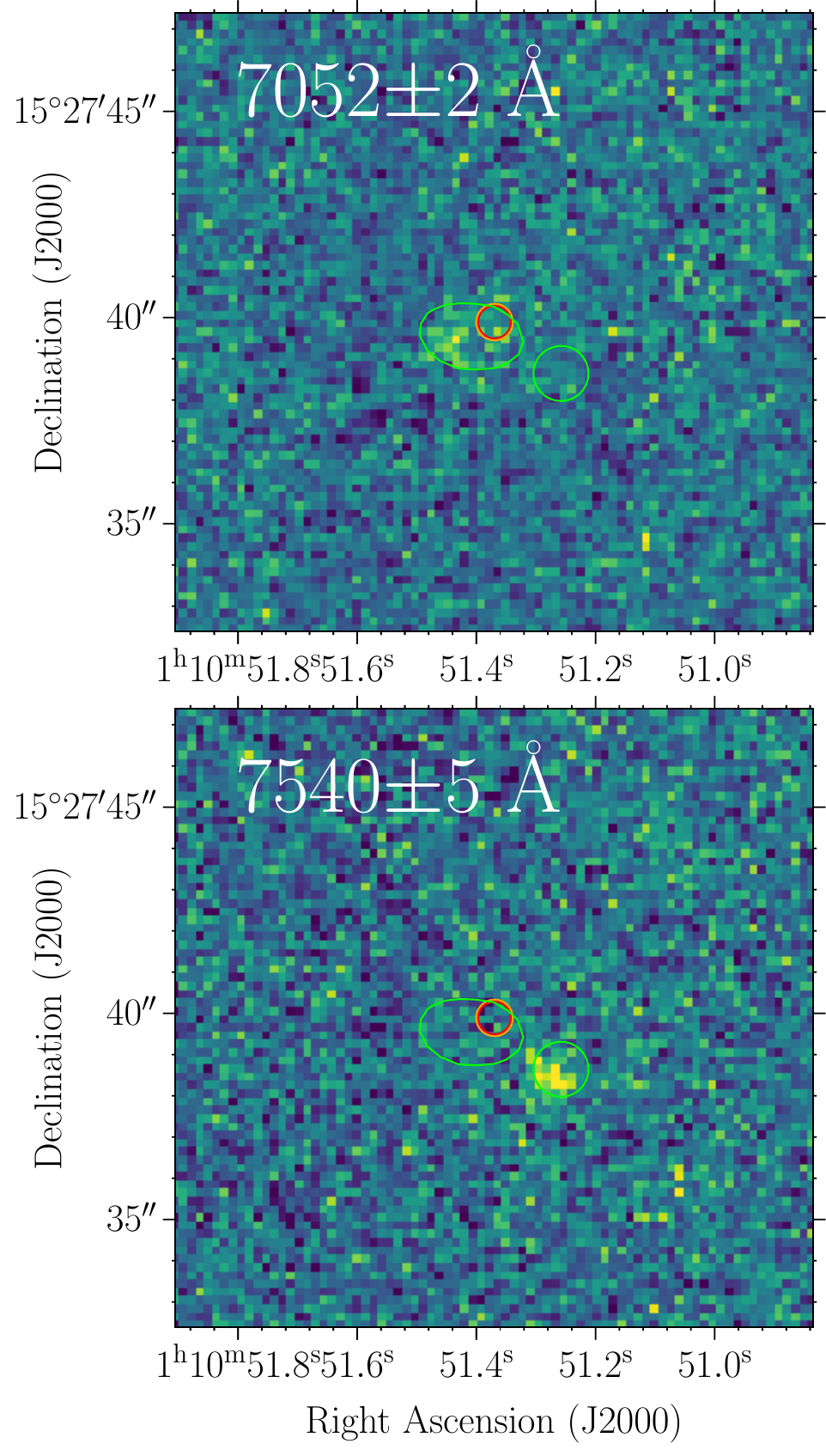}
\includegraphics[width=0.9\columnwidth]{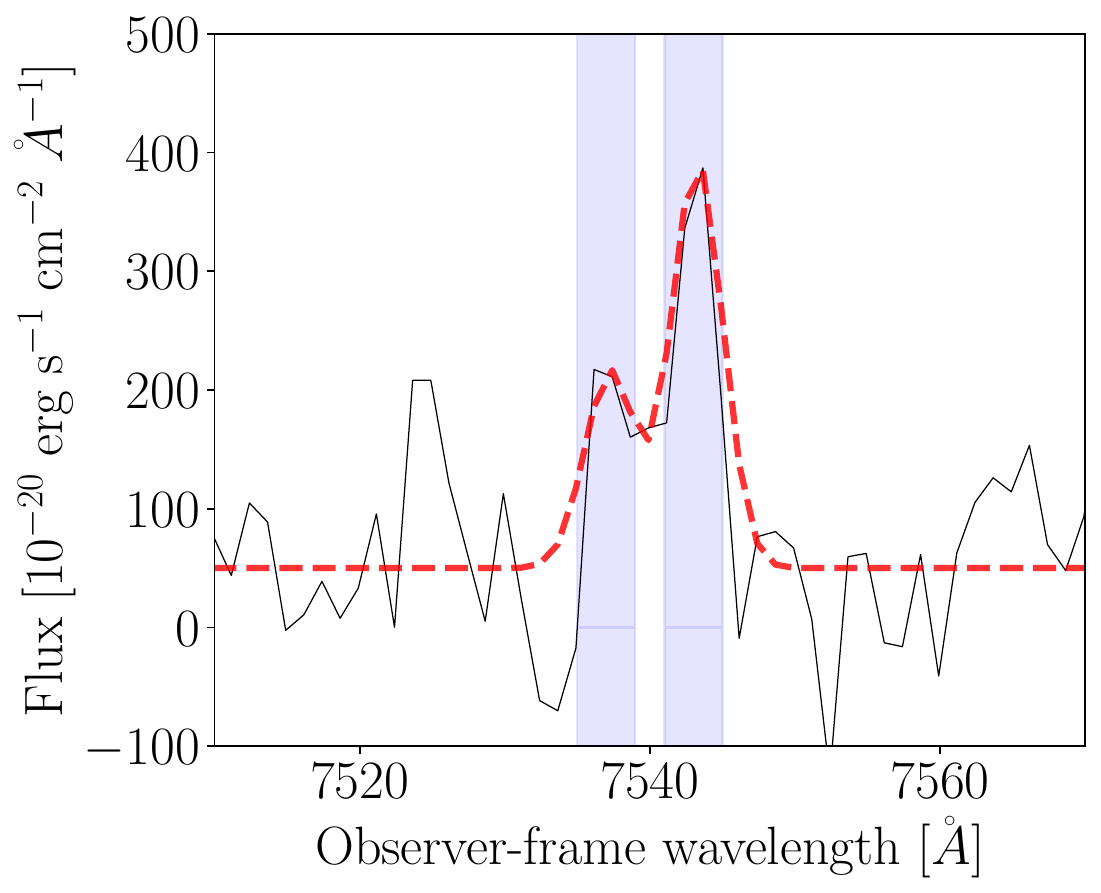}
\caption{{\it Top:} 4 \AA~synthetic narrow-band image centered at 7052 \AA~obtained from the SN~2007if host galaxy MUSE datacube. This wavelength corresponds to the H$\alpha$ emission of SN~2007if host. The red contour marks the SN~2007if location, while the green contours correspond to the SN host galaxy at z=0.0742 and the position of a background higher-z galaxy. {\it Middle:} Similarly, 10 \AA~synthetic narrow-band image centered at 7540 \AA, corresponding to the [\ion{O}{ii}] $\lambda\lambda$ 3727,29 emission line of the background galaxy at z=1.023. {\it Bottom:} Region around the [\ion{O}{ii}] emission line of the high-z galaxy spectrum extracted from the MUSE datacube, where the [\ion{O}{ii}] lines have been fitted with two Gaussian profiles.}
\label{fig:07ifhighz}
\end{figure}

\cite{2011ApJ...733....3C} reported the presence of a high-z galaxy a few arcsec southwest of SN~2007if host galaxy, from a trace of [\ion{O}{ii}] $\lambda\lambda$3727,29 emission present in the 2D spectra. We here confirm the presence of this galaxy at a redshift of $z=1.0228\pm0.0002$ in the MUSE datacube. The top panel of Fig. \ref{fig:07ifhighz} shows a 4 \AA~synthetic narrow-band image centered at 7052 \AA, the wavelength corresponding to the H$\alpha$ emission of the SN~2007if host galaxy, where a bright blob emerges at that location. Below, we show a similar 10 \AA~synthetic narrow-band image centered at 7540 \AA, the corresponding wavelength of the [\ion{O}{ii}] $\lambda\lambda$3727,29 emission line at that redshift, and the high-z galaxy blob appears about 2 arcsec from the SN~2007if host at the same location reported by \cite{2011ApJ...733....3C}. The bottom panel shows a region of the extracted spectrum at that location, and the [\ion{O}{ii}] doublet is clearly seen and fitted with a double Gaussian profile.

\section{Summary tables} \label{sec:tables}

\begin{table*}
\caption{Final global emission-line fluxes. Fluxes are in units of $10^{-17}$ erg s$^{-1}$ cm$^{-2}$. Upper 3$\sigma$ limits are shown without uncertainties. The [S II] entry is the sum of the two fitted doublet components.}
\label{tab:global_fluxes}
\resizebox{\textwidth}{!}{%
\begin{tabular}{lrrrrr}
\hline\hline
SN & H$\beta$ & [O III]$\lambda5007$ & H$\alpha$ & [N II]$\lambda6583$ & [S II]$\lambda\lambda6716,6731$ \\
\hline
2003fg & $117.76\pm10.37$ & $108.51\pm11.53$ & $\cdots$ & $\cdots$ & $\cdots$ \\
2006gz & $5516.98\pm24.52$ & $5708.65\pm19.03$ & $20242.93\pm15.62$ & $5143.69\pm12.42$ & $8866.54\pm23.96$ \\
2007if & $0.49\pm0.33$ & $0.88\pm0.22$ & $1.51\pm0.28$ & $<0.26$ & $<0.93$ \\
2009dc & $391.30\pm70.04$ & $523.66\pm67.53$ & $1209.62\pm8.91$ & $486.00\pm44.00$ & $1278.22\pm16.77$ \\
LSQ12gpw & $792.60\pm4.34$ & $425.40\pm3.81$ & $2715.35\pm3.69$ & $762.51\pm2.72$ & $887.64\pm3.72$ \\
2012dn\tablefootmark{a}  & {\it ($>$9858.48)}  & {\it ($>$7952.16)}  & {\it ($>$32336.01)} & {\it ($>$7941.87)} & {\it ($>$12402.91)} \\
2013ao & $12.03\pm0.13$ & $24.99\pm0.09$ & $35.04\pm0.07$ & $4.82\pm0.06$ & $6.29\pm0.06$ \\
LSQ14fmg & $45.99\pm1.34$ & $33.66\pm1.33$ & $98.47\pm0.78$ & $17.02\pm0.60$ & $50.38\pm0.98$ \\
CSS140126-120307-010132 & $5.66\pm0.78$ & $4.54\pm0.50$ & $8.49\pm0.33$ & $1.90\pm0.58$ & $10.80\pm0.78$ \\
CSS140501-170414+174839 & $236.20\pm3.87$ & $145.79\pm3.23$ & $752.14\pm3.80$ & $203.33\pm3.18$ & $236.64\pm6.21$ \\
2015M & $9.34\pm1.19$ & $9.12\pm2.14$ & $36.43\pm0.17$ & $<4.18$ & $<11.91$ \\
ASASSN-15hy & $310.35\pm1.95$ & $394.04\pm1.62$ & $821.19\pm1.06$ & $67.16\pm0.81$ & $314.75\pm1.67$ \\
ASASSN-15pz & $2558.59\pm5.44$ & $3491.98\pm4.73$ & $7407.66\pm2.75$ & $1012.24\pm1.96$ & $2543.20\pm3.73$ \\
2016gxp & $799.59\pm10.28$ & $988.94\pm7.27$ & $2710.88\pm5.28$ & $2807.77\pm6.07$ & $2282.65\pm10.40$ \\
2020esm & $1589.16\pm6.30$ & $1612.38\pm5.65$ & $4951.31\pm5.48$ & $988.36\pm4.22$ & $1924.56\pm8.49$ \\
2020hvf & $140.45\pm7.45$ & $195.27\pm5.84$ & $242.22\pm6.02$ & $59.01\pm1.93$ & $91.12\pm9.96$ \\
2020krv & $186.38\pm6.44$ & $226.32\pm7.97$ & $1310.92\pm7.22$ & $388.10\pm5.85$ & $478.99\pm6.80$ \\
2020sme & $209.61\pm5.79$ & $129.93\pm5.46$ & $1331.10\pm10.12$ & $587.42\pm6.78$ & $469.02\pm8.21$ \\
2021zny\tablefootmark{a} & {\it ($>$1509.26)} & {\it ($>$420.80)}  & {\it ($>$2084.30)} & {\it ($>$606.20)} & {\it ($>$1020.18)} \\
2022pul & $\cdots$ & $\cdots$ & $\cdots$ & $\cdots$ & $\cdots$ \\
\hline
\end{tabular}
}
\tablefoot{
\tablefoottext{a}{The global aperture extends beyond the IFS field of view. The emission-line fluxes are therefore incomplete and are quoted as lower limits.}
}
\end{table*}

\begin{table*}
\caption{Final local emission-line fluxes. Fluxes are in units of $10^{-17}$ erg s$^{-1}$ cm$^{-2}$. Upper 3$\sigma$ limits are shown without uncertainties. The [S II] entry is the sum of the two fitted doublet components.}
\label{tab:local_fluxes}
\resizebox{\textwidth}{!}{%
\begin{tabular}{lrrrrr}
\hline\hline
SN & H$\beta$ & [O III]$\lambda5007$ & H$\alpha$ & [N II]$\lambda6583$ & [S II]$\lambda\lambda6716,6731$ \\
\hline
2003fg & $0.26\pm0.04$ & $0.19\pm0.03$ & $\cdots$ & $\cdots$ & $\cdots$ \\
2006gz & $1.94\pm0.83$ & $2.66\pm0.53$ & $8.79\pm0.72$ & $1.21\pm0.89$ & $4.98\pm0.84$ \\
2007if & $0.19\pm0.05$ & $<0.19$ & $0.23\pm0.05$ & $<0.14$ & $<0.24$ \\
2009dc & $<1.50$ & $<1.71$ & $1.03\pm0.24$ & $<0.49$ & $<3.46$ \\
LSQ12gpw & $2.23\pm0.11$ & $2.81\pm0.15$ & $11.10\pm0.13$ & $2.58\pm0.08$ & $4.01\pm0.12$ \\
2012dn & $261.19\pm0.49$ & $394.10\pm0.50$ & $899.45\pm0.46$ & $143.99\pm0.27$ & $287.97\pm0.44$ \\
2013ao & $0.40\pm0.06$ & $0.97\pm0.12$ & $0.74\pm0.06$ & $0.16\pm0.04$ & $0.42\pm0.11$ \\
LSQ14fmg & $0.82\pm0.06$ & $0.54\pm0.08$ & $1.59\pm0.03$ & $0.32\pm0.02$ & $0.84\pm0.04$ \\
CSS140126-120307-010132 & $0.19\pm0.05$ & $0.16\pm0.04$ & $0.27\pm0.03$ & $0.15\pm0.06$ & $0.21\pm0.03$ \\
CSS140501-170414+174839 & $0.02\pm0.01$ & $0.06\pm0.01$ & $0.12\pm0.02$ & $0.03\pm0.01$ & $0.06\pm0.03$ \\
2015M & $<0.35$ & $0.36\pm0.08$ & $0.77\pm0.08$ & $0.08\pm0.03$ & $<0.26$ \\
ASASSN-15hy & $0.99\pm0.08$ & $0.76\pm0.06$ & $2.53\pm0.05$ & $0.28\pm0.03$ & $0.94\pm0.05$ \\
ASASSN-15pz & $2.00\pm0.14$ & $1.47\pm0.10$ & $4.28\pm0.07$ & $0.80\pm0.05$ & $4.40\pm0.11$ \\
2016gxp & $4.53\pm0.44$ & $3.46\pm0.38$ & $2.34\pm0.29$ & $0.36\pm0.37$ & $1.21\pm0.20$ \\
2020esm & $1.25\pm0.14$ & $1.72\pm0.15$ & $2.88\pm0.11$ & $0.44\pm0.09$ & $2.98\pm0.26$ \\
2020hvf & $3.33\pm0.96$ & $4.09\pm1.08$ & $19.47\pm1.01$ & $5.23\pm0.95$ & $10.85\pm1.04$ \\
2020krv & $0.65\pm0.19$ & $0.25\pm0.18$ & $4.04\pm0.14$ & $1.28\pm0.13$ & $1.78\pm0.19$ \\
2020sme & $0.30\pm0.11$ & $0.16\pm0.07$ & $1.21\pm0.13$ & $0.07\pm0.06$ & $0.17\pm0.07$ \\
2021zny & $0.75\pm0.23$ & $0.63\pm0.20$ & $1.36\pm0.18$ & $0.53\pm0.22$ & $1.22\pm0.18$ \\
2022pul & $<10.79$ & $<7.99$ & $<16.57$ & $<3.30$ & $<17.40$ \\
\hline
\end{tabular}
}
\end{table*}

\begin{table*}
\caption{Results from global spectra, including: star-formation rate and mass surface densities, specific SFR, H$\alpha$ equivalent width, and oxygen abundance in the O3N2 and D16 scales.} 
\label{tab:global_properties}
\resizebox{\textwidth}{!}{%
\begin{tabular}{lrrrrrr}
\hline\hline
SN & log$_{10}$ SFR ($M_\odot$ yr$^{-1}$) & Stellar Mass ($M_\odot$) & log$_{10}$ sSFR (yr$^{-1}$) & H$\alpha$ EW (\AA) & 12 + log$_{10}$ O/H$_{O3N2}$ & 12 + log$_{10}$ O/H$_{D16}$ \\ 
\hline
2003fg               & $\cdots$         & $(9.01\pm1.87)\times10^{10}$ & $\cdots$          & $\cdots$       & $\cdots$      & $\cdots$      \\ 
2006gz               & $0.353\pm0.004$  & $(1.07\pm0.21)\times10^{10}$ & $-9.677\pm0.030$  & 36.31$\pm$0.04 & $8.54\pm0.10$ & $8.38\pm0.11$ \\ 
2007if               & $-2.851\pm0.907$ & $(3.50\pm0.72)\times10^{7}$  & $-10.395\pm0.907$ & 8.33$\pm$1.78  & $<8.32$       & $\sim8.01$       \\  
2009dc               & $-1.161\pm0.027$.& $(2.04\pm0.44)\times10^{10}$ & $-11.471\pm0.040$ & 1.24$\pm$0.02  & $8.58\pm0.15$ & $8.28\pm0.17$ \\ 
LSQ12gpw                & $0.105\pm0.005$  & $(1.06\pm0.07)\times10^{10}$ & $-9.919\pm0.030$  & 3.86$\pm$2.93  & $8.64\pm0.10$ & $8.56\pm0.11$ \\ 
2012dn\tablefootmark{a}  & {\it ($>-$0.27)} & {\it ($>$8.5$\times$10$^{9}$)} & {\it($-$10.21$\pm$0.03)} & {\it (38.56$\pm$0.01)} & {\it (8.57$\pm$0.10)} & {\it (8.42$\pm$0.11)} \\
2013ao               & $-2.060\pm0.009$ & $(7.82\pm0.54)\times10^{7}$  & $-9.953\pm0.031$  & 40.43$\pm$0.23 & $8.35\pm0.11$ & $8.43\pm0.12$ \\ 
LSQ14fmg                & $-1.243\pm0.004$ & $(1.62\pm0.11)\times10^{9}$  & $-10.452\pm0.030$ & 17.42$\pm$0.16 & $8.53\pm0.11$ & $8.09\pm0.12$ \\ 
CSS140126-120307-010132 & $-2.165\pm0.018$ & $(1.33\pm0.09)\times10^{8}$  & $-10.289\pm0.035$ & 14.12$\pm$0.66 & $8.55\pm0.15$ & $7.81\pm0.26$ \\ 
CSS140501-170414+174839 & $-0.100\pm0.014$ & $(1.74\pm0.12)\times10^{10}$ & $-10.341\pm0.033$ & 19.19$\pm$0.11 & $8.62\pm0.11$ & $8.56\pm0.13$ \\ 
2015M                & $-2.363\pm0.103$ & $(2.47\pm0.17)\times10^{7}$  & $-9.756\pm0.108$  & 7.11$\pm$0.04  & $<8.33$       & $\sim8.07$       \\ 
ASASSN-15hy             & $-1.458\pm0.001$ & $(2.75\pm0.19)\times10^{8}$  & $-9.898\pm0.030$  & 29.18$\pm$0.05 & $8.35\pm0.10$ & $7.76\pm0.11$ \\ 
ASASSN-15pz             & $-0.682\pm0.002$ & $(5.73\pm0.40)\times10^{9}$  & $-10.441\pm0.030$ & 34.32$\pm$0.02 & $8.41\pm0.10$ & $8.14\pm0.10$ \\ 
2016gxp              & $\cdots$         & $(7.17\pm0.50)\times10^{10}$ & $\cdots$          & $\cdots$       & $\cdots$      & $\cdots$      \\ 
2020esm              & $-0.013\pm0.003$ & $(1.06\pm0.07)\times10^{10}$ & $-10.038\pm0.030$ & 40.53$\pm$0.07 & $8.51\pm0.10$ & $8.30\pm0.11$ \\ 
2020hvf              & $-3.001\pm0.011$ & $(1.44\pm0.16)\times10^{8}$  & $-11.161\pm0.032$ & 0.32$\pm$0.01  & $8.49\pm0.11$ & $8.42\pm0.15$ \\ 
2020krv              & $0.397\pm0.028$  & $(4.16\pm0.34)\times10^{9}$  & $-9.222\pm0.041$  & 4.06$\pm$2.77  & $8.55\pm0.13$ & $8.56\pm0.14$ \\ 
2020sme              & $0.194\pm0.023$  & $(3.67\pm0.25)\times10^{10}$ & $-10.371\pm0.038$ & 17.34$\pm$0.60 & $8.69\pm0.12$ & $8.80\pm0.14$ \\ 
2021zny\tablefootmark{a} & {\it ($>-$0.73)} & {\it ($>$2.6$\times$10$^{10}$)} & {\it($-$11.14$\pm$0.03)} & {\it (5.50$\pm$0.05)}  & {\it (8.74$\pm$0.10)} & {\it (8.40$\pm$0.11)} \\
2022pul              & $\cdots$         & $\cdots$                     & $\cdots$          & $\cdots$       & $\cdots$      & $\cdots$      \\ 
\hline
\end{tabular}}
\tablefoot{
\tablefoottext{a}{The global aperture extends beyond the IFS field-of-view, so the global-flux quantities (SFR and stellar mass) are unreliable and are not reported. The sSFR is a ratio, and the H$\alpha$EW and gas-phase oxygen abundances are line-flux ratios, unaffected by the missing flux.}
}
\end{table*}

\begin{table*}
\centering
\caption{Results from the local environment spectra, including: star-formation rate, mass, specific SFR, H$\alpha$ equivalent width, and oxygen abundance in the O3N2 and D16 scales.} 
\label{tab:local_properties}
\resizebox{\textwidth}{!}{%
\begin{tabular}{lrrrrrr}
\hline
SN & log$_{10}$ $\Sigma_{SFR}$ ($M_\odot$ yr$^{-1}$ kpc$^{-2}$) & $\Sigma_{M_\star}$ ($M_\odot\,{\rm kpc}^{-2}$) & log$_{10}$ sSFR (yr$^{-1}$) & H$\alpha$ EW (\AA) & 12 + log$_{10}$ O/H$_{O3N2}$ & 12 + log$_{10}$ O/H$_{D16}$ \\ 
\hline
2003fg               & $\cdots$       & $(1.05\pm0.07)\times10^{8}$ & $\cdots$        & $\cdots$      & $\cdots$      & $\cdots$      \\ 
2006gz               & $-2.84\pm0.37$ & $(1.66\pm0.11)\times10^{6}$ & $-9.06\pm0.37$  & $16.7\pm1.7$  & $8.41\pm0.12$ & $7.97\pm0.26$ \\
2007if               & $-3.77\pm0.10$ & $(1.08\pm0.07)\times10^{7}$ & $-10.80\pm0.11$ & $8.9\pm2.4$   & $<8.66$       & $>8.46$       \\ 
2009dc               & $-4.23\pm0.10$ & $\cdots$                    & $\cdots$        & $\cdots$      & $<8.61$       & $>7.78$       \\
LSQ12gpw                & $-1.98\pm0.04$ & $(4.51\pm0.31)\times10^{6}$ & $-8.63\pm0.05$  & $3.1\pm2.7$   & $8.50\pm0.13$ & $8.42\pm0.06$ \\
2012dn               & $-1.79\pm0.00$ & $(9.28\pm0.64)\times10^{6}$ & $-8.76\pm0.03$  & $254.4\pm0.6$ & $8.42\pm0.12$ & $8.26\pm0.10$ \\ 
2013ao               & $-3.75\pm0.04$ & $(7.71\pm0.53)\times10^{6}$ & $-10.64\pm0.05$ & $28.2\pm2.8$  & $8.40\pm0.14$ & $8.19\pm0.18$ \\ 
LSQ14fmg                & $-3.03\pm0.01$ & $(1.44\pm0.10)\times10^{7}$ & $-10.19\pm0.03$ & $21.9\pm0.5$  & $8.56\pm0.13$ & $8.16\pm0.15$ \\ 
CSS140126-120307-010132 & $-3.66\pm0.04$ & $(1.48\pm0.10)\times10^{7}$ & $-10.83\pm0.05$ & $10.1\pm1.4$  & $8.67\pm0.18$ & $8.56\pm0.22$ \\ 
CSS140501-170414+174839 & $-3.36\pm0.76$ & $(5.34\pm0.37)\times10^{6}$ & $-10.09\pm0.76$ & $34.1\pm7.4$  & $8.40\pm0.15$ & $8.35\pm1.10$ \\ 
2015M                & $-4.29\pm0.09$ & $(9.66\pm0.67)\times10^{5}$ & $-10.27\pm0.10$ & $6.8\pm2.3$   & $<8.41$       & $>7.96$       \\ 
ASASSN-15hy             & $-3.97\pm0.01$ & $(7.55\pm0.52)\times10^{6}$ & $-10.85\pm0.03$ & $14.9\pm0.4$  & $8.46\pm0.12$ & $7.97\pm0.18$ \\ 
ASASSN-15pz             & $-3.93\pm0.01$ & $(4.26\pm0.29)\times10^{5}$ & $-9.56\pm0.03$  & $8.5\pm0.1$   & $8.54\pm0.12$ & $7.79\pm0.15$ \\ 
2016gxp              & $-4.04\pm0.05$ & $(3.06\pm0.21)\times10^{7}$ & $-11.53\pm0.06$ & $0.2\pm0.1$   & $8.51\pm0.14$ & $8.01\pm0.11$ \\ 
2020esm              & $-3.32\pm0.02$ & $(9.18\pm0.63)\times10^{6}$ & $-10.28\pm0.03$ & $26.5\pm1.5$  & $8.43\pm0.14$ & $7.63\pm0.20$ \\ 
2020hvf              & $-3.52\pm0.29$ & $(2.94\pm0.20)\times10^{7}$ & $-10.98\pm0.29$ & $1.2\pm0.4$   & $8.53\pm0.16$ & $8.32\pm0.13$ \\ 
2020krv              & $-2.22\pm0.29$ & $(7.65\pm0.53)\times10^{7}$ & $-10.10\pm0.29$ & $\cdots$      & $8.72\pm0.18$ & $8.51\pm0.13$ \\ 
2020sme              & $-3.22\pm0.41$ & $(1.21\pm0.08)\times10^{7}$ & $-10.30\pm0.41$ & $41.6\pm19.7$ & $8.42\pm0.11$ & $8.06\pm0.10$ \\
2021zny              & $-3.92\pm0.07$ & $(1.12\pm0.08)\times10^{7}$ & $-10.97\pm0.07$ & $3.4\pm0.5$   & $8.62\pm0.10$ & $8.29\pm0.17$ \\ 
2022pul              & $<-4.74$       & $\cdots$                    & $\cdots$        & $\cdots$      & $<8.55$       & $>7.83$       \\ 
\hline
\end{tabular}
}
\end{table*}

\end{appendix}
\end{document}